\documentclass[reprint,aps,twocolumn,nofootinbib]{revtex4-1}
\usepackage{graphicx}
\usepackage[english]{babel}
\usepackage{amssymb,amsfonts,psfrag,amsmath,bm,bbm,indentfirst,subfigure,bbold}
\usepackage{color}
\usepackage[colorlinks,linkcolor=blue,urlcolor=blue,citecolor=blue]{hyperref}
\usepackage{ulem}
\usepackage{natbib}
\usepackage{amsmath}
\usepackage{accents}
\allowdisplaybreaks  

\usepackage{mathtools}

\makeatletter
\newcommand{\raisemath}[1]{\mathpalette{\raisem@th{#1}}}
\newcommand{\raisem@th}[3]{\raisebox{#1}{$#2#3$}}
\makeatother

\definecolor{darkgreen}{rgb}{0,0.5,0}

\newcommand{\Tr}{{\rm Tr}}

\newcommand{\e}{\varepsilon}
\newcommand{\ep}{\varepsilon'}

\newcommand{\rp}{{\color{red} +}}
\newcommand{\rmi}{{\color{red} -}}
\newcommand{\mydot}[1]{\accentset{\mbox{\bfseries .}}{#1}}

\newcommand{\dg}[1]{{\color{darkgreen} \textbf{#1}}}
\newcommand{\re}[1]{{\color{red} #1}}
\newcommand{\bl}[1]{{\color{blue} #1}}

\newcommand{\bgs}[1]{\mbox{\scriptsize{\boldmath$#1$}}}

\definecolor{purple}{rgb}{0.35,0,0.35}
\definecolor{orange}{rgb}{1,0.5,0}
\definecolor{darkred}{rgb}{.7,0,0}
\definecolor{darkgreen}{rgb}{0,.5,0}
\definecolor{darkblue}{rgb}{0.1,0.1,.6}
\definecolor{grey}{rgb}{.6,.6,.6}
\definecolor{dimgreen}{rgb}{0.2,0.6,0.1}
\definecolor{RoyalBlue}{rgb}{0,0.416,0.702}
\definecolor{DGLorange}{rgb}{0.949,0.573,0}

\renewcommand{\emph}[1]{\textit{#1}}

\begin{document}

\title{Keldysh Derivation of Oguri's Linear Conductance Formula for
  Interacting Fermions}

\author{Jan Heyder}
\author{Florian Bauer}
\author{Dennis Schimmel}
\author{Jan von Delft}

\affiliation{Arnold Sommerfeld Center for Theoretical Physics and Center for
NanoScience,
Ludwig-Maximilians-Universit\"at
M\"unchen, Theresienstrasse 37, D-80333 M\"unchen, Germany}

\date{April 12, 2017}
   
\begin{abstract}
We present a Keldysh-based derivation of a formula,
previously obtained by Oguri using the Matsubara formalisum,
 for the linear conductance through a central, interacting
    region coupled to non-interacting fermionic leads.
  Our starting point is the well-known Meir-Wingreen
    formula for the current, whose derivative w.r.t.\ to the 
    source-drain voltage yields the conductance. We perform this
      derivative analytically, by exploiting an exact flow equation
    from the functional renormalization group, which expresses the
    flow w.r.t.\ voltage of the self-energy in terms of the
    two-particle vertex. This yields a Keldysh-based
      formulation of Oguri's formula for the linear conductance,
  which facilitates applying
it in the context of approximation schemes formulated in the Keldysh formalism.
    (Generalizing our approach to the non-linear conductance is
      straightforward, but not pursued here.) -- We illustrate our
      linear conductance formula within the context of a model that
    has previously been shown to capture the essential physics of a
    quantum point contact in the regime of the 0.7 anomaly.  The
    model involves a tight-binding chain with a one-dimensional
    potential barrier and onsite interactions, which we treat using
    second order perturbation theory. We show that numerical
    costs can be reduced significantly by using a non-uniform lattice
    spacing, chosen such that the occurence of artificial bound states
    close to the upper band edge is avoided.   
\end{abstract}

\maketitle

\section{Introduction
\label{sec:cond_derivation}}

Two cornerstones of the theoretical description of transport through a
mesoscopic system are the Landauer-B\"uttiker \cite{Landauer1957} and Meir-Wingreen \cite{Meir1992}
formulas for the conductance. The Landauer-B\"uttiker formula describes
the conductance between two reservoirs connected by a central
region in the absence of interactions. The Meir-Wingreen
formula applies to the more general case that the central region
contains electron-electron interactions: it expresses the current, in
beautifully compact fashion, in terms of the Fermi functions of the
reservoirs, and the retarded, advanced and Keldysh components of the Green's
function for the central region.

To actually apply the Meir-Wingreen formula, these Green's functions
have to be calculated explicitly, which in general is a challenging
task. Depending on the intended application, a wide range of different
theoretical tools have been employed for this purpose. Much attention
has been lavished on the case of non-equilibrium transport through a
quantum dot described by a Kondo or Anderson model, where the central
interacting region consists of just a single localized spin or a
single electronic level, see
Refs. \cite{Eckel2010,Andergassen2010} for reviews. Here we are
interested in the less well-studied case of systems for which the
physics of the interacting region cannot be described by just a single
site, but rather requires an extended model, consisting of many
sites. 

We have recently used a model of this type in a paper that offers
  an explanation for the microscopic origin of the 0.7-anomaly in the
  conductance through a quantum point contact (QPC) \cite{Bauer2013}.  The model involves a
  tight-binding chain with a one-dimensional potential barrier and
  onsite interactions. In Ref.~\cite{Bauer2013} we used two approaches
  to treat interactions: second-order perturbation theory (SOPT) and
  the functional renormalization group (fRG). Our calculations of
  the linear conductance were based on an exact formula derived by
  Oguri \cite{Oguri2001,Oguri2003}. He started from the Kubo formula in the Matsubara formalism
  and performed the required analytical continuation of the
  two-particle vertex function occurring therein using Eliashberg
  theory \cite{Eliashberg1962}.  

  Since Oguri's formula for the linear conductance is exact, it can
  also be used when employing methods different from SOPT, for
  example fRG, to calculate the self-energy and two-particle
  vertex. If this is done in the Matsubara formalism, and if one
    attempts to capture the frequency dependence of the self-energy
  (as for the fRG calculations of Ref.~\cite{Bauer2013}), one is
  limited, in practice, to the case of zero temperature, because
  finite-temperature calculations would require an analytic
  continuation of numerical data from the imaginary to the real
  frequency axis, which is a mathematically ill-defined problem.  This
  problem can be avoided by calculating the self-energy and vertex
  directly on the real axis using the Keldysh formalism
  \cite{Keldysh1964,Kamenev2009}. However, to then calculate the
  linear conductance, the ingredients occuring in Oguri's formula
  would have to be transcribed into Keldysh language, and such a
  transcription is currently not available in the literature in easily
  accessible form.

  The main goal of the present paper is to derive a Keldysh
    version of Oguri's formula for the linear conductance by working
    entirely within the Keldysh formalism.  Our starting point is the
  Meir-Wingreen formula for the current, $J(V)$, with the conductance
  defined by $ \text{g} = \partial_V J$.  Rather than performing this
  derivative numerically, we here perform it \textit{analytically},
  based on the following central observation:
  The voltage derivative
  of the Green's functions that occur in the Meir-Wingreen formula,
  $\partial_V \mathcal{G}$, all involve the voltage derivative of the
  self-energy, $\partial_V \Sigma$. The latter can be expressed in
  terms of the two-particle vertex by using an exact flow equation
  from the fRG. (Analogous
  strategies have been used in the past for the
  dependence of the self-energy on temperature 
\cite{Honerkamp2001} or chemical potential \cite{Sauli2006,Jakobs2006}.)
 We show that it is
  possible to use this observation to derive Oguri's formula for the
  linear conductance, expressed in Keldysh notation, provided that the
  Hamiltonian is symmetric and conserves particle number.  Our
  argument evokes a Ward identity \cite{Ward1950}, following from
  $U(1)$-symmetry, which provides a relation between components of the
  self-energy and components of the vertex. 

  As an application of our Keldysh version of Oguri's conductance
  formula, we use Keldysh-SOPT to calculate the conductance through a
  QPC using the model of Ref.~\cite{Bauer2013}. Some
    results of this type were already presented in
    Ref.~\cite{Bauer2013}, but without offering a detailed
    account of the underlying formalism.  Providing these detail is
    one of the goals of the present paper. We also discuss some
    details of the numerical implementation of these calculations. In
    particular, we show that it is possible to greatly reduce the
  numerical costs by using a non-monotonic lattice spacing when
  formulating the discretized model.  We present results for the
  conductance as function of barrier height for different choices of
  interaction strength $U$, magnetic field $B$ and temperature $T$ and
  discuss both the successes and limitations of the SOPT scheme.

The paper is organized as follows: After introducing the general
interacting model Hamiltonian in Sec.~\ref{sec:model}, we present the
Keldysh derivation of Oguri's conductance formula in
Sec.~\ref{sec:conductance}. We set the stage for explicit conductance
calculations by expressing the self-energy and the two-particle vertex
within Keldysh SOPT in Sec.~\ref{sec:SOPT}. We introduce our the
1D-model of a QPC and discuss results for the conductance in
Sec.~\ref{sec:QPC}.  A detailed collection of definitions and
properties of both Green's and vertex functions in the Keldysh formalism
can be found in Appendix~\ref{App:Keldysh} and in
Ref.~\cite{Jakobs2010b} (in fact our paper closely follows the
notation used therein). A diagrammatic derivation of the fRG
flow-equation for the self-energy is given in Appendix~\ref{App:fRG}
and the Ward identity resulting from particle conservation is
presented in Appendix~\ref{App:ward}. In Appendix~\ref{App:FDT} we
perform an explicit calculation to verify the fluctuation-dissipation
theorem for the vertex-functions within SOPT. Finally, we apply the
method of finite differences in Appendix~\ref{App:MoFD}, to discretize
the continuous Hamiltonian using a non-constant discretization scheme.

\section{Microscopic model \label{sec:model}}

Within this work we consider a system composed of a finite central interacting region
coupled to two non-interacting semi-infinite fermionic leads, a left lead, with chemical potential $\mu^l$,
temperature $T^l$ and Fermi-distribution function $f^l$, and a right lead, with chemical potential $\mu^r$, temperature
$T^r$ and Fermi-distribution function $f^r$. The two leads are not directly connected to each other, but only via the central
region.  A similar setup was considered in Ref.~\cite{Meir1992} and
Ref.~\cite{Oguri2001}.

The general form of the model Hamiltonian reads
\begin{equation}
  {H} = {H}_{\dg{0}} + {H}_{\rm{int}} = \sum_{\color{blue}ij} h_{{\color{blue}ij}} d_{{\color{blue}i}}^{\dagger} d_{{\color{blue}j}}^{} +\sum_{{\color{blue}ij}}
  U_{{\color{blue}ij}} n_{{\color{blue}i}} n_{{\color{blue}j}} \; ,
 \label{eq:ModelGeneral}
\end{equation}
where $h_{{\color{blue}ij}}$ is a hermitian matrix, and
$U_{{\color{blue}ij}}$ is a real, symmetric matrix, non-zero only for
states $\color{blue}i$,$\color{blue}j$ within the central
region. $d_{{\color{blue}i}}^{\dagger}$/$d_{{\color{blue}i}}$
creates/destroys an electron in state $\color{blue} i$ and
$n_{{\color{blue}i}} =
d_{{\color{blue}i}}^{\dagger}d_{{\color{blue}i}}$ counts the number of
electrons in state $\color{blue} i$. While in general the index
$\color{blue} i$ can represent any set of quantum numbers we will
regard it as a composite index, referring, e.g. to the site and spin of an electron for a spinful lattice model. Note,
that the Hamiltonian conserves particle number, which is crucial in order to formulate a continuity equation for the charge current in the system.

We use a block representation of the matrix $h$ of the single-particle Hamiltonian
\begin{align}
{h} \!=\! 
 \left( \begin{array}{ccc}
 {h}_l &  {h}_{l c} & 0 \\
~\!\!{h}_{c l} & {h}_{\dg{0},c} & {h}_{c r}\\
0 & {h}_{rc} & {h}_r \\ 
 \end{array}
 \right),
 \label{eq:quadratic_H}
 \end{align}
where the indices $l$, $r$, and $c$ stand for the left lead, right lead, and central region, respectively.
For example, the spatial indices of the matrix ${h}_{\dg{0},c}$ both take values only within the central region, while the first spatial index of ${h}_{c l}$ takes a value within the central region and the second spatial index takes a value within the left lead.
The subscript $0$ emphasizes the absence of interactions in the definition of ${h}_{\dg{0},c}$ (the leads and the coupling between the leads and the central region are assumed non-interacting throughout the whole paper).

\section{Transport formulas\label{sec:conductance}}

We henceforth work in the Keldysh formalism. Our notation for Keldysh indices, which mostly follows that of Ref.~\cite{Jakobs2010b}, is set forth in detail in Appendix~\ref{App:Keldysh}, to allow the main text to focus only on the essential steps of the argument.

\subsection{Current formula}

We begin by retracing the derivation of the Meir-Wingreen formula. In
steady state the number of particles in the central region is
constant. Hence, the particle current from the left lead into the
central region is equal to the particle current from the central
region into the right lead, $J:=J_{l\to c} = J_{c\to r}$. [We remark
that this continuity equation can also be obtained by imposing the
invariance of the partition sum under a gauged $U(1)$ transformation,
following from particle conservation of the Hamiltonian, see
Appendix~\ref{App:ward}]. This allows us to focus on the current
through the interface between left lead and central region. Expressing
the current in terms of the time-derivative of the total particle
number operator of the left lead,
$n_l = \sum_{\bl{i}\in L}n_{\bl{i}}$, we obtain the Heisenberg
equation of motion
$J = -e \langle \dot{n}_l\rangle = -ie/\hbar \langle[{H},n_l]
\rangle$,
where $e$ is the electronic charge and $\hbar$ is Planck's
constant. For the Hamiltonian of Eq.~(\ref{eq:ModelGeneral}), the
current thus reads
\begin{align}
J & = -\frac{ie}{\hbar}\! \sum_{\underset{{\color{blue}j} \in C}{ {\color{blue}i} \in L}}\left[ h_{\bl{ij}}\langle  d_{\raisemath{-2.5pt}{{\color{blue}j}}}^{{\color{red}-}}(t) [d_{{\color{blue}i}}^{\color{red}+}]^{\dagger}(t) \rangle 
- h_{\bl{ji}}\langle  d_{\raisemath{-2.5pt}{{\color{blue}i}}}^{{\color{red}-}}(t) [d_{{\color{blue}j}}^{\color{red}+}]^{\dagger}(t) \rangle\right] \notag \\
& =\! \frac{e}{\hbar}\left[{\rm{Tr}}\{({h}_{l c} - {h}_{c l})G^{\re{-}|\re{+}}\}\right],
\label{eq:current_left_start}
\end{align}
with the interacting equal-time lesser Green's function $G_{\bl{i}|\bl{j}}^{\re{-}|\re{+}}\! =\! G_{\bl{i}|\bl{j}}^{\re{-}|\re{+}}(t|t)\! =\! -i \langle  d_{\raisemath{-2.5pt}{{\color{blue}i}}}^{{\color{red}-}}(t) [d_{{\color{blue}j}}^{\color{red}+}]^{\dagger}(t) \rangle 
$ (here we used time-translational invariance of the steady-state). Fourier transformation of Eq.~(\ref{eq:current_left_start}) yields 
\begin{align}
J  = 
 \frac{e}{h}\!\int \!\!d\e\! ~{\rm{Tr}}\big\{({h}_{l c} - {h}_{c l})\mathcal{G}^{\re{-}|\re{+}}(\e)\big\},
 \label{eq:MW_start}
\end{align}
with $h \!=\! 2\pi\hbar$. We introduced the symbol $\mathcal{G}$ for a Green's function that depends on a single frequency only (as opposed to the Fourier transform of the time-dependent Green's function $G$, which, in general, depends on two frequencies, see Appendix~\ref{App:Keldysh}, Eq.~(\ref{eq:used_functions}), for details).

Following the strategy of Ref.~\cite{Meir1992}, we use Dyson's equation, Eq.~(\ref{eq:Dyson_offdiagonal}), to express the current in terms of the central region Green's function $\mathcal{G}_c$ and rotate from the contour basis into the Keldysh basis (the explicit Keldysh rotation is given by Eq.~(\ref{eq:transformation}) and Eq.~(\ref{eq:keldysh_greens_relations})). This yields
\begin{equation}
J\! =\! \frac{i e}{2h}\! \int\!\! d \e
~\Tr \{\Gamma^l [\mathcal{G}_c^{\re{2}|\re{2}} - (1-2f^l)(\mathcal{G}_c^{\re{2}|\re{1}} -\mathcal{G}_c^{\re{1}|\re{2}} )]\},
\label{eq:current_mid}
\end{equation}
with retarded, $\mathcal{G}_c^{\re{2}|\re{1}}(\e)$, advanced,
$\mathcal{G}_c^{\re{1}|\re{2}}(\e)$, and Keldysh central region
Green's function, $\mathcal{G}_c^{\re{2}|\re{2}}(\e)$, and the
hybridization function
$\Gamma^l(\e)\! =\! i~ h_{cl}(g_l^{\re{2}|\re{1}}(\e)\! -\!
g_l^{\re{1}|\re{2}}(\e))h_{lc}$,
where $g_l(\e)$ is the Green's function of the isolated left
lead. Here and below we omit the frequency argument for all quantities
that depend on the integration variable
only. Eq.~(\ref{eq:current_mid}) is the celebrated Meir-Wingreen
formula for the current (c.f.\ Eq.~(6) in Ref. \cite{Meir1992}
for a symmetrized version thereof).

We now present a version of the Meir-Wingreen formula in terms of the interacting one-particle irreducible self-energy $\Sigma$ (with retarded, $\Sigma^{\re{1}|\re{2}}$, advanced, $\Sigma^{\re{2}|\re{1}}$ and Keldysh component $\Sigma^{\re{1}|\re{1}}$ [Eq.~(\ref{eq:Dyson_time}), Eq.~(\ref{eq:used_functions}), Eq.~(\ref{eq:matrix_representation})]). It can be derived by means of Dyson's equation, Eq.~(\ref{eq:Dyson}), which enables a reformulation of the Green's functions in Eq.~(\ref{eq:current_mid}) in terms of the hybridization functions $\Gamma$, the lead distribution functions $f$ and the self-energy $\Sigma$: 
\begin{flalign}
& \mathcal{G}_c^{\re{2}|\re{1}}\! -\! \mathcal{G}_c^{\re{1}|\re{2}}  = \mathcal{G}_c^{\re{2}|\re{1}}\big(\big[\mathcal{G}_c^{\re{1}|\re{2}}\big]^{-1}\!-\big[\mathcal{G}_c^{\re{2}|\re{1}}\big]^{-1}\big)\mathcal{G}_c^{\re{1}|\re{2}} & \notag \\
&\quad\quad\quad\quad~= \mathcal{G}_c^{\re{2}|\re{1}}\big(-i(\Gamma^l + \Gamma^r) +\! \Sigma^{\re{1}|\re{2}} - \Sigma^{\re{2}|\re{1}}\big) \mathcal{G}_c^{\re{1}|\re{2}},& \notag \\
&\mathcal{G}_c^{\re{2}|\re{2}}\!  =\! \mathcal{G}_c^{\re{2}|\re{1}}\!\big(-i\sum_{k=l,r}(1-2f^k)\Gamma^k +\! \Sigma^{\re{1}|\re{1}}\big)\mathcal{G}_c^{\re{1}|\re{2}}.& 
\end{flalign}
Hence, the current formula can be written as the sum of two terms,
\begin{align}
J = &  \frac{e}{h} \int \! d \varepsilon 
\Bigl[(f^l-f^r) \Tr \lbrace \Gamma^l \mathcal{G}_c^{\re{2}|\re{1}} \Gamma^r \mathcal{G}_c^{\re{1}|\re{2}}\rbrace + 
      \label{eq:current_end} \\
    & \; \; 
      \Bigl.  +  \frac{i}{2} \Tr\lbrace \Gamma^l \mathcal{G}_c^{\re{2}|\re{1}}\! \left(\! \Sigma^{\re{1}|\re{1}} - 
      (1\!-\!2f^l) ( \Sigma^{\re{1}|\re{2}} - \Sigma^{\re{2}|\re{1}})\right)\! \mathcal{G}_c^{\re{1}|\re{2}} \rbrace  \Bigr] . 
      \nonumber 
\end{align}
In equilibrium, i.e.\! $f\!:=\!f^l\!=\!f^r$, the current must fulfill $J\!=\!0$. With the first term of Eq.~(\ref{eq:current_end}) vanishing trivially, this imposes the fluctuation-dissipation theorem (FDT) for the self-energy at zero bias voltage, $\Sigma^{\re{1}|\re{1}} \!=\! (1\!-\!2f) ( \Sigma^{\re{1}|\re{2}}\! -\! \Sigma^{\re{2}|\re{1}})$. Note that a similar FDT can be formulated for the Green's function in Eq.~(\ref{eq:current_mid}).

\subsection{Differential conductance formula\label{sec:diff_cond}}

Differentiating Eq.~(\ref{eq:current_mid}) w.r.t.\ the source-drain voltage $V\!=\!(\mu_l-\mu_r)/e$, i.e.\ the voltage drop from the left to the right lead, provides the differential conductance $\text{g}_V = \partial_V J$. We denote derivatives w.r.t. frequency by a prime, e.g.\ ${f^l}' := \partial_\e f^l$, and derivatives w.r.t.\ the source-drain voltage by a dot, $\mydot{\mathcal{G}}_c := \partial_V \mathcal{G}_c$. Using Dyson's equation [Eq.~(\ref{eq:Dyson})], we can express the derivative of the Green's function in terms of derivatives of the self-energy:
\begin{align}
\mydot{\mathcal{G}}_c^{\re{\alpha}|\re{\alpha'}}& \!=\! \sum_{\re{\beta},\re{\beta'}} \mathcal{G}_c^{\re{\alpha}|\re{\beta'}} \mydot{\Sigma}^{\re{\beta'}|\re{\beta}}\mathcal{G}_c^{\re{\beta}|\re{\alpha'}} + S^{\re{\alpha}|\re{\alpha'}} \notag , \\
S^{\re{1}|\re{1}} &\! =\! S^{\re{1}|\re{2}} \!=\! S^{\re{2}|\re{1}}\! =\! 0 ~,~ S^{\re{2}|\re{2}} = \mathcal{G}_c^{\re{2}|\re{1}}\mydot{\Sigma}_{\rm{lead}}^{\re{1}|\re{1}}\mathcal{G}_c^{\re{1}|\re{2}}.
\end{align}
Here we introduced the socalled single scale propagator $S$ and the
lead self-energy
$\Sigma_{{\rm{lead}}}^{\re{1}|\re{1}}\!=\! -i\sum_{k =
  l,r}(1-2f^k)\Gamma^k$
[Eq.~(\ref{eq:lead_selfenergy})].  Hence, we can write the
differential conductance in the form
\begin{align}
\text{g}_V \! & = \! \frac{i e}{2h}\! \int\!\! d \e \Tr \Biggl\{
\Gamma^l \Biggl[
 \sum_{\re{\beta},\re{\beta'}} \mathcal{G}_c^{\re{\alpha}|\re{\beta'}} \mydot{\Sigma}^{\re{\beta'}|\re{\beta}}\mathcal{G}_c^{\re{\beta}|\re{\alpha'}} + S^{\re{2}|\re{2}} 
\Biggr.\Biggr.
\notag 
\\
& \hspace{5em} - (1-2f^l)(\mathcal{G}_c^{\re{2}|\re{1}} \mydot{\Sigma}^{\re{1}|\re{2}}\mathcal{G}_c^{\re{2}|\re{1}} - 
\mathcal{G}_c^{\re{1}|\re{2}} \mydot{\Sigma}^{\re{2}|\re{1}}\mathcal{G}_c^{\re{1}|\re{2}}) \notag \\ 
& \hspace{5em} \Biggl. \Biggl.  
+ 2\mydot{f^l}(\mathcal{G}_c^{\re{2}|\re{1}}-\mathcal{G}_c^{\re{1}|\re{2}}) 
\Biggr] \Biggr\}.
\label{eq:differential_conductance}
\end{align}
We specify the voltage via the chemical potentials in the leads,
$\mu^l=\mu +\alpha e V$ and $\mu^r=\mu +(\alpha-1)eV$, with
$\alpha\!\in\![0,1]$. This yields
\begin{align}
S^{\re{2}|\re{2}} = -2ie ~\mathcal{G}_c^{\re{2}|\re{1}}\!\left[\alpha {f^l}'\Gamma^l + (\alpha-1){f^r}'\Gamma^r\right]\mathcal{G}_c^{\re{1}|\re{2}}.
\label{eq:SSP}
\end{align} 
Note that in the special case $\alpha\!=\! 0$, i.e.\ if the voltage is
applied to the right lead only, the last term in
Eq.~(\ref{eq:differential_conductance}) vanishes and the differential
conductance takes a particularly simple form. This is a consequence of
our initial choice to express the current via the time derivative of
the left lead's occupation. 

Eq.~\eqref{eq:differential_conductance} for the differential
conductance of an interacting Fermi system involves derivatives of all
self-energy components, $\mydot{\Sigma}$. Below, we show how these can
be expressed in terms of the irreducible two-particle vertex
$\mathcal{L}$ and the single scale propagator $S$ using the fRG flow
equation for the self-energy. In this paper we apply this scheme 
to derive a Keldysh Kubo-type formula for the \textit{linear}
conductance (i.e.\ taking the limit $V \to 0$), which for a symmetric
Hamiltonian yields a Keldysh version of Oguri's formula. However, we
emphasize that an extension to finite bias $(V \neq 0$) is trivial;
for that case, too, Eq.~(\ref{eq:differential_conductance}) can
  be written in terms of the two-particle vertex, following the
strategy discussed below.

In Ref.~\cite{Bauer2013} we used
Eq.~(\ref{eq:differential_conductance}) (with $\alpha\!=\!1/2$) to
calculate the differential conductance (linear and non-linear) for a
model designed to describe the lowest transport mode of a quantum
point contact (QPC).  The model involves a 1D parabolic potential
barrier in the presence of an onsite electron-electron interaction
(see Sec.~\ref{sec:QPC} for details of the model). In
Ref.~\cite{Bauer2013} we used Keldysh-SOPT (details are
presented in Sec.~\ref{sec:SOPT}) to evaluate both the self-energy and
its derivative with respect to voltage.  The results qualitatively
reproduce the main feature of the 0.7 conductance anomaly, including
its typical dependence on magnetic field and temperature, as well as
the zero-bias peak in the non-linear conductance. For the
remainder of this paper, though, we will consider only the
linear conductance.

\subsection{Linear conductance formula}

In linear response, i.e.\ $V\!\to\! 0$, the linear conductance $g_0$
does not depend on the specific choice of $\alpha$. For the sake of
simplicity we use $\alpha \!=\! 1$, which corresponds to a voltage
setup $\mu^l= \mu + eV$ and $\mu^r = \mu$.  Henceforth, a dot implies
the derivative \textit{at zero bias}, e.g.\
$\mydot{f}^l = \left. \partial_V f^l\right|_{V=0}$, and we have
$\mydot{f}^l\!=\!-ef'$ and $\mydot{f}^r\! =\! 0$. Differentiating
Eq.~(\ref{eq:current_end}) w.r.t. the voltage, followed by setting
$V\!=\!0$, yields the following formla for the \textit{linear} conductance:
\begin{align}
\text{g}_0  =  & \left. \partial_V J\right|_{V=0}  \notag \\
 = & -  \frac{e^2}{h}\!\int\! d \e f' \Tr \lbrace 
\Gamma^l \mathcal{G}_c^{\re{2}|\re{1}}\!( \Gamma^r\! +\! i (\Sigma^{\re{1}|\re{2}}\!-\!\Sigma^{\re{2}|\re{1}})) \mathcal{G}_c^{\re{1}|\re{2}}\rbrace  \nonumber \\ 
& + \! \frac{e^2}{h}
\int \!d\e \Tr \lbrace 
\Gamma^l \mathcal{G}_c^{\re{2}|\re{1}}\Phi^l \mathcal{G}_c^{\re{1}|\re{2}} \rbrace.
\label{eq:conductance_start}
\end{align}
\color{black}
All quantities in the integrand are evaluated in equilibrium. The voltage derivatives of the self-energy are combined in the expression
\begin{align}
\Phi^l \!=\!  \frac{i}{2e}\left[\mydot{\Sigma}^{\re{1}|\re{1}}\! -\!(1-2f)\!\left(\mydot{\Sigma}^{\re{1}|\re{2}}\!-\mydot{\Sigma}^{\re{2}|\re{1}}\right)\right].
\label{eq:sigma_derivatives}
\end{align}
Provided that all components of the self-energy and its derivative in
Eq.~(\ref{eq:sigma_derivatives}) are known at zero bias,
Eq.~(\ref{eq:conductance_start}) is sufficient to calculate the linear
conductance. But, as is shown below, it is possible to express the
voltage derivatives of $\Sigma$ directly in terms of the two-particle
vertex $\mathcal{L}$, i.e.\! the rank-four tensor defined as the sum
of all one-particle irreducible diagrams with four external amputated
legs (see Appendix~\ref{App:Keldysh}). This not only reduces the
numbers of objects to be calculated, but more importantly, it
completely eliminates the voltage from the linear conductance formula:
whereas the derivative $\mydot{\Sigma}$ needs information of the
self-energy at finite bias, the two-particle vertex does not.

To this end we use the fact that an exact expression for the
derivative of the self-energy w.r.t. some parameter $\Lambda$ can
  be related to the two-particle vertex via an exact relation, the
  socalled flow equation of the functional renormalization group
(fRG) (for a diagrammatic derivation of this equation see
Appendix~\ref{App:fRG} and Ref.~\cite{Jakobs2010a}. A rigorous
functional derivation of the full set of coupled fRG equations for all
1PI vertex functions is given in e.g.\
Ref.~\cite{Metzner2012}). For example, this type of
  relation was exploited in Ref. \cite{Oguri2001a} and
  \cite{Kopietz2010} to derive non-equilibrium properties of
  the single impurity Anderson model.  Though $\Lambda$ is usually
taken to be some high-energy cut-off, it can equally well be a
physical parameter of the system, such as temperature
\cite{Honerkamp2001}, chemical potential \cite{Sauli2006,Jakobs2006}
or, as in the present case, voltage: $\Lambda\!=\!V$. If only the
  quadratic part of the bare action depends explicitely on the flow
  parameter, as is the case here, the general flow equation reads
\begin{align}
\partial_\Lambda \Sigma_{\bl{i}|\bl{j}}^{\re{\alpha'}|\re{\alpha}}(\e) = \frac{1}{2\pi i}\int d\ep  \sum_{\underset{\bl{kl} \in C}{\re{\beta\beta'}}}S_{\Lambda,\bl{k}|\bl{l}}^{\re{\beta}|\re{\beta'}}(\ep)\mathcal{L}_{\Lambda,\bl{ik}|\bl{jl}}^{\re{\alpha' \beta'}|\re{\alpha \beta}}(\e',\e;0) ,
\label{eq:fRG_flow}
\end{align}
where $\mathcal{L}(\e',\e;0)$ is the irreducible two-particle vertex,
defined via Eq.~(\ref{eq:partitioning_Green4}) and
Eq.~(\ref{eq:used_functions}). The specific form of this equation for
a given flow-parameter $\Lambda$ is encoded in the single-scale
propagator $S$, which is given by
\begin{align}
S_\Lambda = -\mathcal{G}_c \partial_{\Lambda} \left[\mathcal{G}_{\dg{0},c}\right]^{-1}\mathcal{G}_c = \mathcal{G}_c \mathcal{G}_{\dg{0},c}^{-1}\left[\partial_{\Lambda} \mathcal{G}_{\dg{0},c}\right]\mathcal{G}_{\dg{0},c}^{-1}\mathcal{G}_c ,
\label{eq:SSprop}
\end{align}
with bare central region Green's function $\mathcal{G}_{0,c}(\e)$. According to Eq.~(\ref{eq:Dyson_bare}) only its Keldysh component, $\mathcal{G}_{\dg{0},c}^{\re{2}|\re{2}}$, depends explicitly on the voltage. Additionally, we use $\left[\mathcal{G}_{\dg{0},c}^{-1}\right]^{\re{2}|\re{2}}\!=\!0$, following from causality, Eq.~(\ref{eq:causality}), which yields:
\begin{align}
 S_{V=0}^{\re{2}|\re{2}} & =  \mathcal{G}_c^{\re{2}|\re{1}}\left[\mathcal{G}_{\dg{0},c}^{-1}\right]^{\re{1}|\re{2}}\partial_{V=0} \mathcal{G}_{\dg{0},c}^{\re{2}|\re{2}}\left[\mathcal{G}_{\dg{0},c}^{-1}\right]^{\re{2}|\re{1}}\mathcal{G}_c^{\re{1}|\re{2}} \notag \\
&  = ~ -2ief' \mathcal{G}_c^{\re{2}|\re{1}}\Gamma^l \mathcal{G}_c^{\re{1}|\re{2}}, \notag \\
 S_{V=0}^{\re{1}|\re{1}}  & = S_{V=0}^{\re{1}|\re{2}} = S_{V=0}^{\re{2}|\re{1}} = 0.
 \label{eq:SS_evaluation}
\end{align}
It is instructive to realize that this is indeed the single-scale
propagator already introduced in the derivation of the differential
conductance via Eq.~(\ref{eq:SSP}). The trivial Keldysh structure of
$S$ now implies, that the $\re{\alpha'}|\re{\alpha}$- dependence of
the self-energy derivatives only enters via that of the two-particle
vertex:
\begin{equation}
\mydot{\Sigma}_{\bl{i}|\bl{j}}^{\re{\alpha'}|\re{\alpha}}(\e)\!=\!  \frac{1}{2\pi i} \!\int\! d\e'\!\! \sum_{\bl{kl} \in C}S_{V=0,\bl{k}|\bl{l}}^{\re{2}|\re{2}}(\e')\mathcal{L}_{\bl{il}|\bl{jk}}^{\re{\alpha' 2}|\re{\alpha 2}}(\e',\e;0).
\label{eq:dSigma_V}
\end{equation}
This allows us to write Eq.~(\ref{eq:sigma_derivatives}) in the form
\begin{align}
\Phi_{\bl{i}|\bl{j}}^l(\e) = 
& 
\frac{1}{2\pi i}\!\int \!d\e' f'(\e')
\\ \nonumber
& \times 
\sum_{\bl{kl} \in C}\! \!  \left[\mathcal{G}_c^{\re{2}|\re{1}}(\e')\Gamma^l(\e') \mathcal{G}_c^{\re{1}|\re{2}}(\e')\right]_{\bl{k}|\bl{l}}K_{\bl{il}|\bl{jk}}(\e',\e;0),
\end{align}
with vertex response part
\begin{align}
K_{\bl{il}|\bl{jk}}(\e',\e;0) = & 
\mathcal{L}_{\bl{il}|\bl{jk}}^{\re{12}|\re{12}}(\e',\e;0) -(1-2f(\e))
\label{eq:K}
\\ & \times 
(\mathcal{L}_{\bl{il}|\bl{jk}}^{\re{12}|\re{22}}(\e',\e;0) - 
\mathcal{L}_{\bl{il}|\bl{jk}}^{\re{22}|\re{12}}(\e',\e;0)).
\nonumber 
\end{align}
We use the invariance of the trace under a cyclic permutation, ${\rm{Tr}}\lbrace\Gamma^l \mathcal{G}_c^{\re{2}|\re{1}}\Phi^l \mathcal{G}_c^{\re{1}|\re{2}} \rbrace = {\rm{Tr}}\lbrace\Phi^l\mathcal{G}_c^{\re{1}|\re{2}}\Gamma^l \mathcal{G}_c^{\re{2}|\re{1}} \rbrace$, and interchange the frequency labels, $\e\leftrightarrow\e'$, to obtain the linear conductance formula
\begin{align}
\text{g}_0 = & -  \frac{e^2}{h}\int d \e f'
\Bigl[ \Tr \lbrace 
\Gamma^l \mathcal{G}_c^{\re{2}|\re{1}}\left( \Gamma^r + i (\Sigma^{\re{1}|\re{2}}-\Sigma^{\re{2}|\re{1}})\right) \mathcal{G}_c^{\re{1}|\re{2}}\rbrace \Bigr. 
\nonumber \\ 
& \qquad \Bigl. 
-  \Tr \lbrace \Gamma^l\mathcal{G}_c^{\re{1}|\re{2}} \tilde{\Phi}^l \mathcal{G}_c^{\re{2}|\re{1}}\rbrace \Bigr],
\label{eq:lin_con_final_hermitsch}
\end{align}
with the rearranged vertex correction term
\begin{align}
\tilde{\Phi}_{\bl{l}|\bl{k}}^l(\e) & = \frac{1}{2\pi i }\int d\e' 
\nonumber \\ 
& 
\sum_{{\color{blue}ij}\in C}\left[\mathcal{G}_c^{\re{1}|\re{2}}(\e')\Gamma^l(\e')\mathcal{G}_c^{\re{2}|\re{1}}(\e')\right]_{\bl{j}|\bl{i}}K_{{\color{blue}il|jk}}(\e,\e';0).
\label{eq:P}
\end{align}
In Appendix~\ref{App:ward} we show that particle conservation implies
that the imaginary part of the self-energy and the vertex correction
are related by the following Ward identity:
\begin{equation}
i[\Sigma^{\re{1}|\re{2}}(\e)-\Sigma^{\re{2}|\re{1}}(\e)] = \tilde{\Phi}^l+\tilde{\Phi}^r.
\label{eq:ward_final}
\end{equation} 
This result is obtained by demanding the invariance of the physics
under a gauged, local $U(1)$ transformation, which must hold for any
Hamiltonian that conserves the particle number in the system. This
symmetry implies an infinite hierarchy of relations connecting
different Green's functions. The first equation in this hierarchy
reproduces the continuity equation used in the beginning of the above
derivation. The second equation in the hierarchy is
Eq.~(\ref{eq:ward_final}), which connects parts of one-particle and
two-particle Green's function. Inserting the Ward identity in
Eq.~(\ref{eq:lin_con_final_hermitsch}) yields
\begin{align}
\nonumber
  \text{g}_0 = & - \frac{e^2}{h}\int d \e f'(\e)
\\ 
\nonumber 
&  \times \Bigl[ \Tr \lbrace 
  \Gamma^l(\e) \mathcal{G}_c^{\re{2}|\re{1}}(\e)\big[\Gamma^r(\e) + \tilde{\Phi}^l(\e) + \tilde{\Phi}^r(\e) \big] \mathcal{G}_c^{\re{1}|\re{2}}(\e)\rbrace \Bigr.
\\
& \qquad \Bigl. -
 \Tr \lbrace \Gamma^l(\e)\mathcal{G}_c^{\re{1}|\re{2}}(\e) \tilde{\Phi}^l(\e) \mathcal{G}_c^{\re{2}|\re{1}}(\e)\rbrace  \Bigr] .
\label{eq:conductance_final}
\end{align}
This formula is the central result of this paper. It expresses the
linear conductance in terms of the two-particle vertex $\mathcal{L}$,
which enters via the vertex part $\tilde{\Phi}$ [Eq.~(\ref{eq:P})] and
the response vertex $K$ [Eq.~(\ref{eq:K})]. Note that the two terms in
Eq.~(\ref{eq:conductance_final}) differ in their Keldysh structure via
the Keldysh indexing of the full Green's functions, which prevents
further compactification of Eq.~(\ref{eq:conductance_final}) for a
non-symmetric Hamiltonian (e.g.\ in the presence of finite spin-orbit
interactions, see. e.g.\ Ref.~\cite{Goulko2014}). If, in
contrast, the Hamiltonian of Eq.~(\ref{eq:ModelGeneral}) is symmetric
(i.e.\ $h_{\bl{ij}}\! =\! h_{\bl{ji}}$),
Eq.~(\ref{eq:conductance_final}) can be compactified significantly
using the following argument: A symmetric Hamiltonian implies that the
Green's function $\mathcal{G}$, the self-energy $\Sigma$ and the
hybridization $\Gamma$ are symmetric, too. This in turn gives a
symmetric $\tilde{\Phi}$ via Eq.~(\ref{eq:ward_final}). Hence, the
trace in the first term of Eq.~(\ref{eq:conductance_final}) is taken
over the product of four symmetric matrices, and transposing yields
$\Tr\lbrace \Gamma^l\mathcal{G}_c^{\re{2}|\re{1}}\big[ \Gamma^r +
\tilde{\Phi}^l + \tilde{\Phi}^r \big]
\mathcal{G}_c^{\re{1}|\re{2}}\rbrace = \Tr\lbrace
\Gamma^l\mathcal{G}_c^{\re{1}|\re{2}}\big[ \Gamma^r + \tilde{\Phi}^l +
\tilde{\Phi}^r \big] \mathcal{G}_c^{\re{2}|\re{1}}\rbrace$.
Hence, all contributions involving $\tilde{\Phi}^l$ cancel in
Eq.~(\ref{eq:conductance_final}) and the linear conductance now simply
reads
\begin{align}
\text{g}_0  = & -\frac{e^2}{h} \int_{-\infty}^{\infty} d\e
 f'(\e)
\nonumber 
\\
& \times {\rm{Tr}} \{\Gamma^l(\e)\mathcal{G}_c^{\re{1}|\re{2}}(\e)[\Gamma^r(\e)+ \tilde{\Phi}^r(\e)]\mathcal{G}_c^{\re{2}|\re{1}}(\e) \}. 
 \label{eq:g_sym_ham}
\end{align}
This equation constitutes a Keldysh version of Oguri's formula
  for the linear conductance for a symmetric Hamiltonian (Eq.~(2.35)
  in Ref.~\cite{Oguri2001}).  Oguri worked in the Matsubara
  formalism and used Eliashberg theory to perform the analytic
  continuation of the vertex from Matsubara frequencies to real
  frequencies. By comparing our formula (\ref{eq:g_sym_ham}) to
  Oguri's version, a connection between the three Keldysh vertex
  components in Eq.~(\ref{eq:K}) and the ones used in Oguri's
  derivation can be established, if desired.

All calculations of the linear conductance reported in
Ref.~\cite{Bauer2013} using Matsubara-fRG and SOPT,
and in Ref.~\cite{Schimmel2017} using Keldysh-fRG,
were based on  Eq.~\eqref{eq:g_sym_ham}.

\subsection{Linear thermal conductance formula}

We end this section with some considerations regarding thermal
conductance, i.e.\ the conductance induced by a temperature difference
between the leads. In the following we assume zero bias voltage,
$V\!=\!0$. The left lead is in thermal equilibrium with
$T^l\! =\! T+\tilde{T}$ and the right lead in thermal equilibrium with
temperature $T^r\! =\! T$. Thus, the temperature gradient between the
leads will provide a charge current through the central
region. Similar to above, we are now interested in the linear response
thermal conductance formula,
$\text{g}_{0,T} = \partial_{\tilde{T}=0}{J}$, which we could calculate
in similar fashion as the linear conductance $\text{g}_0$. Much easier
is the following though: all terms in Eq.~(\ref{eq:conductance_final})
were obtained by once time taking the derivative of the Fermi
distribution $f^l$ w.r.t. the voltage, partly explicitly in
Eq.~(\ref{eq:current_end}) and partly from evaluating the single-scale
propagator in Eq.~(\ref{eq:SS_evaluation}). Now note, that
$\partial_{\tilde{T}=0} f^l = \frac{\e-\mu}{T} f' =
-\frac{(\e-\mu)}{eT}\partial_{V=0} f^l$.
For a symmetric Hamiltonian this directly implies, that the linear
thermal conductance is given by
\begin{align}
\text{g}_{0,T}  = & \frac{e}{hT} \int_{-\infty}^{\infty}\!\! d\e
 (\e-\mu)f'(\e)
\nonumber 
\\ 
&  \times {\rm{Tr}}\{\Gamma^l(\e)\mathcal{G}_c^{\re{1}|\re{2}}(\e) [\Gamma^r(\e)\!+\! \tilde{\Phi}^r(\e)]\mathcal{G}_c^{\re{2}|\re{1}}(\e) \}. 
 \label{eq:gT_sym_ham}
\end{align}

\section{Vertex functions in SOPT
\label{sec:SOPT}}

In Ref.~\cite{Bauer2013} we calculated the linear
  conductance of our QPC model [Sec.\ref{sec:QPC}] using
  Eq.~(\ref{eq:g_sym_ham}), and the non-linear differential conductance
  using Eq.~(\ref{eq:differential_conductance}). There we used fRG
  (within the coupled ladder approximation) to calculate the linear
  conductance at $T=V=0$, and SOPT to calculate both the linear
  conductance at $T\neq 0$ and the non-linear $(V \neq 0)$
  differential conductance at $T=0$.  The details of the fRG approach
  can be found in Ref.~\cite{Bauer2014}. The purpose of the
  present section is to present the details of the SOPT calculations.

In order to apply the conductance formulas derived above we calculate
the self-energy $\Sigma$ and the two-particle vertex $\mathcal{L}$ in
second order perturbation theory (SOPT). Both are defined in
Eq.~(\ref{eq:used_functions}) and needed when evaluating the
  conductance formulas (\ref{eq:conductance_final}) or
  \eqref{eq:g_sym_ham}. The SOPT strategy is to approximate them by a
  diagrammatic series truncated beyond second order in the bare
  interaction vertex $\nu$, defined below.

  Within this section the compact composite index notation used above
  is dropped in favor of a more explicit one. We henceforth use blue
  \textit{roman} subscripts ($\bl{i_1},\bl{i_2},...$) for site indices
  only and explicitly denote spin dependencies using
  $\sigma \in \{\uparrow,\downarrow\} \!=\! \{+,-\}$. A green number
  subscript denotes an object's order in the interaction, e.g.\
  $\Sigma_{\dg{2}}$ is the desired self-energy to second order in the
  bare vertex $\nu$.

Below, the quadratic part of the model Hamiltonian, Eq.~(\ref{eq:ModelGeneral}), is
is represented by a real matrix that is symmetric in position basis and diagonal in spin space
\begin{equation}
h_{\bl{ij}}^\sigma = h_{\bl{ji}}^\sigma \in \mathbb{R}~~,~~ h = h^\uparrow + h^{\downarrow}.
\label{eq:h_quadratic_sym}
\end{equation}
In consequence, the bare Green's function, too, is diagonal in spin space and symmetric in position space:
\begin{align}
\mathcal{G}_{\dg{0},\bl{i}\sigma|\bl{j}\sigma'} = \delta_{\sigma \sigma'} \mathcal{G}_{\dg{0},\bl{i}|\bl{j}}^\sigma ~~,~~\mathcal{G}_{\dg{0},\bl{i}|\bl{j}}^\sigma = \mathcal{G}_{\dg{0},\bl{j}|\bl{i}}^\sigma.
\label{eq:properties_bare_Green}
\end{align}
We distinguish between composite quantum numbers including contour indices $k_n\! =\!
(\re{a_n},\bl{i_n},\sigma_n)$ and composite quantum numbers including Keldysh indices
$\kappa_n\! =\! (\re{\alpha_n},\bl{i_n},\sigma_n)$. The noninteracting
Green's function is represented by a directed line
\begin{align}
\mathcal{G}_{\dg{0},k_1|k_1'}(\e) & = 
\begin{matrix}
\includegraphics[width=2cm]{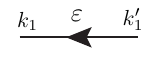}
\end{matrix}.
\label{eq:diagram_bare_G}
\end{align}
We choose an onsite interaction, which reduces the quartic term in Eq.~(\ref{eq:ModelGeneral}) to a single sum
\begin{equation}
\mathcal{H}_{\rm{int}} = \sum_{\bl{i}\in C} U_{\bl{i}} n_{\bl{i}\uparrow} n_{\bl{i}\downarrow},
\end{equation}
i.e.\ we evaluate the vertex functions for the case of an onsite electron-electron interaction.
Since the two-particle interaction is instantaneous in time, we construct the anti-symmetrized bare interaction vertex as
\begin{align}
\nu_{k_1',k_2'|k_1,k_2}(t_1' &, t_2'|t_1, t_2) \notag \\
&  =   U_{\bl{i_1}} \delta_{\bl{i_1}\bl{i_2}}\delta_{\bl{i_1 i_1'}}\delta_{\bl{i_1 i_2'}} (-{\color{red} a_1}) \delta_{{\color{red}a_1 a_2 }}\delta_{{\color{red}a_1 a_1' }}\delta_{{\color{red}a_1 a_2'}} \notag \\
&\quad \times \delta(t_1-t_2)\delta(t_1-t_1')\delta(t_1-t_2') \notag \\
& \quad \times \delta_{\sigma_1 \bar{\sigma}_2}\delta_{\sigma_1'
\bar{\sigma}_2'}(\delta_{\sigma_1'\sigma_1} - \delta_{\sigma_1'\sigma_2}) \, ,
\label{eq:bare_vertex_contour}
\end{align}
with $\bar{\sigma} \!=\! -\sigma$. Note that its spin-dependence is determined by Pauli's exclusion principle and the Slater-determinant character of the fermionic state. After Fourier transformation [ Eq.~(\ref{eq:Fourier_transformation}), Eq.~(\ref{eq:used_functions})] and Keldysh rotation [Eq.~(\ref{eq:transformation}), Eq.~(\ref{eq:Kel_transformation})] we find
\begin{align}
\nu_{\kappa_1',\kappa_2'|\kappa_1\kappa_2} &  (\e_1',\e_2'|\e_1,\e_2)
\!=\!  2 \pi \delta(\e_1\!+\!\e_2\!-\!\e_1'\!-\!\e_2') \bar{u}_{\kappa_1',\kappa_2'|\kappa_1\kappa_2},
\end{align}
where we introduced the bare vertex
\begin{align}
\bar{u}_{\kappa_1',\kappa_2'|\kappa_1\kappa_2} &
  = u_{\bl{i_1}}\delta_{\bl{i_1 i_2}}\delta_{\bl{i_1 i_1'}}\delta_{\bl{i_1 i_2'}}  \xi^{\re{\alpha_1' \alpha_2'}|\re{\alpha_1}\re{\alpha_2}} \notag \\
& \quad\times \delta_{\sigma_1 \bar{\sigma}_2}\delta_{\sigma_1' \bar{\sigma}_2'}(\delta_{\sigma_1'\sigma_1} - \delta_{\sigma_1'\sigma_2}) \notag \\
& = 
\begin{matrix}
\includegraphics[width=1.5cm]{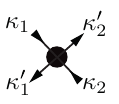}
\end{matrix},
\label{eq:bare_vertex_keldysh}
\end{align}
with $u_{\bl{i}} = U_{\bl i}/2$ and the modulo operation
\begin{equation}
\xi^{\re{\alpha_1' \alpha_2'}|\re{\alpha_1 \alpha_2}} = \begin{cases}1,~~~\rm{if}~ \re{\alpha_1'}+\re{\alpha_2'}+\re{\alpha_1}+\re{\alpha_2}=\rm{odd} \\
0,~~~\rm{else}. \end{cases} \notag 
\label{eq:modulo}
\end{equation}

\subsection{The two-particle vertex in SOPT}

Our goal is to approximate the vertex part, Eq.~(\ref{eq:K}), to second order in the interaction. The fully interacting two-particle vertex, $\mathcal{L}(\e,\e';0)$, has the following diagrammatic representation:
\begin{align}
\mathcal{L}_{\kappa_1' \kappa_2'|\kappa_1 \kappa_2}(\e',\e;0)  &~~ = 
\begin{matrix}
\includegraphics[width=2.3cm]{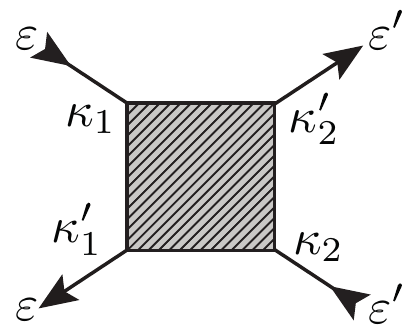} 
\end{matrix}
\end{align}

In SOPT, the vertex $\mathcal{L}_{\dg{2}}$ is given by the sum of all 1PI diagrams with four external amputated legs and not more than two bare vertices. Defining the frequencies
\begin{align}
p = \e+\e'~~,~~x = \e-\e',
\end{align}
the vertex reads
\begin{equation}
\mathcal{L}_{\dg{2}}(\e',\e;0) = \bar{u} + \mathcal{L}_{\dg{2}}^p(p) + \mathcal{L}_{\dg{2}}^x(x) + \mathcal{L}_{\dg{2}}^d(0),
\end{equation}
with particle-particle channel $\mathcal{L}_{\dg{2}}^p$, particle-hole channel $\mathcal{L}_{\dg{2}}^x$ and direct channel $\mathcal{L}_{\dg{2}}^d$ defined as 
\begin{widetext}
\begin{subequations}
\begin{align}
\mathcal{L}_{\dg{2},\kappa_1' \kappa_2'|\kappa_1 \kappa_2}^p(p)  & = 
\begin{matrix}
\includegraphics[width=2cm]{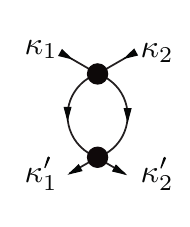} 
\end{matrix}
 = ~\frac{i}{2\pi}\int_{-\infty}^{\infty}\! d\e''\!\sum_{q_1 q_2 q_1' q_2'}\bar{u}_{\kappa_1' \kappa_2'|q_1 q_2}\mathcal{G}_{\dg{0},q_1|q_1'}(p-\e'')\mathcal{G}_{\dg{0},q_2|q_2'}(\e'')\bar{u}_{q_1' q_2'|\kappa_1 \kappa_2}, \\
 \mathcal{L}_{\dg{2},\kappa_1' \kappa_2'|\kappa_1 \kappa_2}^x(x) & = 
\begin{matrix}
\includegraphics[width=2cm]{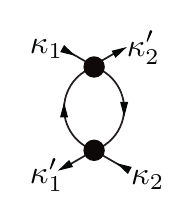} 
\end{matrix}
 = ~\frac{i}{2\pi}\int_{-\infty}^{\infty}\!d\e''\! \sum_{q_1 q_2 q_1' q_2'} \bar{u}_{\kappa_1'  q_2'| q_1 \kappa_2}\mathcal{G}_{\dg{0},q_1|q_1'}(\e'')\mathcal{G}_{\dg{0},q_2|q_2'}(\e''+x)\bar{u}_{q_1' \kappa_2'| \kappa_1 q_2}, \\
\mathcal{L}_{\dg{2},\kappa_1' \kappa_2'|\kappa_1\kappa_2}^d(0) & = 
\begin{matrix}
\includegraphics[width=2cm]{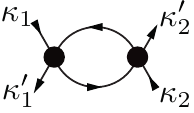} 
\end{matrix}
 = ~\frac{-i}{2\pi}\int_{-\infty}^{\infty}\!d\e''\! \sum_{q_1 q_2 q_1' q_2'} \bar{u}_{\kappa_1' q_2'| \kappa_1 q_1}\mathcal{G}_{\dg{0},q_1|q_1'}(\e'')\mathcal{G}_{\dg{0},q_2|q_2'}(\e'')\bar{u}_{q_1' \kappa_2'| q_2 \kappa_2}.
\end{align}
 \label{eq:channel_definitions}
\end{subequations}
\end{widetext}
These expressions can be derived by a straightforward perturbation theory.

Using Eq.~(\ref{eq:properties_bare_Green}) and
Eq.~(\ref{eq:bare_vertex_keldysh}), we can identify the only
non-vanishing components in spin- and real space,
\begin{subequations}
\begin{align}
&  \Pi_{\bl{ij}}^{\sigma\bar{\sigma}}(p)    = \mathcal{L}_{\dg{2},\bl{i}\sigma\bl{i}\bar{\sigma}|\bl{j}\sigma\bl{j}\bar{\sigma}}^p(p), \\
& X_{\bl{ij}}^{\sigma\sigma'}(x)  = \mathcal{L}_{\dg{2},\bl{i}\sigma\bl{j}\sigma'|\bl{j}\sigma\bl{i}\sigma'}^x(x), \\
& \Delta_{\bl{ij}}^{\sigma\sigma'}(0)  = \mathcal{L}_{\dg{2},\bl{i}\sigma\bl{j}\sigma'|\bl{i}\sigma'\bl{j}\sigma}^d(0).
\end{align}
\end{subequations}

Eq.~(\ref{eq:h_quadratic_sym}) and the channel definitions, Eq.~(\ref{eq:channel_definitions}), imply the symmetries
\begin{subequations}
\begin{align}
& \Pi_{\bl{ij}}  = \Pi_{\bl{ji}}~,~ X_{\bl{ij}}  = X_{\bl{ji}}~,~\Delta_{\bl{ij}}  = \Delta_{\bl{ji}}, \\
& \Pi(p) = \Pi^{\sigma\bar{\sigma}}(p)   = \Pi^{\bar{\sigma}\sigma}(p), \\
& X^{\sigma\sigma'}(x)   = X^{\sigma'\!\sigma}(-x), \\
& \Delta^{\sigma \sigma'}(0)   = \Delta^{\sigma'\!\sigma}(0).
\label{eq:bubble_symmetries}
\end{align}
\end{subequations}

Moreover, and directly following from the Keldysh structure of the
bare vertex in Eq.~(\ref{eq:bare_vertex_keldysh}), we are left with
only four non-zero components per channel in Keldysh space. This is
best seen from realizing, that the internal Keldysh structure of the
diagrams in Eq.~(\ref{eq:channel_definitions}) only depends on whether
the sum of external indices belonging to the same bare vertex is
even/odd.  Furthermore, from the Keldysh structure of the bare vertex,
combined with $\mathcal{G}^{\re{1}|\re{1}}\!=\!0$ and the analytic
properties of $\mathcal{G}$, it follows that
$\mathcal{L}^{\re{22}|\re{22}}=0$. Hence, SOPT preserves the theorem
of causality, Eq.~(\ref{eq:causality}), as it should. (this has also
been shown for a wide range of approximation schemes in
Ref.~\cite{Jakobs2009a}).  Thus, the Keldysh structure of the
channels $Y=\Pi,X,\Delta$ is given by the matrix representation
\begin{equation}
Y = 
\left( \begin{array}{cc}
Y^\re{K} & Y^\re{R} \\
Y^{\re{A}} & 0 \\
\end{array}\right)
=
\left( \begin{array}{cc}
Y^{\re{1}|\re{1}} & Y^{\re{1}|\re{2}} \\
Y^{\re{2}|\re{1}} & Y^{\re{2}|\re{2}} \\
\end{array}\right).
\label{eq:channel_keldysh_structure}
\end{equation}
We define the individual components according to the Keldysh structure of the full vertex,
\begin{align}
\mathcal{L}_{\dg{2}}^{\re{\alpha_1'\alpha_2'}|\re{\alpha_1\alpha_2}} = ~ & \Pi^{\psi(\re{\alpha_1'},\re{\alpha_2'})|\psi(\re{\alpha_1},\re{\alpha_2})} \notag \\
+ & X^{\psi(\re{\alpha_1'},\re{\alpha_2})|\psi(\re{\alpha_1},\re{\alpha_2'})} \notag \\
+ & \Delta^{\psi(\re{\alpha_1'},\re{\alpha_1})|\psi(\re{\alpha_2},\re{\alpha_2'})},
\end{align}
where we introduced the modified modulo operation 
\begin{equation}
\psi(\re{\alpha_1},\re{\alpha_2},...,\re{\alpha_n}) = \begin{cases}\re{1},~\rm{if}~ \sum_{i=1,...,n}\re{\alpha_i}=\rm{odd} \\
\re{2},~\rm{else}. \end{cases} \notag 
\label{eq:modulo2}
\end{equation}
That leaves us with the following explicit formulas 
\begin{subequations}
\begin{align}
& \Pi_{\bl{ij}}^{\re{1}|\re{2}}(p)  
  = -\frac{u_{\bl{i}}u_{\bl{j}}}{2\pi i} 
\int \!d\e\! \left[
\mathcal{G}_{\dg{0},\bl{i}|\bl{j}}^{\sigma,\re{2}|\re{1}}(p\!-\!\e)
\mathcal{G}_{\dg{0},\bl{i}|\bl{j}}^{\bar{\sigma},\re{2}|\re{2}}(\e)
\right. \notag \\
 &  \left.
 \quad\quad\quad\quad\quad\quad\quad\quad\quad +  \mathcal{G}_{\dg{0},\bl{i}|\bl{j}}^{\sigma,\re{2}|\re{2}}(p\!-\!\e)
\mathcal{G}_{\dg{0},\bl{i}|\bl{j}}^{\bar{\sigma},\re{2}|\re{1}}(\e) 
\right], 
\label{eq:bubble_Pr} \\
& \Pi^{\re{2}|\re{1}}  =  \left[\Pi^{\re{1}|\re{2}}\right]^*, 
\label{eq:bubble_Pa}\\
& \Pi_{\bl{ij}}^{\re{1}|\re{1}}(p)   
  = -\frac{u_{\bl{i}}u_{\bl{j}}}{2\pi i} 
\int \!d\e\! \left[
\mathcal{G}_{\dg{0},\bl{i}|\bl{j}}^{\sigma,\re{2}|\re{2}}(p\!-\!\e)
\mathcal{G}_{\dg{0},\bl{i}|\bl{j}}^{\bar{\sigma},\re{2}|\re{2}}(\e)
\right. \notag \\
 &  \left.
 \quad\quad\quad\quad\quad\quad\quad\quad\quad + \mathcal{G}_{\dg{0},\bl{i}|\bl{j}}^{\sigma,\re{2}|\re{1}}(p\!-\!\e)
\mathcal{G}_{\dg{0},\bl{i}|\bl{j}}^{\bar{\sigma},\re{2}|\re{1}}(\e) 
\right. \notag \\
 &  \left.
 \quad\quad\quad\quad\quad\quad\quad\quad\quad + \mathcal{G}_{\dg{0},\bl{i}|\bl{j}}^{\sigma,\re{1}|\re{2}}(p\!-\!\e)
\mathcal{G}_{\dg{0},\bl{i}|\bl{j}}^{\bar{\sigma},\re{1}|\re{2}}(\e) 
\right], 
\label{eq:bubble_Pk}\\
& \Pi^{\re{1}|\re{1}}(p)\Big|_{V=0}  
   =  [1+2b(p-\mu)]\left[ \Pi^{\re{1}|\re{2}}(p) - \Pi^{\re{2}|\re{1}}(p)\right]_{V=0},
\label{eq:bubble:definition1}
\end{align}
\label{eq:pchannel}
\end{subequations}
\begin{subequations}
\begin{align}
& X_{\bl{ij}}^{\sigma\sigma',\re{1}|\re{2}}(x)   
  = -\frac{u_{\bl{i}}u_{\bl{j}}}{2\pi i} 
\int \!d\e\! \left[
\mathcal{G}_{\dg{0},\bl{i}|\bl{j}}^{\bar{\sigma},\re{1}|\re{2}}(\e)
\mathcal{G}_{\dg{0},\bl{i}|\bl{j}}^{\bar{\sigma}',\re{2}|\re{2}}(\e\!+x)
\right. \notag
\label{eq:X_channel_formula} \\
 &  \left.
 \quad\quad\quad\quad\quad\quad\quad\quad\quad\quad~~ + \mathcal{G}_{\dg{0},\bl{i}|\bl{j}}^{\bar{\sigma},\re{2}|\re{2}}(\e)
\mathcal{G}_{\dg{0},\bl{i}|\bl{j}}^{\bar{\sigma}',\re{2}|\re{1}}(\e\!+x) 
\right], \\
& X^{\re{2}|\re{1}}  =  \left[X^{\re{1}|\re{2}}\right]^*, \\
&  X_{\bl{ij}}^{\sigma\sigma',\re{1}|\re{1}}(x)  
  = -\frac{u_{\bl{i}}u_{\bl{j}}}{2\pi i} 
\int \!d\e\! \left[
\mathcal{G}_{\dg{0},\bl{i}|\bl{j}}^{\bar{\sigma},\re{2}|\re{2}}(\e)
\mathcal{G}_{\dg{0},\bl{i}|\bl{j}}^{\bar{\sigma}',\re{2}|\re{2}}(\e\!+x)
\right. \notag \\
 &  \left.
 \quad\quad\quad\quad\quad\quad\quad\quad\quad\quad~~ + \mathcal{G}_{\dg{0},\bl{i}|\bl{j}}^{\bar{\sigma},\re{2}|\re{1}}(\e)
\mathcal{G}_{\dg{0},\bl{i}|\bl{j}}^{\bar{\sigma}',\re{1}|\re{2}}(\e\!+x) 
\right] \notag \\
 &  \left.
 \quad\quad\quad\quad\quad\quad\quad\quad\quad\quad~~  +\mathcal{G}_{\dg{0},\bl{i}|\bl{j}}^{\bar{\sigma},\re{1}|\re{2}}(\e)
\mathcal{G}_{\dg{0},\bl{i}|\bl{j}}^{\bar{\sigma}',\re{2}|\re{1}}(\e\!+x) \right], \\
& X^{\re{1}|\re{1}}(x)\Big|_{V=0} 
= [1+2b(x+\mu)]\big[ X^{\re{1}|\re{2}}(x) - X^{\re{2}|\re{1}}(x)\big]_{V=0},
\label{eq:bubble:definition2}
\end{align}
\end{subequations}
\begin{subequations}
\begin{align}
& \Delta_{\bl{ij}}^{\sigma\sigma',\re{1}|\re{2}}(0)  
   = \frac{u_{\bl{i}}u_{\bl{j}}}{2\pi i} 
\int \!d\e\! \left[
\mathcal{G}_{\dg{0},\bl{i}|\bl{j}}^{\bar{\sigma},\re{1}|\re{2}}(\e)
\mathcal{G}_{\dg{0},\bl{i}|\bl{j}}^{\bar{\sigma}',\re{2}|\re{2}}(\e) 
\right. \notag \\
 &  \left.
 \quad\quad\quad\quad\quad\quad\quad\quad\quad~~ +
\mathcal{G}_{\dg{0},\bl{i}|\bl{j}}^{\bar{\sigma},\re{2}|\re{2}}(\e)
\mathcal{G}_{\dg{0},\bl{i}|\bl{j}}^{\bar{\sigma}',\re{2}|\re{1}}(\e) 
\right], \\
& \Delta = \Delta^{\re{2}|\re{1}}   = \Delta^{\re{1}|\re{2}}, \\
& \Delta^{\re{1}|\re{1}}   = 0.
\label{eq:bubble:definition3}
\end{align}
\label {eq:vertex_final}
\end{subequations}
Here, we introduced the Bose distribution function, $b(z) = 1/(e^{(z-\mu)/T}-1)$, with chemical potential $\mu$ and temperature $T$. $[~]^*$ denotes the complex conjugate. Note that the components of every individual channel fulfill a fluctuation dissipation theorem (FDT) in equilibrium [Eqs.(\ref{eq:bubble:definition1},\ref{eq:bubble:definition2},\ref{eq:bubble:definition3})], warranting the choice of notation introduced in Eq.~(\ref{eq:channel_keldysh_structure}). We derive this FDT  in detail in Appendix~\ref{App:FDT}.

Finally we write down the three components of the SOPT two-particle vertex that occur in the vertex-correction part, Eq.~(\ref{eq:K}):
\begin{subequations}
\begin{align}
& \mathcal{L}_{\dg{2},\bl{i}\sigma,\bl{l}\sigma'|\bl{j}\sigma,\bl{k}\sigma'}^{\re{12}|\re{22}}(\e',\e;0) = \notag \\
&~~~~~~~ \delta_{\sigma\bar{\sigma}'}\delta_{\bl{ij}}\delta_{\bl{ik}}\delta_{\bl{il}} u_{\bl{i}} 
+ \delta_{\sigma\bar{\sigma}'}\delta_{\bl{il}}\delta_{\bl{jk}}\Pi_{\bl{ij}}^{\re{1}|\re{2}}(p) \nonumber \\
&~~~ + \delta_{\bl{ik}}\delta_{\bl{jl}}X_{\bl{ij}}^{\sigma\sigma',\re{1}|\re{2}}(x)
+ \delta_{\sigma\sigma'}\delta_{\bl{ij}}\delta_{\bl{kl}}\Delta_{\bl{ik}}^{\sigma\sigma'}(0), \\
& \mathcal{L}_{\dg{2}}^{\re{22}|\re{12}}  = \bar{u} + \Pi^{\re{2}|\re{1}} + X^{\re{2}|\re{1}} + \Delta, \\
& \mathcal{L}_{\dg{2}}^{\re{12}|\re{12}}   = \Pi^{\re{1}|\re{1}} + X^{\re{1}|\re{1}}.
\label{eq:vertex_SOPT}
\end{align}
\end{subequations}

Utilizing the equilibrium's FDT for the $\Pi$-, and $X$-channel [Eq.~(\ref{eq:bubble:definition1}), Eq.~(\ref{eq:bubble:definition2})], we find
\begin{align}
&  K_{{\bl{i}\sigma,\bl{l}\sigma'|\bl{j}\sigma,\bl{k}\sigma'}}(\e',\e;0)  = \nonumber \\
& \quad \delta_{\sigma\bar{\sigma}'}\delta_{\bl{il}}\delta_{\bl{jk}}\left[2f(\e)+2b(p-\mu)\right](\Pi_{\bl{ij}}^{\re{1}|\re{2}}(p) - \Pi_{\bl{ij}}^{\re{2}|\re{1}}(p)) \nonumber \\
 & + \delta_{\bl{ik}}\delta_{\bl{jl}}\left[2f(\e)+2b(x+\mu)\right](X_{\bl{ij}}^{\sigma\sigma',\re{1}|\re{2}}(x) - X_{\bl{ij}}^{\sigma\sigma',\re{2}|\re{1}}(x)).
 \end{align}

We note, that this result (for $\mu\!=\!0$) has been obtained before by Oguri (see Eq.~(4.7) of
Ref.~\cite{Oguri2001}) using Matsubara formalism and an analysis of the two-particle
vertex following Eliashberg \cite{Eliashberg1962}.

\subsection{The self-energy in SOPT}

Our goal is to approximate the self-energy to second order in the
interaction. The fully interacting self-energy, $\Sigma(\e)$, has the
following diagrammatic representation:
\begin{align}
\Sigma_{\kappa_1'|\kappa_1}(\e)  &~~ = 
\begin{matrix}
\includegraphics[width=1.5cm]{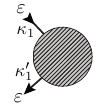} 
\end{matrix}
\end{align}
In SOPT, the self-energy $\Sigma_{\dg{2}}$ is given by the sum of all 1PI diagrams with two external amputated legs and not more than two bare vertices. This amounts to three topologically different diagrams:
\begin{align}
\Sigma_{\dg{2},\kappa_1'|\kappa_1}(\e) = &  
\; \begin{matrix}
\includegraphics[width=6.5cm]{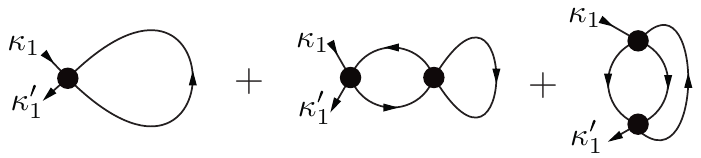} 
\end{matrix} 
  \notag \\
  = &\, \frac{-i}{2\pi}\int_{-\infty}^{\infty} \! \! d\e'\sum_{q_1 q_1'}\Big[ \bar{u}_{k_1' q_1'|k_1 q_1} +\gamma_{\dg{2},k_1' q_1'|k_1 q_1}^d(0)
\nonumber \\ 
&  +\gamma_{\dg{2},k_1' q_1'|k_1 q_1}^p(\e+\e')\Big]\mathcal{G}_{\dg{0},q_1|q_1'}(\e').
\label{eq:Sigma_diagrammatic}
\end{align}
We note that, equivalently, the third 
diagram can also be expressed via either spin configuration, $X^{\sigma\sigma}$ or $X^{\sigma\bar{\sigma}}$, [Eq.~(\ref{eq:X_channel_formula}), Eq.~(\ref{eq:SIgma_r})] of the particle-hole vertex channel $\gamma_{\dg{2}}^x$ instead of the particle-particle channel $\gamma_{\dg{2}}^p$. 

As a consequence of the spin-dependence of both the noninteracting
Green's function and the bare vertex,
Eq.~(\ref{eq:properties_bare_Green}) and
Eq.~(\ref{eq:bare_vertex_keldysh}), as well as the real space symmetry
of the Hamiltonian, Eq.~(\ref{eq:h_quadratic_sym}), the self-energy,
too, is spin-diagonal and symmetric in real space:
\begin{align}
\Sigma_{\bl{i}\sigma|\bl{j}\sigma'} = \delta_{\sigma\sigma'}\Sigma_{\bl{i}|\bl{j}}^\sigma~~,~~ \Sigma_{\bl{i}|\bl{j}}^\sigma = \Sigma_{\bl{j}|\bl{i}}^\sigma.
\end{align} 
The Keldysh structure of the self-energy is given by matrix structure
[Eq.~(\ref{eq:matrix_representation})] with
$\Sigma^{\re{R}}\!=\!\Sigma^{\re{1}|\re{2}}$. The theorem of causality
demands $\Sigma^{\re{2}|\re{2}}\!=\!0$ [Eq.~(\ref{eq:causality})].
Finally, explicit evaluation of the diagrams in
Eq.~(\ref{eq:Sigma_diagrammatic}) yields
\begin{widetext}
\begin{subequations}
\begin{align}
& \Sigma_{\dg{2},\bl{i}|\bl{j}}^{\sigma,\re{1}|\re{2}}(\e)\!  =\!
\frac{-i}{2\pi} \!
\int \!d\e' \Big[ \delta_{\bl{i}\bl{j}} u_\bl{i} \mathcal{G}_{\dg{0},\bl{i}|\bl{i}}^{\bar{\sigma},\re{2}|\re{2}}(\e') +  \delta_{\bl{i}\bl{j}}\!
\sum_{\bl{k}} \mathcal{G}_{\dg{0},\bl{k}|\bl{k}}^{\sigma,\re{2}|\re{2}}(\e')\Delta_{\bl{i}\bl{k}}^{\sigma\sigma}(0) + 
\mathcal{G}_{\dg{0},\bl{i}|\bl{j}}^{\sigma,\re{2}|\re{2}}(\e') X_{\bl{ij}}^{\sigma\sigma,\re{1}|\re{2}}(\e-\e') +  
\mathcal{G}_{\dg{0},\bl{i}|\bl{j}}^{\sigma,\re{2}|\re{1}}(\e') X_{\bl{ij}}^{\sigma\sigma,\re{1}|\re{1}}(\e-\e') \Big],
\label{eq:SIgma_r} \\
& \Sigma_{\dg{2}}^{\re{21}}   = \left[\Sigma^{\re{12}}\right]^{*}, 
\label{eq:SIgma_a}\\
& \Sigma_{\dg{2},\bl{i}|\bl{j}}^{\sigma,\re{1}|\re{1}}(\e)    = \frac{-i}{2 \pi} \int d\e' \left[ 
\mathcal{G}_{\dg{0},\bl{i}|\bl{j}}^{\sigma,\re{2}|\re{2}}(\e') X_{\bl{ij}}^{\sigma\sigma,\re{1}|\re{1}}(\e-\e') +   
\mathcal{G}_{\dg{0},\bl{i}|\bl{j}}^{\sigma,\re{2}|\re{1}}(\e') X_{\bl{ij}}^{\sigma\sigma,\re{1}|\re{2}}(\e-\e') + 
\mathcal{G}_{\dg{0},\bl{i}|\bl{j}}^{\sigma,\re{1}|\re{2}}(\e') X_{\bl{ij}}^{\sigma\sigma,\re{21}}(\e-\e')\right], \\
& \Sigma_{\dg{2},\bl{i}|\bl{j}}^{\sigma,\re{1}|\re{1}}(\e)|_{V=0} 
= (1-2f(\e))\left[\Sigma_{\dg{2},\bl{i}|\bl{j}}^{\sigma,\re{1}|\re{2}}(\e) - \Sigma_{\dg{2},\bl{i}|\bl{j}}^{\sigma,\re{2}|\re{1}}(\e)\right]_{V=0}.
\label{eq:Sigma_k}
\end{align}
\label{eq:Sigma}
\end{subequations}
\end{widetext}
We derive the FDT, Eq.~(\ref{eq:Sigma_k}), in Appendix~\ref{App:FDT}.

\subsection{Voltage derivative of the self-energy in SOPT}

In order to calculate the differential conductance via Eq.~(\ref{eq:differential_conductance}) we now provide explicit formulas for the voltage derivative of the self-energy components. In principle we could use the natural approach and differentiate the r.h.s. of the self-energy expressions, Eq.~(\ref{eq:Sigma}), with the corresponding vertex components given by Eqs.(\ref{eq:pchannel})-(\ref{eq:vertex_final}). To illustrate the power of the fRG flow equation we choose an alternative, more direct route, by expanding Eq.~(\ref{eq:dSigma_V}) up to second order in the bare interaction and allow for arbitrary values of the voltage $V$. 

To first order in the interaction the single-scale propagator, Eq.~(\ref{eq:SSprop}), reads
\begin{align}
S_{\dg{1},V}^{\re{2}|\re{2}} & = 
\mydot{\mathcal{G}}_{\dg{0}}^{\re{2}|\re{2}} + \mathcal{G}_{\dg{0}}^{\re{2}|\re{1}}\Sigma_{{\dg{1}}}^{\re{1}|\re{2}}\mydot{\mathcal{G}}_{\dg{0}}^{\re{2}|\re{2}} + 
\mydot{\mathcal{G}}_{\dg{0}}^{\re{2}|\re{2}}\Sigma_{{\dg{1}}}^{\re{2}|\re{1}}\mathcal{G}_{\dg{0}}^{\re{1}|\re{2}}.
\label{SS_SOPT}
\end{align}
Inserting both Eq.~(\ref{SS_SOPT}) and the SOPT vertex, Eq.~(\ref{eq:vertex_SOPT}), in Eq.~(\ref{eq:dSigma_V}) directly yields
\begin{widetext}
\begin{align}
& \mydot{\Sigma}_{\dg{2},\bl{i}|\bl{j}}^{\sigma,\re{1}|\re{2}}(\e)   = \! 
\frac{-i}{2\pi}  \int d\e'  \Big[\delta_{\bl{ij}} u_{\bl{i}}
\mydot{\mathcal{G}}_{0,\bl{i}|\bl{i}}^{\bar{\sigma},\re{2}|\re{2}} 
+ \delta_{\bl{ij}}\!\!\sum_{\bl{k}}\! \left[
u_{\bl{i}}\!\left(\!\mathcal{G}_{0,\bl{i}|\bl{k}}^{\bar{\sigma}\re{2}|\re{1}}\Sigma_{\dg{1},\bl{k}|\bl{k}}^{\bar{\sigma}\re{1}|\re{2}}\mydot{\mathcal{G}}_{0,\bl{k}|\bl{i}}^{\bar{\sigma}\re{2}|\re{2}} 
\!+\!
\mydot{\mathcal{G}}_{0,\bl{i}|\bl{k}}^{\bar{\sigma},\re{2}|\re{2}}\Sigma_{\dg{1},\bl{k}|\bl{k}}^{\bar{\sigma},\re{2}|\re{1}}\mathcal{G}_{0,\bl{k}|\bl{i}}^{\bar{\sigma},\re{1}|\re{2}}
\right) \!\!+\!\mydot{\mathcal{G}}_{\dg{0},\bl{k}|\bl{k}}^{\sigma,\re{2}|\re{2}}\Delta_{\bl{ik}}^{\sigma\sigma}(0)\!\right]
\notag \\
&  \hspace{9em} + \mydot{\mathcal{G}}_{\dg{0},\bl{i}|\bl{j}}^{\sigma,\re{2}|\re{2}} X_{\bl{ij}}^{\sigma\sigma,\re{1}|\re{2}}(x)
+ \mydot{\mathcal{G}}_{\dg{0},\bl{i}|\bl{j}}^{\bar{\sigma},\re{2}|\re{2}} \big(X_{\bl{ij}}^{\sigma\bar{\sigma},\re{1}|\re{2}}(x) + \Pi_{\bl{ij}}^{\re{1}|\re{2}}(p) \big)\Big], \notag \\
& \mydot{\Sigma}_{\bl{i}|\bl{j}}^{\sigma,\re{2}|\re{1}}(\e)   =  \left[\mydot{\Sigma}_{\bl{i}|\bl{j}}^{\sigma,\re{1}|\re{2}}(\e)\right]^{*}, \notag \\
& \mydot{\Sigma}_{\bl{i}|\bl{j}}^{\sigma,\re{1}|\re{1}}(\e)   = \! \frac{-i}{2\pi} \int d\e' 
\Bigg[ 
\mydot{\mathcal{G}}_{\dg{0},\bl{i}|\bl{j}}^{\sigma,\re{2}|\re{2}} X_{\bl{ij}}^{\sigma\sigma,\re{1}|\re{1}}(x)
+ \mydot{\mathcal{G}}_{\dg{0},\bl{i}|\bl{j}}^{\bar{\sigma},\re{2}|\re{2}} \left(X_{\bl{ij}}^{\sigma\bar{\sigma},\re{1}|\re{1}}(x) + \Pi_{\bl{ij}}^{\re{1}|\re{1}}(p) \right)
\Bigg],
\label{eq:SOPT_diff}
\end{align}
\end{widetext}
where the derivative of the Keldysh bare Green's function is given by [e.g.\ Eq.~(\ref{eq:Dyson_bare})]
\begin{equation}
\mydot{\mathcal{G}}_{\dg{0}}^{\re{2}|\re{2}} = \mathcal{G}_{\dg{0}}^{\re{2}|\re{1}} \mydot{\Sigma}_{{\rm{lead}}}^{\re{1}|\re{1}}\mathcal{G}_{\dg{0}}^{\re{1}|\re{2}}\! =\! 2i \mathcal{G}_{\dg{0}}^{\re{2}|\re{1}}\Big(\sum_{k\in l,r} \mydot{{f^k}}\Gamma^k\Big)\mathcal{G}_{\dg{0}}^{\re{1}|\re{2}}.
\end{equation}
For compactness, we dropped all arguments that match the integration frequency in Eq.~(\ref{eq:SOPT_diff}).
 
It is important to note that the energy integral $\int\! d\e'$ in
Eq.~(\ref{eq:SOPT_diff}) can be performed trivially for the special
case of zero temperature, $T\!=\!0$: Then the derivative of the Fermi
functions in $\mydot{\mathcal{G}}_{\dg{0}}^{\re{2}|\re{2}}$ are Dirac
delta functions [for the definition of the voltage see
Sec.(\ref{sec:diff_cond})]:
\begin{align}
\mydot{f}^l(\e') &\overset{T=0}{=} e \alpha\cdot \delta(\e' - \mu - e \alpha V) \notag \\
\mydot{f}^r(\e') &\overset{T=0}{=} e (\alpha-1)\cdot \delta(\e' - \mu - e (\alpha-1) V).
\end{align} 
This reduces the integration in Eq.~(\ref{eq:SOPT_diff}) to evaluating
the integrand at the chemical potentials of the left and right lead,
respectively. Naturally, this simplification proves extremely
beneficial: we can express the self-energy at arbitrary voltage as
\begin{align}
\Sigma(V) = \Sigma(0) + \int_{0}^{V} \!\!d V'~ \mydot{\Sigma}(V').
\end{align}
Numerically calculating this voltage integration provides both the self-energy $\Sigma(V')$ and its derivative $\mydot{\Sigma}(V')$ within the whole intervall $0\!\leq\! V'\! \leq\! V$. Hence, this procedure can save orders of magnitude of calculation time compared to the direct evaluation of the self-energy and its voltage derivative via Eq.~(\ref{eq:Sigma}) and Eq.~(\ref{eq:SOPT_diff}), respectively. 

\section{1D Model of a QPC\label{sec:QPC}}

As an application of the above formalism, we now study the influence
of electron-electron interactions on the linear conductance of a
one-dimensional symmetric potential barrier of height $V_c$
(measured w.r.t. the chemical potential $\mu$) and parabolic near the
top,
\begin{equation}
V(x) = V_{c} + \mu - \frac{m \Omega_x^2}{2 \hbar^2} x^2,
\label{eq:potential_definition}
\end{equation}
where $m$ is the electron's mass. The geometry of the barrier is
  determined by the energy scale $\Omega_x$ and the length scale
  $l_x\!=\!\hbar/\sqrt{2m\Omega_x}$.  While the system extends to
infinity, the potential is non-zero only within the central region
$C$, defined by $-\ell/2 \!<\! x\! <\! \ell/2$, and drops smoothly to
zero as $|x|$ approaches $|\ell|/2$. We call the outer homogeneous
regions the left lead $L$ ($x\!<\!-\ell/2$) and the right lead $R$
($x\!>\!\ell/2$).

Numerics cannot deal with the infinite Hilbert space of this
continuous system. Hence, we discretize real space using the method of
finite differences (see Appendix~\ref{App:MoFD} for details),
which maps the system onto a discrete set of space points
$\{ x_{\bl j}\}$. This results in the tight-binding representation
\begin{align}
 \hspace{-3.5mm} H\!=\! \sum_{\bl{j}\sigma} [E_{\bl{j}}^{\sigma} n_{\bl{j}\sigma} \! - \! \tau_\bl{j}
 (d^\dagger_{\bl{j}\sigma} d^{}_{\bl{j+1}\sigma} + {\rm h.c.})] \!+\! \sum_{\bl{j}\in C} U_{\bl{j}} n_{\bl{j}\uparrow} n_{\bl{j}\downarrow},
 \label{eqModelChain}
\end{align}
with spin-dependent onsite energy $E_{\bl{j}}^{\sigma}\! = \! E_{\bl{j}} \! -\! \sigma B/2 \! = \!
V_{\bl{j}}\! +\!  \tau_{\bl{j-1}}\! +\!  \tau_{\bl{j}}\! -\! \sigma B/2$, site-dependent hopping amplitude $\tau_{\bl{j}} =
\hbar^2/(2ma_{\bl{j}}^2)$, spacing $a_\bl{j}\! =\! x_{\bl{j+1}}-x_\bl{j}$ and
potential energy $V_\bl{j} = V(x_\bl{j})$.  Note that we included a homogeneous Zeeman-field
$B$ to investigate magnetic field dependencies, as well as an onsite-interaction, whose
strength is tuned by the site-dependent parameter $U_\bl{j}$. 

In Ref.~\cite{Bauer2013} we have used this model to investigate the physics of a quantum point contact (QPC), a short one-dimensional constriction
We  showed that the model suffices to reproduce the main features of the 0.7 anomaly, including the strong reduction of conductance as function of magnetic field, temperature and source-drain voltage in a \textit{sub-open} QPC (see below). We argued, that the appearance of the 0.7 anomaly is due to an interplay of a maximum in the local density of states (LDOS) just above the potential barrier (the ``\textit{van-Hove} ridge'') and electron-electron interactions. 

In Ref.~\cite{Bauer2013} we have introduced a real space discretization scheme that dramatically minimizes numerical costs. Here, we discuss this scheme in more detail.
We discuss both the noninteracting physics of the model as well as the magnetic field and temperature dependence of the linear conductance in the presence of interactions using SOPT.

\subsection{The choice of discretization}

For a proper description of the continuous case it is essential to choose the spacing much smaller than the length scale on which the potential changes (condition of adiabatic discretization). We model the central region by $N \!=\! 2N'\!+\!1$ sites, located at the space points $\{x_{\bl{-N'}},x_{\bl{-N'+1}},...,x_{\bl{N'-1}},x_{\bl{N'}}\}$, where $N\!\gtrsim\!100$ proves sufficient for a potential of the form Eq.~(\ref{eq:potential_definition}). 
Due to the parity symmetry of the barrier we always choose $x_{\bl 0}\!=\!0$ and $x_{\bl{j}}\!=\! - x_{\bl{-j}}$. 

The discretization of real space introduces an upper bound, $E^{\rm{max}} = {\rm{max}}(V_{\bl j}\!+2\tau_{\bl{j-1}} + 2\tau_{\bl j})$,
 for the eigenenergies of the bare Hamiltonian. In addition, it causes the formation of a site-dependent energy band, defined as the energy intervall where the local density of states (LDOS) is non-negligible, i.e. where eigenstates have non-negligible weight. In case of an adiabatic discretization this energy band follows the shape of the potential. At a site $\bl{j}$ it is defined within the upper and lower band edge
   \begin{figure*}
 ~\hspace{-3mm}
\includegraphics[width = 183mm]{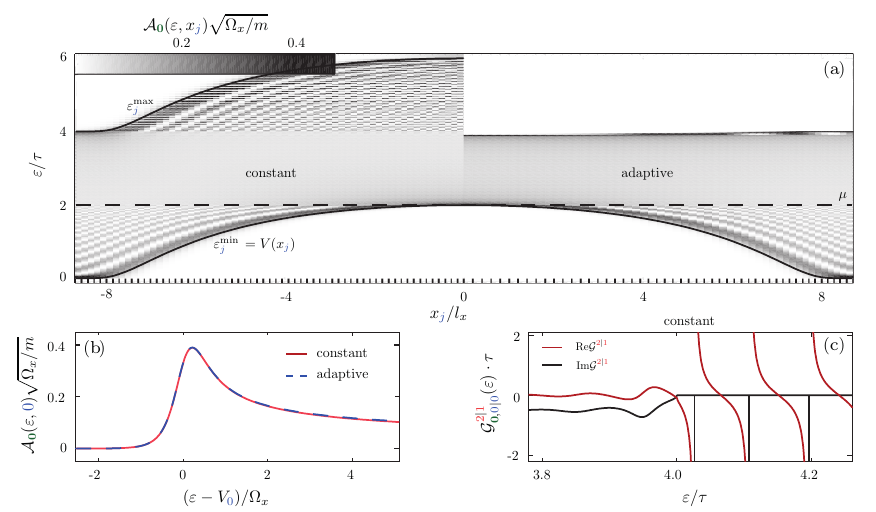}\\
\caption{\small {(a), left half: The non-interacting LDOS of the central region, $\mathcal{A}_{\dg{0}}(\e,x_{\bl{j}})$, resulting from a constant real-space discretization. The position of the discrete points $x_{\bl{j}}$ is indicated by the x-axis ticks. Both the lower and upper band edge follow the shape of the potential: $\e_{\bl{j}}^{{\rm{min}}}\!=\!V(x_{\bl{j}})$ and $\e_{\bl{j}}^{{\rm{max}}}\!=\!V(x_{\bl{j}})+4\tau$. The local maximum of $\e_{\bl{j}}^{{\rm{max}}}$ at $\bl{j}\!=\!0$ causes the formation of bound states for energies $\e\!>\!4\tau$. (c), their discrete spectrum shows up as poles in the non-interacting Green's function $\mathcal{G}_{\dg{0},\bl{0}|\bl{0}}(\e)$. (a), right half: The non-interacting LDOS of the central region resulting from an adaptive real-space discretization with $c\!=\! 0.55$ [Eq.~(\ref{eq:non-monotonic_hopping})], i.e. the spacing $a_{\bl{j}}$ increases towards the barrier center (see x-axis ticks). Hence, the band width decreases with increasing barrier height, resulting in a local minimum of $\e_{\bl{j}}^{{\rm{max}}}$ at $\bl{j}\!=\!0$. (b), the LDOS at the central site, $\mathcal{A}_{\dg{0}}(\e,\bl{0})$, for both schemes. }}
\label{fig:band_constant_a}
\end{figure*}
\begin{equation}
\varepsilon_{\bl j}^{\rm{min}}  = V_{\bl{j}}, \qquad 
\varepsilon_{\bl j}^{\rm{max}} = V_{\bl{j}} + w_{\bl j},
\label{eq:band_edges}
\end{equation}
where the band width depends on the local spacing, i.e.\ on the choice of discretization (see Appendix~\ref{App:MoFD} for additional information):
\begin{equation}
 w_{\bl{j}} = 2\tau_{\bl{j-1}}  + 
 2\tau_{\bl{j}} = \frac{\hbar^2}{m}\left(\frac{1}{a_{\bl{j-1}}^2}+\frac{1}{a_{\bl j}^2}\right).
 \label{eq:band_width_model}
 \end{equation}
 Note that a larger distance between successive sites leads to a narrowing of the energy band and vice versa; while the lower band edge is, for any adiabatic discretization, directly given by the potential, the upper band edge depends sensitively on the applied discretization scheme. 

In the following we discuss and compare two different discretization procedures:  The standard approach of equidistant  discretization (constant hopping $\tau$) causes a local maximum $\varepsilon_{\bl{0}}^{{\rm{max}}}\!=\! V_{\bl{0}}+2\tau$ of the upper band edge in the vicinity of the barrier center. This approach leads to artificial bound states far above the potential barrier, which complicate numerical implementation and calculation. Hence, we recommend and apply an alternative adaptive scheme 
where the spacing increases (the band width decreases) with increasing potential, i.e.\ towards $\bl{j}\!=\!0$. Note that this still implies a constant hopping $\tau_{|\bl{j}|>N'}\!=\!\tau$ in the leads.

\subsubsection{Constant discretization}

We discuss the case of constant spacing $a\!=\!a_{\bl{j}}$, implying
grid points $x_\bl{j}\!=\! a\bl{j}$ and a constant hopping $\tau \!=\! \hbar^2/(2ma^2)$. In a homogeneous system, $V(x_{\bl{j}})\!=\!0$, the energy eigenstates are Bloch waves $\psi_{k}(x_{\bl{j}})= e^{ika\bl{j}}$, which form an energy band $\e_k = 2\tau[1-\cos(ka)]$
of width $w\!=\!4\tau$. Adding the parabolic potential, 
\begin{align}
V(x_\bl{j}) = V_c + \mu - \frac{\Omega_x^2}{4\tau}\bl{j}^2,
\label{eq:potential_discrete_constant_spacing}
\end{align} 
these states are now subject to scattering at the barrier which causes the formation of standing wave patterns for energies $\varepsilon\!<\!V_{\bl 0}=\!V(\bl{0})\!=\!V_c\!+\!\mu$ below the barrier top. The left half ($x_{\bl j}\!<\!0$) of Fig.~\ref{fig:band_constant_a}(a) shows the noninteracting central region's local density of states (LDOS), $\mathcal{A}_{\dg{0}}^{\sigma}(x_{\bl{j}},\e) \!=\! -1/(\pi a)\! \cdot\!
{\rm{Im}}\mathcal{G}_{\dg{0},\bl{j}|\bl{j}}^{\sigma,\re{2}|\re{1}}(\e)$ at $B\!=\!0$, as a function of position
$x_{\bl{j}}$ and energy $\e$. Due to the condition of
adiabaticity the energy band smoothly follows the shape of the potential, implying a
site-dependent upper band edge, $\e^{{\rm{max}}}(x_{\bl{j}}) \!= \!V_\bl{j}\!+4\tau$.  

The local maximum of $\e^{\rm{max}}(x_{\bl{j}})$ in the central region's center generates
artificial bound states, owed to the discretization scheme, in the energy interval $\e\! \in\!
[4\tau,4\tau+\!V_{\bl{0}}]$. This is illustrated in Figure \ref{fig:band_constant_a}(c), where
the real and imaginary parts of the bare Green's function of the central
site, $\mathcal{G}_{\dg{0},\bl{0}|\bl{0}}^{\re{2}|\re{1}}(\e)$, are plotted. These bound states result from the shape of the upper band edge: Since the band in the homogeneous leads is restricted to energies
below $4\tau$ (unlike in the continuous case), all states with higher energy are spacially confined to within the central region, have an infinite
lifetime and form a discrete spectrum, determined by the shape of the applied
potential $V(x_{\bl{j}})$. 

The calculation of
self-energy and two particle vertex, Eq.~(\ref{eq:Sigma}) and
Eq.~(\ref{eq:vertex_final}), is performed by ad-infinitum frequency integrations over
products of Green's functions.  Thus, the energy region of the upper band edge and the local bound states must be included in their
calculation with adequate care. This involves determining the
exact position and weigth of the bound states, which requires high numerical
effort, as well as dealing with the numerical evaluation of principal value
integrals and convolutions, where one function has poles and
the other one is continuous. While all this is doable with sufficient dedication, we can avoid such complications
entirely by adapting the discretization scheme, discussed next.

 \subsubsection{Adaptive discretization} According to
  Eq.~(\ref{eq:band_edges}) and Eq.~(\ref{eq:band_width_model}) we can
  modify the band width locally by choosing non-equidistant
    discretization points. In the following we discuss a non-constant
  discretization scheme that reduces the band width within the central
  region enough so that the upper band edge exhibits a local minimum
  at $x_{\bl{0}}$ rather than a local maximum (as in the case of
  constant spacing). In consequence the Green's functions are
  continuous within the whole energy band, which facilitates a
  numerical treatment of interactions.

For a non-constant real space discretization it proves useful to first define the onsite energy $E_{\bl{j}}$ and the hopping $\tau_{\bl{j}}$ of the discrete tight-binding Hamiltonian Eq.~(\ref{eqModelChain}) and then use these expressions to calculate the geometry of the corresponding physical barrier, i.e.\ its height $V_c$ and curvature $\Omega_x$.

We specify the onsite energy to be quadratic near the top with
\begin{align}
E_\bl{j}  = \tilde{E}_\bl{j} + 2\tau \simeq \tilde{E}_\bl{0}\left[1-\frac{\bl{j}^2}{N'^2}\right] + 2\tau,
\label{eq:onsite_potential_non-monotonic_hopping}
\end{align}
where $\tilde{E}_{\bl{0}}$ is positive.  We use the shape of
$\tilde{E}_\bl{j}$ within $C$ (which, apart from its height and the
quadratic shape around the top does not influence transport
properties, as long as $\tilde{E}_\bl{j}$ goes adiabatically to zero
upon approaching $\bl{j}=|N'|$) to define a site-dependent hopping
(amounting to a site-dependent spacing)
\begin{align}
\tau_\bl{j} = \tau\left[ 1-\frac{c}{2\tau}\left(\tilde{E}_\bl{j}+\tilde{E}_{\bl{j+1}}\right)\right],
\label{eq:non-monotonic_hopping}
\end{align}
where we have introduced a dimensionless positive  parameter $c <\tau/\tilde{E}_{\bl{0}}$ that determines how strongly the band width is to be reduced. Note that Eq.~(\ref{eq:non-monotonic_hopping}) describes a hopping, that is constant ($=\tau$) in the leads, where $\tilde{E}_{\bl{j}}\!=\!V_{\bl{j}}\!=\!0$, and decreases with increasing $\tilde{E}_{\bl{j}}$ in the central region. This corresponds to a site-dependent lattice spacing $a_{\bl{j}}\!=\! a \sqrt{\tau/\tau_{\bl{j}}}$, which increases towards the center of the central region. The real space position $x_{\bl{j}}$ that corresponds to a site $\bl{j}$ is given by
\begin{align}
x_{\bl{j}} = \operatorname{sgn}(\bl{j})\sum_{\bl{j'}=1}^{|\bl{j}|}a_{\bl{j'}} = a \sqrt{\tau}\operatorname{sgn}(\bl{j})\sum_{\bl{j'}=1}^{|\bl{j}|}\frac{1}{\sqrt{\tau_{\bl j}}},
\end{align}
where $\operatorname{sgn}(x)$ is the sign function.
Following Eq.~(\ref{eq:band_edges}), the construction introduced in Eq.~(\ref{eq:onsite_potential_non-monotonic_hopping}) and Eq.~(\ref{eq:non-monotonic_hopping}) leads to an upper band edge given by
\begin{align}
\e_{\bl{j}}^{\rm{max}} \simeq E_\bl{j} + \tau_{\bl{j-1}} + \tau_\bl{j} \simeq 4\tau +  (1-2c)\tilde{E}_\bl{j},
\end{align}
which for the choice $c\!>\!0.5$ indeed exhibits a smooth local minimum at $\bl{j}\!=\!0$, thus avoiding the bound states discussed above for the constant discretization, $c\!=\!0$.

\begin{figure*}
\includegraphics[width = 183mm]{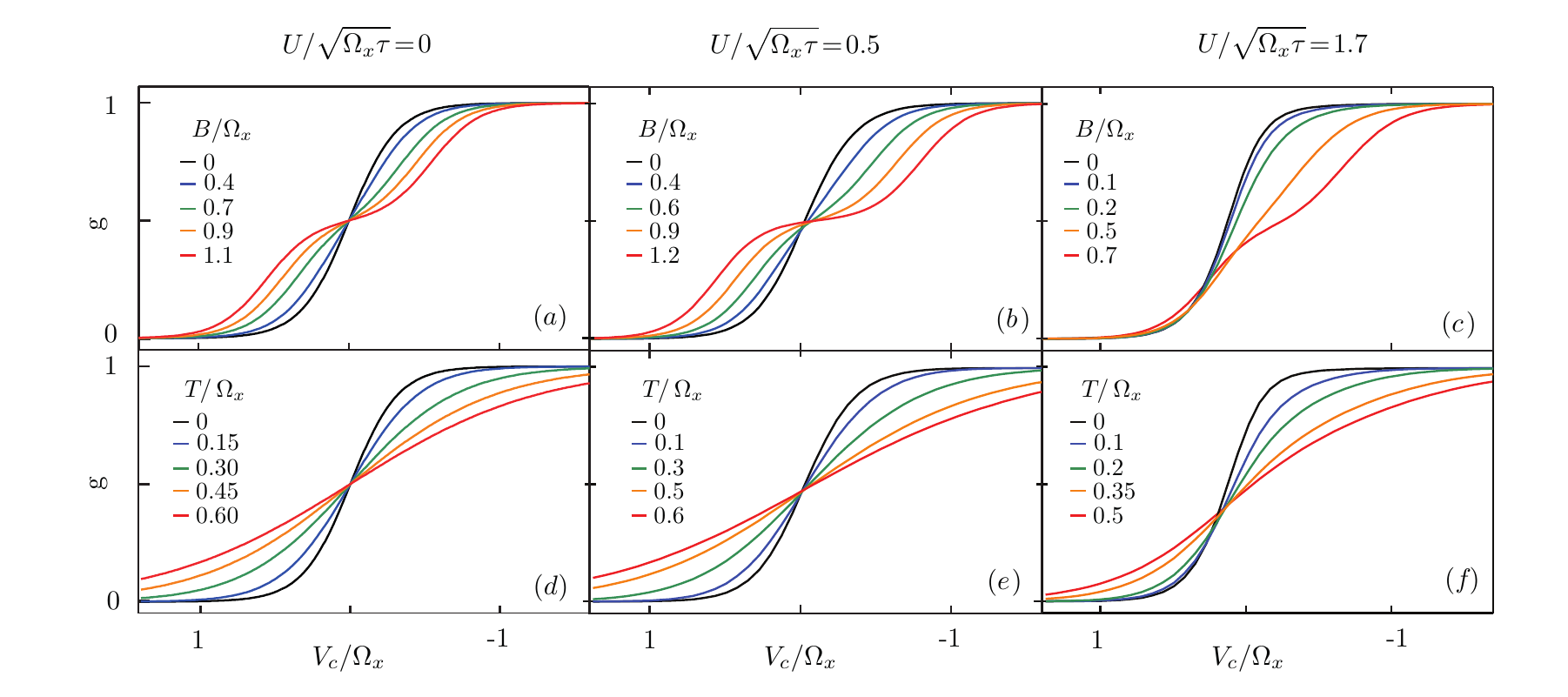}\\
\caption{\small: (a)-(c), Linear conductance as a function of barrier height $V_c$ for some values of magnetic field $B$ with interaction strength $U$ increasing from left to right. (d)-(f), Linear conductance as a function of barrier height $V_c$ for some values of temperature $T$ with interaction strength $U$ increasing from left to right. Interactions cause an asymmetric evolution of conductance with magnetic field and temperature due to the interaction-enhanced reduction of conductance in the \textit{sub-open} regime -- the 0.7 anomaly.}
\label{fig:results}
\end{figure*}

Despite the drastic manipulation of $\e_{\bl{j}}^{\rm{max}}$, the lower band edge still serves as a proper potential barrier,
\begin{align}
\e_{\bl{j}}^{\rm{min}}\! =\!V_\bl{j} \simeq  (1+2c)\tilde{E}_\bl{j},
\end{align}
with a quadratic potential barrier top whose height now depends on the compensation factor $c$:
\begin{align}
V_\bl{j} \simeq (1+2c)\tilde{E}_\bl{j}\left[1-\frac{\bl{j}^2}{N'^2}\right].
\end{align}
Finally, we write the potential barrier in the form given in Eq.~(\ref{eq:potential_discrete_constant_spacing}), i.e.\ express the curvature $\Omega_x$ in units of the constant lead-hopping $\tau$. By comparison we find
\begin{equation}
V_c  = V_{\bl{0}} - \mu, \qquad
\Omega_x  = \frac{2}{N'}\sqrt{V_{\bl{0}}\tau_{\bl{0}}}.
\label{eq:Wx_Vc_definitions}
\end{equation}
The right half ($x_{\bl{j}}\!>\!0$) of
Fig.~\ref{fig:band_constant_a}(a) shows the LDOS of the central region
for an adaptive discretization with $c\!=\!0.55$. All additional
parameters are chosen such that the resulting potential barrier
matches the case of constant discretization (plotted for
$x_{\bl{j}}\!<\!0$). Most importantly, the minimum of
$\e_{\bl{j}}^{\rm{max}}$ at $\bl{j}\!=\!0$ prevents the occurance of
bound states above the barrier, which allows for a faster numerical
evaluation of the vertex functions. Importantly, both discretization
schemes approximate the same physical system; their differences are
non-neglegible only for energies far above the barrier, i.e. far away
from the energies relevant for transport. This can be seen from the
matching grey scale at the interface $\bl{j}\!=\!0$ for energies
$\e<V_\bl{0}+\mathcal{O}(\Omega_x)$, as well as from comparison of the
central site's LDOS in Fig.~\ref{fig:band_constant_a}(c).

\vspace{3em}

\subsection{The choice of system parameters}     

To ensure that the discrete model reflects the transport properties of the continuous barrier, Eq.~(\ref{eq:potential_definition}), the chemical potential of the system (or of both leads in non-equilibrium) must be chosen far enough below the global minimum of $\e^{\rm{max}}(x_{\bl{j}})$. Only in this case the unphysical upper band edge does not contribute to the results. The onsite-energy is chosen as
\begin{align}
\tilde{E}_\bl{j} & = 
\theta(N'-|\bl{j}|)\tilde{E}_\bl{0}  \exp\!\left(-\frac{\left(\frac{\bl{j}}{N'}\right)^2}{1-\left(\frac{\bl{j}}{N'}\right)^2}\right),
\end{align}
where $\theta(x)$ is the Heavyside step function. Note, that this definition is consistent with Eq.~(\ref{eq:onsite_potential_non-monotonic_hopping}). In order to calculate the site-dependent coupling we use $c\!=0.55$ in Eq.~(\ref{eq:non-monotonic_hopping}). Hence, for a barrier height $V_\bl{0}\!=\!\mu$ (corresponding to a noninteracting transmission $\mathcal{T}_{\dg{0}}\!=\!0.5$, see Eq.~(\ref{eq:Buettiker}) below),  we get a potential curvature $\Omega_x\!=\!0.039\tau$. Finally, the shape of the onsite interaction is chosen as
\begin{align}
U_\bl{j} & = 
\theta(N'-|\bl{j}|)U_\bl{0}  \exp\!\left(-\frac{\left(\frac{\bl{j}}{N'}\right)^6}{1-\left(\frac{\bl{j}}{N'}\right)^2}\right).
\label{eq:U_definition}
\end{align}

\subsection{Non-interacting properties of the model}

In Ref.~{\cite{Bauer2013}} we argued that the model of Eq.~(\ref{eqModelChain}), combined with a potential with parabolic barrier top, Eq.~(\ref{eq:potential_definition}), is sufficient to describe the physics of the lowest subband of a QPC: Making a saddle-point ansatz for the electrostatic potential caused by voltages applied to a typical QPC gate structure provides an effective 1D-potential of the form Eq.~(\ref{eq:potential_definition}). Information about the transverse geometry of the QPCs potential can be incorporated into the site-dependent effective interaction strength $U_\bl{j}$, see Eq.~(\ref{eq:U_definition}).

The non-interacting, spin-dependent transmission through a quadratic barrier of height $V_{\bl{0}}=V_c+\mu$ and curvature $\Omega_x$, Eq.~(\ref{eq:potential_definition}), in the presence of a magnetic field $B$ can be derived analytically \cite{Buettiker1990} and is given by
\begin{align}
\mathcal{T}_{\dg{0}}^\sigma(\e) = 
\frac{1}{e^{-2\pi (\e-V_\bl{0}+\sigma B/2)/\Omega_x} + 1}.
\label{eq:Buettiker}
\end{align}
Hence, according to the Landauer-B\"uttiker formula, the
non-interacting (bare) linear conductance,
\begin{align}
\text{g}_{\dg{0}} = -\frac{e^2}{h}\sum_\sigma \int_{-\infty}^{\infty} f'(\e) \mathcal{T}_{\dg{0}}^\sigma(\e),
\label{eq:Landauer_Buettiker}
\end{align}
is a step function of width $\Omega_x$ at $B\!=\!T\!=\!0$, changing
from $0$ to $1$, when the barrier top is shifted through $\mu$ from
above. This step gets broadened with temperature [see Figure
\ref{fig:results}(d)] and develops a double-step structure with
magnetic field [see Figure \ref{fig:results}(a)]. For all $B$ and $T$
the bare conductance obeys the symmetry
$\text{g}_{\dg{0}}(V_c) = 1-\text{g}_{\dg{0}}(-V_c)$.

Furthermore, an analytic expression for the non-interacting LDOS at the chemical potential in the barrier center as function of barrier height $V_c$ can be calculated [see e.g.\ Ref.~\cite{Sloggett2008}],

\begin{align}
\mathcal{A}_{\dg{0}}(\e\!=\!\mu,\bl{0}) = \frac{\left|\Gamma\left(1/4+i V_c/(2\Omega_x)\right)\right|^2}{4\sqrt{2}\pi^2e^{\pi V_c/(2\Omega_x)}} ,
\label{eqLDOSU0center}
\end{align}
where $\Gamma(z)$ is the complex gamma-function. This is a smeared and shifted version of the 1D van Hove singularity [see Ref.~{\cite{Bauer2013}} for further details], peaked at 
$V_c\!=\! -\mathcal{O}(\Omega_x)$, i.e.\ if the barrier top lies sightly below the chemical potential. Here, the value of the noninteracting conductance is given by $\text{g}_{\dg{0}} \approx 0.8$. Hence, we call this parameter regime \textit{sub-open}.

\subsection{Interacting results}

As was discussed in Ref.~\cite{Bauer2013}, the shape of the LDOS in the barrier center lies at the heart of the mechanism causing the 0.7 conductance anomaly: Semiclassically, the LDOS can be interpreted as being inversely proportional to the velocity $v$ of the charge carriers, $\mathcal{A}_{\dg{0}}(\e,x_{\bl{j}}) \propto 1/v_{\bl{j}}(\e)$. Hence, the average time that a non-interacting electron with energy $\e=\mu$ spends in the vicinity of the barrier center is maximal in the \textit{sub-open} regime (where $\mathcal{A}_{\dg{0}}(\mu,\bl{0})$ is maximal, see Eq.~(\ref{eqLDOSU0center}) and its subsequent discussion), resulting in an enhanced scattering probability and thus a strong reduction of conductance at finite interaction strength in this parameter regime. 

Figure \ref{fig:results} compares the bare conductance, calculated via the Landauer-B\"uttiker formula [Eq.~(\ref{eq:Landauer_Buettiker})], with the conductance obtained by taking into account interactions using SOPT, calculated via the Keldysh version of Oguri's formula [Eq.~(\ref{eq:g_sym_ham})], as a function of barrier height $V_c$ for several values of magnetic field (panels (a)-(c)) and temperature (panels (d)-(f)), for three interaction strengths increasing from left to right. For small but finite interactions, $U\sqrt{\Omega_x\tau}=0.5$, the shape of the LDOS causes a slight asymmetry in the conductance curves at (b) finite magnetic field or (e) finite temperature: A finite magnetic field induces an imbalance of spin-species in the vicinity of the barrier center. This imbalance is enhanced by exchange interactions via Stoner-type physics, where the disfavoured spin species (say spin down) is pushed out of the center region by the coulomb blockade of the the favoured spin-species (say spin up). Hence, transport is dominated by the spin-up channel, resulting in a strong reduction of total conductance in the \textit{sub-open} regime even for a small magnetic field. A finite temperature, on the other hand, opens phase-space for inelastic scattering, which, again, is strongest for large LDOS, again resulting in the reduction of conductance in the \textit{sub-open} regime. This interaction-induced trend continues with increasing interactions, and gives rise to a weak 0.7 anomaly at $B\neq 0$, Figure \ref{fig:results}(c), or $T\neq 0$, Figure \ref{fig:results}(f), for intermediate interaction strength, $U\sqrt{\Omega_x\tau}=1.7$. Upon a further increase of interactions, SOPT breaks down (see below), and more elaborate methods are needed to obtain qualitatively correct results. This was done in Ref.~\cite{Bauer2013} and Ref.~\cite{Bauer2014}, where we used fRG to reach interaction strength of up to $U\sqrt{\Omega_x\tau}=3.5$; they yielded a pronounced 0.7 anomaly even at $B=T=0$ and its typical magnetic field development into the spin-resolved conductance steps at high field.

The main limitations of SOPT when treating the inhomogeneous system,
introduced in Eq.~(\ref{eq:potential_definition}), can be explained as
follows: Upon increasing interactions, the LDOS is shifted towards
higher energy, as Hartree contributions cause an effective higher
potential barrier compared to the non-interaction case. As a
consequence, a proper description of interactions requires information
about this shift to be incorporated into the calculation of the vertex
functions via feed-back of the self-energy into all
propagators. However, SOPT calculates the self-energy and the
two-particle vertex [Section \ref{sec:SOPT}] using only bare
propagators, which only carry information of the bare LDOS. Together
with the drastic truncation of the perturbation series beyond second
order, this limits the quantitative validity of SOPT to weak
interaction strength and the qualitative validity of SOPT to
intermediate interaction strength. In particular, the skewing of
  the conductance curves with increasing temperature is typically much
  stronger for measured data curves than seen in
  Fig.~\ref{fig:results}(f).  Nevertheless, SOPT does serve as a
useful too for illustrating the essential physics involved in the
appearance of the 0.7 conductance anomaly.

\section{Conclusion and Outlook}

In this paper we discuss electronic transport through an interacting
region of arbitrary shape using the Keldysh formalism. Starting from
the well-established Meir-Wingreen formula for the system's current we
derive exact formulas for both the differential and linear conductance. In
the latter case we use the fRG flow-equation for the self-energy as
well as a Ward identity, following from the Hamiltonian's particle
conservation, to obtain a Keldysh version of Oguri's linear
conductance formula. As an application, we use
SOPT to calculate the conductance of
the lowest subband of a QPC, which we model by a one-dimensional
parabolic potential barrier and onsite interactions -- a setup we have
recently used to explore the microscopic origin of the 0.7 conductance
anomaly \cite{Bauer2013}. We present detailed discussion of the model's
properties and argue that an adaptive, non-constant real space
discretization scheme greatly facilitates numerical effort. We treat
the influence of interactions using SOPT, presenting all details that
are necessary to employ the derived conductance formulas. Our
SOPT-results for the linear conductance as function of magnetic field
and temperature illustrate that the anomalous reduction of conductance
in the \textit{sub-open} regime of a QPC is due to an interplay of the
\textit{van-Hove} ridge and electron-electron interactions.

A logical next step would be to go beyond SOPT by treating
  interactions using Keldysh-fRG. Work in this direction is currently
  in progress. For example, in Ref.~\cite{Schimmel2017} 
  the conductance formula \eqref{eq:g_sym_ham} was used 
to compute the finite-temperature linear conductance through an
interacting QPC using Keldysh-fRG.

\section*{Acknowledgements} 

We thank S.\ Andergassen, S.\ Jakobs, V.\ Meden and H.\ Schoeller for very
helpful discussions.  We acknowledge support from the DFG via SFB-631,
SFB-TR12, De730/4-3, and the Cluster of
Excellence \emph{Nanosystems Initiative Munich}.

\appendix
\numberwithin{equation}{section}
\section{Properties of Green's and vertex functions in Keldysh formalism}
\label{App:Keldysh}

To investigate transport properties of the system in and out of
equilibrium, we apply the well-established Keldysh formalism
\cite{Keldysh1964,Kamenev2009}. Here we collect some of its standard
ingredients. We mostly follow the definitions and conventions given in
Ref.~\cite{Jakobs2010b}.

All operators carry Keldysh time-contour indices,
${\color{red}a_1},{\color{red}a_1'},\re{a_2},... = \{\re{+},\re{-}\}$,
marking the position of the time argument $t$ of an operator as lying
on the forward (${\color{red}-}$) or backward ({\color{red}+}) branch
of the Keldysh contour. We use Keldysh indices with or without a
prime, $\re{a}$ or $\re{a'}$, to label the time arguments of
annihilation or creation operators, respectively. Since the model
Hamiltonian, Eq.~(\ref{eq:ModelGeneral}), is time-independent, the
only non-zero matrix elements of the Hamiltonian in contour space have
equal contour indices:
\begin{align}
\mathcal{H}_{\dg{0}}^{\re{a_1}|\re{a_1'}} & = -{\color{red}a_1}\cdot \delta_{\color{red} a_1 a_1'} H_{\dg{0}}, \nonumber \\
\mathcal{H}_{\rm{int}}^{{{\color{red}a_1 a_2}|{\color{red} a_1' a_2'}}} & = -{\color{red}a_1}\cdot\delta_{\color{red}a_1 a_2}\delta_{\color{red}a_1 a_1'}\delta_{\color{red}a_1 a_2'} H_{\rm{int}},
\end{align}
with $\{ \re{a}\}$ labeling the time arguments of annihilation operators and $\{\re{a'}\}$ labeling the time arguments of creation operators. Note that a calligraphic $\mathcal{H}$ carries contour indices, while a capital $H$ does not.

We define time-dependent, $n$-particle Keldysh Green's functions as the expectation values
\begin{align}
 & G_{\mbox{\scriptsize{\boldmath$\bl{i}$}}|\mbox{\scriptsize{\boldmath$\bl{i'}$}}}^{n,\mbox{\scriptsize{\boldmath$ \re{a}$}}|
 \mbox{\scriptsize{\boldmath$ \re{a'}$}}\mbox{\boldmath{}}}(\mbox{\normalsize{\boldmath$t$}}|\mbox{\normalsize{\boldmath$t'$}}\mbox{\boldmath{}}) = G_{\bl{i_1},...,\bl{i_n}|\bl{i_1'},...\bl{i_n'}}^{{\color{red}a_1},...,{\color{red}a_n}|{\color{red}a_1'},...{\color{red}a_n'}}(t_1,...,t_n|t_n',...,t_1')  = \notag \\
& (-i)^n\langle\mathcal{T}_c d_{\bl{i_1}}^{\color{red}a_1}(t_1) ... d_{\bl{i_n}}^{\color{red}a_n}(t_n) [d_{\bl{i_n'}}^{\color{red}a_n'}]^\dagger(t_n') ... [d_{\bl{i_1'}}^{\color{red}a_1'}]^\dagger(t_1')\rangle,
\label{eq:define_Green}
\end{align}
where we use boldface notation for multi-indices $\mbox{\boldmath{$x$}} = (x_1,...,x_n)$. The operator $d_{\bl i}^{\re{a}}(t)/\left[d_{\bl i}^{\re{a}}\right]^\dagger\!(t)$ destroys/creates an electron at time $t$ on contour branch $\re{a}$ in quantum state $\bl{i}$, and the time-ordering operator $\mathcal{T}_c$ moves later contour times to the left. In case of equal time arguments, annihilation operators are always arranged to the right of creation operators. The bare, non-interacting Green's function, whose time-dependence is governed by the quadratic part of the Hamiltonian, $\mathcal{H}_{\dg{0}}$, carries an additional subscript, $G_{\dg{0}}$.

We define anti-symmetrized, irreducible, $n$-particle vertex functions, $\gamma_{\mbox{\scriptsize{\boldmath$\bl{i'}$}}|\mbox{\scriptsize{\boldmath$\bl{i}$}}}^{n,\mbox{\scriptsize{\boldmath$ \re{a'}$}}|
 \mbox{\scriptsize{\boldmath$ \re{a}$}}\mbox{\boldmath{}}}(\mbox{\normalsize{\boldmath$t'$}}|\mbox{\normalsize{\boldmath$t$}}\mbox{\boldmath{}})$, as the sum of all $1$-particle irreducible ($1$PI)  diagrams with $n$ amputated ingoing and $n$ amputated outgoing legs. For an explicit series representation of the one- and two-particle vertex, see Eq.~(\ref{eq:vertex_series}). A formula for the prefactor of every single diagram is given by Eq.~(20) of Ref.~\cite{Jakobs2010b}. 
 
 The Dyson equation provides a direct relation between the one-particle Green's and vertex function:
 \begin{equation}
G(t_1|t_1') \!=\! G_{\dg{0}}(t_1|t_1') \!-\! \int\! d\tau_1d\tau_1' G_{\dg{0}}(t_1|\tau_1')\gamma(\tau_1'|\tau_1)G(\tau_1|t_1').
\label{eq:Dyson_time}
\end{equation}
Here and below, whenever quantum state indices $\bl{i}$ and contour
indices $\re{a}$/Keldysh indices $\re{\alpha}$ are implicit, they are
understood to be summed over in products.

Decomposing the two-particle Green's yields a connection to the
two-particle vertex function via
\begin{align}
& G(t_1,t_2|t_1',t_2') =  G(t_1|t_1')G(t_2|t_2') - G(t_1|t_2')G(t_2|t_1') \notag \\
& -\!  i\! \int\! d\mbox{\boldmath$\tau$} G(t_1|\tau_1')G(t_2|\tau_2') \gamma(\tau_1',\tau_2'|\tau_1,\tau_2)G(\tau_1|t_1')G(\tau_2 | t_2').
\label{eq:partitioning_Green4}
\end{align}

Our choice of sign for $\gamma$ is opposite to that of Ref.~\cite{Jakobs2010b}.

Since the Hamiltonian, Eq.~(\ref{eq:ModelGeneral}), is
time-independent, the Green's/vertex functions are translationally
invariant in time, implying that $n$-particle functions depend on
$2n-1$ time arguments only:
\begin{align}
 G(t_1,...,t_n|t_1',...t_n')  & = G(0,...,t_n-t_1|t_1'-t_1,...,t_n'-t_1), \nonumber \\
 \gamma(t_1',...,t_n'|t_1,...t_n) &  = \gamma(0,...,t_n'-t_1'|t_1-t_1',...,t_n-t_1').
 \label{eq:time_translational_invariance}
\end{align}
As a consequence, the Fourier-transform,
\begin{align}
G(\mbox{\normalsize{\boldmath$\e$}}|\mbox{\normalsize{\boldmath$\e'$}}\mbox{\boldmath{}})  & = \int \!d\mathbf{t}d\mathbf{t'}~ 
e^{i\mbox{\footnotesize{\boldmath$\e t$}}\mbox{\boldmath{}}} 
e^{-i\mbox{\footnotesize{\boldmath$\e' t'$}}\mbox{\boldmath{}}
}
G(\mathbf{t} | \mathbf{t'}), \nonumber \\
\gamma(\mbox{\normalsize{\boldmath$\e'$}}|\mbox{\normalsize{\boldmath$\e$}}\mbox{\boldmath{}})  & = \int \!d\mathbf{t}d\mathbf{t'}~ 
e^{i\mbox{\footnotesize{\boldmath$\e' t'$}}\mbox{\boldmath{}}} 
e^{-i\mbox{\footnotesize{\boldmath$\e t $}}\mbox{\boldmath{}}}
\gamma(\mathbf{t'} | \mathbf{t}),
\label{eq:Fourier_transformation}
\end{align}
fulfills energy conservation. In particular, this allows for the following representation for the one- and two-particle functions, where calligraphic letters $\mathcal{G}$ and $\mathcal{L}$ are used when a $\delta$-function has been split off:
\begin{align}
& G(\e_1|\e_1') = 2\pi\delta(\e_1-\e_1')\mathcal{G}(\e_1), \nonumber \\
& G(\e_1,\e_2|\e_1',\e_2') = 2\pi\delta(\e_1+\e_2-\e_1'-\e_2')\mathcal{G}(\e_2,\e_1';\e_1-\e_1'), \nonumber \\
&\gamma(\e_1'|\e_1) = - 2\pi\delta(\e_1'-\e_1)\Sigma(\e_1'), \nonumber \\
&\gamma(\e_1',\e_2'|\e_1,\e_2) = 2\pi\delta(\e_1'+\e_2'-\e_1-\e_2)\mathcal{L}(\e_2',\e_1;\e_1' - \e_1).
\label{eq:used_functions}
\end{align}
The one-particle vertex-function $\Sigma$, introduced above, is called the interacting irreducible self-energy. We Fourier-transform Dyson's equation, Eq.~(\ref{eq:Dyson_time}), which provides
\begin{align}
\mathcal{G}(\e) \! & =\! \mathcal{G}_{\dg{0}}(\e)\! +\! \mathcal{G}_{\dg{0}}(\e) \Sigma(\e)\mathcal{G}(\e)\! 
  =\! \left[\left[\mathcal{G}_{\dg{0}}(\e)\right]^{-1} - \Sigma(\e)\right]^{-1}. 
 \label{eq:Dyson_frequency}
\end{align}
Note that this is a matrix equation in both Keldysh and position space.

The four single-particle Green's functions and self-energies in
contour space are called chronological
($\mathcal{G}^{{\color{red} -|-}}$, $\Sigma^{{\color{red} -|-}}$),
lesser ($\mathcal{G}^{{\color{red} -|+}}$,
$\Sigma^{{\color{red} -|+}}$), greater
($\mathcal{G}^{{\color{red} +|-}}$, $\Sigma^{{\color{red} +|-}}$) and
anti-chronological ($\mathcal{G}^{{\color{red} +|+}}$,
$\Sigma^{{\color{red} +|+}}$). As a consequence of the definition,
Eq.~(\ref{eq:define_Green}), the single-particle Green's functions
fulfill the contour-relation
\begin{equation}
\mathcal{G}^{\rp|\rp} + \mathcal{G}^{\rmi|\rmi} = \mathcal{G}^{\rmi|\rp} + \mathcal{G}^{\rp|\rmi}.
\label{eq:pm_green}
\end{equation} 

We define the transformation from contour space
(${\color{red}a} = \{{\color{red}-},{\color{red}+}\}$) into Keldysh
space (${\color{red}\alpha} = \{{\color{red}1},{\color{red}2}\}$) by
the rotation
\begin{align}
R =  \left( \begin{array}{cc}
R^{\re{-}|\re{1}} & R^{\re{-}|\re{2}}  \\
R^{\re{+}|\re{1}} & R^{\re{+}|\re{2}} \end{array} \right)
 = \frac{1}{\sqrt{2}}
  \left( \begin{array}{cc}
1 & 1  \\
-1 & 1 \end{array} \right).
\label{eq:transformation}
\end{align}
Hence, any $n$-th rank tensor
$A^{n,\re{\bgs{\alpha'}}|\bgs{\re{\alpha}}\mbox{\boldmath{}}}$ in
Keldysh space is represented in contour space by
\begin{align}
  A^{n,\bgs{\re{\alpha}}|\bgs{\re{\alpha'}}\mbox{\boldmath{}}} = \sum_{\bgs{\re{a}},\bgs{\re{a'}}\mbox{\boldmath{}}} \left[R^{-1}\right]^{\bgs{\re{\alpha}}|\bgs{\re{a}}\mbox{\boldmath{}}}A^{n,\bgs{\re{a}}|\bgs{\re{a'}}\mbox{\boldmath{}}} R^{\bgs{\re{a'}}|\bgs{\re{\alpha'}}\mbox{\boldmath{}}}.
\label{eq:Kel_transformation}
\end{align}
As can be shown explicitly (see Chapter 4.3 of
Ref.~\cite{Jakobs2010b}) the Green's and vertex functions
fulfill a theorem of causality:
\begin{align}
  \mathcal{G}^{{{\color{red}1}...{\color{red}1}|{\color{red}1}...{\color{red}1}}} = 0 , \nonumber \\
  \mathcal{L}^{{{\color{red}2}...{\color{red}2}|{\color{red}2}...{\color{red}2}}} = 0.
\label{eq:causality}
\end{align}
The remaining three non-zero Keldysh components of the single-particle
functions are called retarded ($\mathcal{G}^{\re{2}|\re{1}}$,
$\Sigma^{\re{1}|\re{2}}$), advanced ($\mathcal{G}^{\re{1}|\re{2}}$,
$\Sigma^{\re{2}|\re{1}}$) and Keldysh ($\mathcal{G}^{\re{2}|\re{2}}$,
$\Sigma^{\re{1}|\re{1}}$):
\begin{align}
\mathcal{G} & =  \left( \begin{array}{cc}
0 & \mathcal{G}^{\re{A}}  \\
\mathcal{G}^{\re{R}} & \mathcal{G}^{\re{K}} \end{array} \right) 
=
\left( \begin{array}{cc}
0 & \mathcal{G}^{\re{1}|\re{2}}  \\
\mathcal{G}^{\re{2}|\re{1}} & \mathcal{G}^{\re{2}|\re{2}} \end{array} \right), \notag \\
\Sigma & =  \left( \begin{array}{cc}
\Sigma^{\re{K}} & \Sigma^{\re{R}} \\
\Sigma^{\re{A}} & 0 \end{array} \right)
=
\left( \begin{array}{cc}
\Sigma^{\re{1}|\re{1}} & \Sigma^{\re{1}|\re{2}} \\
\Sigma^{\re{2}|\re{1}} & 0 \end{array} \right).
\label{eq:matrix_representation}
\end{align}
The transformation, Eq.~(\ref{eq:Kel_transformation}), provides the identities 
\begin{subequations}
\begin{align}
\mathcal{G}^{{\rmi}|{\rp}} &= \frac{1}{2} \left[
\mathcal{G}^{{{\color{red}2}|{\color{red}2}}}-\left( \mathcal{G}^{{{\color{red}2}|{\color{red}1}}} - \mathcal{G}^{{{\color{red}1}|{\color{red}2}}} \right)
\right], \\
  \mathcal{G}^{{\rp}|{\rmi}} - \mathcal{G}^{{\rmi}|{\rp}} &= \mathcal{G}^{{{\color{red}2}|{\color{red}1}}}- \mathcal{G}^{{{\color{red}1}|{\color{red}2}}}, \\
  \mathcal{H}_{\dg{0}}^{\re{1}|\re{2}} & = \mathcal{H}_{\dg{0}}^{\re{2}|\re{1}} = H_{\dg{0}}~,~   \mathcal{H}_{\dg{0}}^{\re{1}|\re{1}} = \mathcal{H}_{\dg{0}}^{\re{2}|\re{2}} = 0 , 
\label{eq:keldysh_greens_relations}
\end{align}
\end{subequations}
all of which are used in the derivation of the conductance formula in
Sec.~\ref{sec:cond_derivation}. Note that a calligraphic $\mathcal{H}$
carries Keldysh indices, while a capital $H$ does not. The
retarded/advanced components are analytic in the upper/lower half
plane of the complex frequency plane. Hence, the following notation is
always implied,
\begin{align}
\mathcal{G}^{\re{2}|\re{1}}(\e) & = \mathcal{G}^{\re{2}|\re{1}}(\e+i\delta)~,~ \Sigma^{\re{1}|\re{2}}(\e) = \Sigma^{\re{1}|\re{2}}(\e+i\delta), \\
\mathcal{G}^{\re{1}|\re{2}}(\e) & = \mathcal{G}^{\re{1}|\re{2}}(\e-i\delta)~,~ \Sigma^{\re{2}|\re{1}}(\e) = \Sigma^{\re{2}|\re{1}}(\e-i\delta),
\end{align}
 with real, infinitesimal, positive $\delta$. In contrast, the Keldysh component describes fluctuations, restricted to the real frequency axis. In equilibrium, the single-particle functions fulfill a fluctuation dissipation theorem (FDT):
\begin{subequations}
\begin{align}
\Sigma^{\re{1}|\re{1}}(\e) = (1 - 2f(\e))\big[\Sigma^{\re{1}|\re{2}}(\e) - \Sigma^{\re{2}|\re{1}}(\e)\big]
\label{eq:FDT_sigma}, \\
\mathcal{G}^{\re{2}|\re{2}}(\e) = (1 - 2f(\e))\big[ \mathcal{G}^{\re{2}|\re{1}}(\e) - \mathcal{G}^{\re{1}|\re{2}}(\e)\big],
\end{align}
\label{eq:FDT_green}
\end{subequations}
where $f(\e) = 1/(1+\exp[(\e-\mu)/T])$ is the Fermi distribution function.

Within this work we consider a system composed of a finite central interacting region
coupled to two non-interacting fermionic leads: The left lead, with chemical potential $\mu_L$ and temperature $T_L$, and the right lead, with chemical potential $\mu_R$ and temperature $T_R$. 
We can represent the quadratic part of the Hamiltonian in block-matrix form as
\begin{align}
{h}_{\dg{0}} \!=\! 
 \left( \begin{array}{ccc}
 h_l & h_{lc} & 0 \\
~\!\!h_{cl} & h_{\dg{0},c} & h_{cr}\\
0 & h_{rc} & h_r \\ 
 \end{array}
 \right).
 \label{eq:quadratic_H1}
 \end{align}
 where the matrices $h_l$ and $h_r$ fully define the properties of the isolated leads, and the matrix $h_{\dg{0},c}$ describes the non-interacting part of the isolated central region. Finally, $h_{cl}$ and $h_{cr}$ specify the coupling of the central region to the corresponding lead.  
Similarly, we write the system's Green's function, $\mathcal{G}(\e)$ [Eq.~(\ref{eq:Dyson_frequency})], in the same basis (for the bare, non-interacting Green's function $\mathcal{G}_{\dg{0}}$ we set $\Sigma \!=\! 0$):
\begin{align}
\mathcal{G} \!=\! 
 \left( \begin{array}{ccc}
\mathcal{G}_l & \mathcal{G}_{lc} & \mathcal G_{lr} \\
\mathcal{G}_{cl} & \mathcal{G}_c & \mathcal{G}_{cr}\\
\mathcal G_{rl}&\mathcal{G}_{rc} & \mathcal{G}_r \\ 
 \end{array}
 \right).
 \label{eq:matrix_Green}
\end{align}

We use the \textit{small letter} $g$ to denote the Green's function of an isolated subsystem, e.g.\ $g_{l}(\e)$ is the Green's function of the isolated left lead $L$. The non-interacting Green's function of the central region is given by Dyson's equation
\begin{align}
\mathcal{G}_{\dg{0},c} & = g_{\dg{0},c} \!+\! g_{\dg{0},c}\Sigma_{{\rm{lead}}}\mathcal{G}_{\dg{0},c}  = \left[\left[g_{\dg{0},c}\right]^{-1}\!-\!\Sigma_{{\rm{lead}}}\right]^{-1}.
\end{align}
Again note that this is a matrix equation in Keldysh and position space. We incorporated environment contributions into the lead self-energy
\begin{align}
\Sigma_{{\rm{lead}}} = \sum_{k = l,r} h_{ck} g_k h_{kc}.
\label{eq:lead_selfenergy}
\end{align}
The individual Keldysh components of the non-interacting Green's function are given by
\begin{subequations}
\begin{align}
\mathcal{G}_{\dg{0},c}^{\re{1}|\re{2}}(\e) & =  \left(\e- h_{c}- \Sigma_{{\rm{lead}}}^{\re{2}|\re{1}}(\e)\right)^{-1}, \\
\mathcal{G}_{\dg{0},c}^{\re{2}|\re{1}}(\e) & =  \left(\e- h_{c} - \Sigma_{{\rm{lead}}}^{\re{1}|\re{2}}(\e)\right)^{-1}, \\
\mathcal{G}_{\dg{0},c}^{\re{2}|\re{2}}(\e) & = \mathcal{G}_{\dg{0},c}^{\re{2}|\re{1}}(\e) \Sigma_{{\rm{lead}}}^{\re{1}|\re{1}}(\e)\mathcal{G}_{\dg{0},c}^{\re{1}|\re{2}}(\e) \notag \\ & = -i\sum_{k=l,r} [1-2f^k(\e)]\mathcal{G}_{\dg{0},c}^{\re{2}|\re{1}}(\e)\Gamma^k(\e)\mathcal{G}_{\dg{0},c}^{\re{2}|\re{1}}(\e),
\end{align}
\label{eq:Dyson_bare}
\end{subequations}
where we introduced the hybridization function, $\Gamma^k(\e)\! =\! i h_{ck}(g_k^{\re{2}|\re{1}}(\e)\! -\! g_k^{\re{1}|\re{2}}(\e))h_{kc}$.

With the interaction being restricted to the central region we use the notation $\Sigma = \Sigma_c = C\Sigma C$ for the interacting self-energy. Dyson's equation, Eq.~(\ref{eq:Dyson_frequency}), and the real space structure, Eq.~(\ref{eq:matrix_Green}), yields
\begin{align}
\mathcal{G}_c(\e) = \left[\left[\mathcal{G}_{\dg{0},c}(\e)\right]^{-1}-\Sigma(\e)\right]^{-1}.
\label{eq:Dyson_full}
\end{align}
The matrix representation of its Keldysh structure is given by
\begin{align}
\left( \begin{array}{cc}
0  & \mathcal{G}_c^{\re{1}|\re{2}}  \\
\mathcal{G}_c^{\re{2}|\re{1}}  & \mathcal{G}_c^{\re{2}|\re{2}} \\
 \end{array}
 \right)
 \!=\!
\left[
\left( \begin{array}{cc}
0 & \mathcal{G}_{\dg{0},c}^{\re{1}|\re{2}}  \\
\mathcal{G}_{\dg{0},c}^{\re{2}|\re{1}}  & \mathcal{G}_{\dg{0},c}^{\re{2}|\re{2}} \\
 \end{array}
 \right)^{-1}
 -
 \left( \begin{array}{cc}
\Sigma^{\re{1}|\re{1}}  & \Sigma^{\re{1}|\re{2}}  \\
\Sigma^{\re{2}|\re{1}}  & 0 \\
 \end{array}
 \right)
 \right]^{-1}.
 \label{eq:quadratic_H2}
 \end{align}
Block matrix inversion then provides the components
\begin{subequations}
\begin{align}
\mathcal{G}_c^{\re{1}|\re{2}}(\e) &= \left(\e- h_{c}- \Sigma_{{\rm{lead}}}^{\re{2}|\re{1}}(\e) - \Sigma^{\re{2}|\re{1}}(\e)\right)^{-1}, \\
\mathcal{G}_c^{\re{2}|\re{1}}(\e) &= \left(\e- h_{c}- \Sigma_{{\rm{lead}}}^{\re{1}|\re{2}}(\e) - \Sigma^{\re{1}|\re{2}}(\e)\right)^{-1}, \\
\mathcal{G}_c^{\re{2}|\re{2}}(\e) &= \mathcal{G}_c^{\re{2}|\re{1}}(\e)\left[ \Sigma_{{\rm{lead}}}^{\re{1}|\re{1}}+\Sigma^{\re{1}|\re{1}}\right] \mathcal{G}_c^{\re{1}|\re{2}}(\e).
\end{align}
\label{eq:Dyson}
\end{subequations}
From Eq.~(\ref{eq:Dyson_frequency}), we can show, that the off-diagonal components of the full Green's function, are given by
\begin{align}
\mathcal{G}_{kc} = g_k \mathcal{H}_{kc}\mathcal{G}_c ~,~\mathcal{G}_{ck} = \mathcal{G}_c \mathcal{H}_{ck} g_k,
\label{eq:Dyson_offdiagonal}
\end{align}
where in this single case, $\mathcal H_{kc}$ is the matrix-element $h_{kc}$ with additional Keldysh-structure.
In general, one has
\begin{subequations}
\begin{align}
\mathcal{G}_{\dg{0},\bl{i}|\bl{j}}^{\re{1}|\re{2}}  & \! =\! \left[\mathcal{G}_{\dg{0},\bl{j}|\bl{i}}^{\re{2}|\re{1}}\right]^*\!, 
\mathcal{G}_{\bl{i}|\bl{j}}^{\re{1}|\re{2}}   \! =\! \left[\mathcal{G}_{\bl{j}|\bl{i}}^{\re{2}|\re{1}}\right]^*\!, 
\Sigma_{\bl{i}|\bl{j}}^{\re{1}|\re{2}}   \! =\! \left[\Sigma_{\bl{j}|\bl{i}}^{\re{2}|\re{1}}\right]^*\!, \\
\mathcal{G}_{\dg{0},\bl{i}|\bl{j}}^{\re{2}|\re{2}} & \!=\! -\left[\mathcal{G}_{\dg{0},\bl{j}|\bl{i}}^{\re{2}|\re{2}} \right]^* \!,
\mathcal{G}_{\bl{i}|\bl{j}}^{\re{2}|\re{2}}  \!=\! -\left[\mathcal{G}_{\bl{j}|\bl{i}}^{\re{2}|\re{2}} \right]^* \!,
\Sigma_{\bl{i}|\bl{j}}^{\re{1}|\re{1}} \!=\! -\left[\Sigma_{\bl{j}|\bl{i}}^{\re{1}|\re{1}} \right]^*.
\end{align}
\label{eq:G_Sigma_site_structure}
\end{subequations}
For a symmetric, real Hamiltonian, the following additional symmetries hold in equilibrium
\begin{subequations}
\begin{align}
\mathcal{G}_{\dg{0},\bl{i}|\bl{j}}^{\re{1}|\re{2}} &\! =\! \mathcal{G}_{\dg{0},\bl{j}|\bl{i}}^{\re{1}|\re{2}}~,~
\mathcal{G}_{\bl{i}|\bl{j}}^{\re{1}|\re{2}} \!=\!  \mathcal{G}_{\bl{j}|\bl{i}}^{\re{1}|\re{2}}~,~  
\Sigma_{\bl{i}|\bl{j}}^{\re{1}|\re{2}} \!=\!  \Sigma_{\bl{j}|\bl{i}}^{\re{1}|\re{2}},\\
\mathcal{G}_{\dg{0},\bl{i}|\bl{j}}^{\re{2}|\re{1}} &\! =\!  \mathcal{G}_{\dg{0},\bl{j}|\bl{i}}^{\re{2}|\re{1}}~,~
\mathcal{G}_{\bl{i}|\bl{j}}^{\re{2}|\re{1}} \!=\!  \mathcal{G}_{\bl{j}|\bl{i}}^{\re{2}|\re{1}}~,~
\Sigma_{\bl{i}|\bl{j}}^{\re{2}|\re{1}} \!=\! \Sigma_{\bl{j}|\bl{i}}^{\re{2}|\re{1}}.
\end{align}
\label{eq:G_Sigma_site_structure1}
\end{subequations}
\pagebreak
\begin{widetext}

\section{\, Diagrammatic discussion of the fRG flow-equation of the self-energy}
\label{App:fRG}

In this appendix we provide a diagrammatic plausibility argument for the fRG
flow-equation for the self-energy, Eq.~(\ref{eq:fRG_flow}). A detailed diagrammatic derivation may be found in Ref.~\cite{Jakobs2010a}. We use the
observation, that every diagram in the diagrammatic series of the
self-energy contains a sub-diagram which appears in the diagrammatic
series of the two-particle vertex. As a consequence, taking the
derivative of the self-energy, $\partial_\Lambda\Sigma$, w.r.t. some
parameter $\Lambda$ allows for a resummation of diagrams, such that
the full two-particle vertex series can be factorized. Hence, we get
an equation which can formally be written as
$\partial_\Lambda \Sigma_\Lambda = \int
\!S_\Lambda\mathcal{L}_\Lambda$,
with the socalled single-scale propagator $S$ and the two-particle
vertex $\mathcal{L}$, both depending on the parameter $\Lambda$.

\vspace{1em} 

The self-energy $\Sigma$ and two-particle vertex $\mathcal{L}$ are diagrammatically defined as the sum of all one-particle irreducible diagrams with two and four amputated external legs, respectively. Using the graphical representation of the bare Green's function, Eq.~(\ref{eq:diagram_bare_G}), and the bare vertex, Eq.~(\ref{eq:bare_vertex_keldysh}), the first terms of their perturbation series are (we omit the arrows for the sake of simplicity)
\begin{subequations}
\begin{align}
\begin{matrix}
\includegraphics[width=16cm]{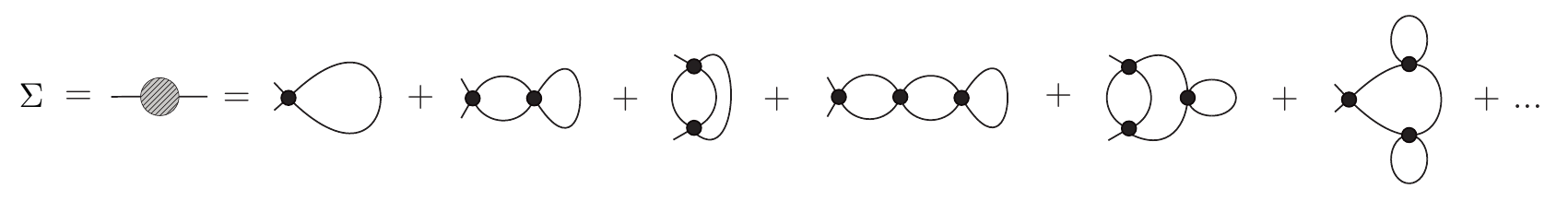} 
\end{matrix}
\label{eq:Sigma_series}\\
\begin{matrix}
\includegraphics[width=16cm]{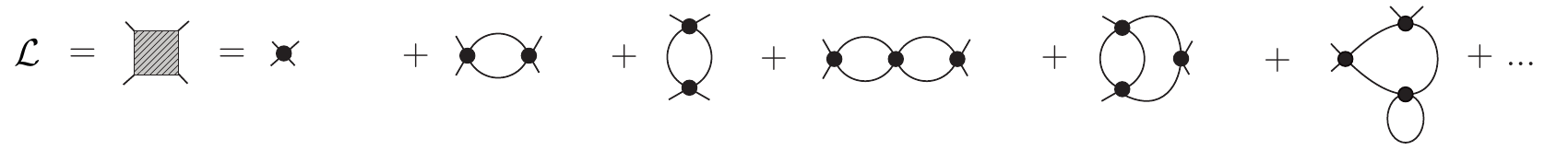} 
\end{matrix} 
\label{eq:Gamma_series}
\end{align}
\label{eq:vertex_series}
\end{subequations}
We introduce a parameter $\Lambda$  into the bare propagator, $\mathcal{G}_{\dg{0}} \rightarrow \mathcal{G}_{\dg{0}}^\Lambda$, and represent its derivative w.r.t. $\Lambda$ by a crossed-out line, $
\partial_\Lambda \mathcal{G}_{\dg{0}}^\Lambda = 
\begin{matrix}
\includegraphics[width=1cm]{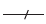} 
\end{matrix}.
$ 
Hence, the derivative of the self-energy is given by
\begin{align}
\begin{matrix}
\includegraphics[width=18cm]{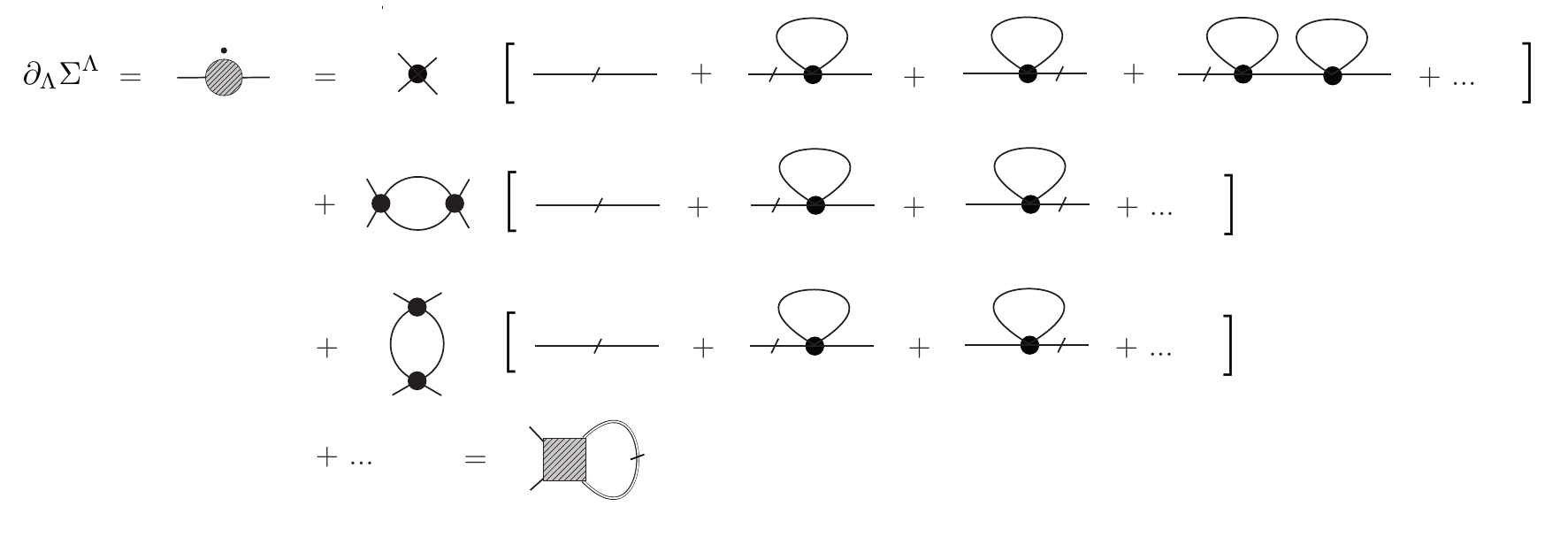} 
\end{matrix}
\label{eq:Sigma_derivative}
\end{align}
where we introduced the socalled single scale propagator 
\begin{flalign}
& \includegraphics[width=15cm]{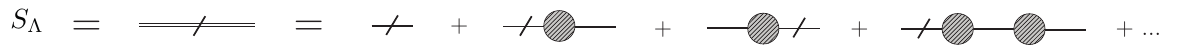} \nonumber\\
& \phantom{S_\Lambda}= \left[ 1+ \mathcal{G}_{\dg{0}} \Sigma + \mathcal{G}_{\dg{0}} \Sigma \mathcal{G}_{\dg{0}} \Sigma + \dots \right] \left( \partial_\Lambda \mathcal{G}_{\dg{0}}^\Lambda \Sigma \right) \left[ 1+ \mathcal{G}_{\dg{0}} \Sigma + \mathcal{G}_{\dg{0}} \Sigma \mathcal{G}_{\dg{0}} \Sigma + \dots \right] \nonumber \\
& \phantom{S_\Lambda}= \mathcal G \left[ \mathcal{G}_{\dg{0}} \right]^{-1} \left( \partial_\Lambda \mathcal{G}_{\dg{0}}^\Lambda \Sigma \right) \left[ \mathcal{G}_{\dg{0}} \right]^{-1} \mathcal G 
\end{flalign}
Finally, we fix the prefactor of the diagram on the r.h.s. in Eq.~(\ref{eq:Sigma_derivative}) by the following argument: The first order self-energy, $\Sigma_{\dg{1}}$,  is of Hartree-type and hence purely determined by the local density $n_{\bl{j}}$ and the local interaction strength $U_{\bl{j}}$. 
We calculate the first order self-energy in equilibrium
\begin{align}
\label{eq:Hartree_fix_prefactor_fRG}
\left. \Sigma^R_{\dg{1},\bl{j}} \right|_{V=0} = n_{\bl{j}}U_{\bl{j}} = 
\frac{1}{\pi i}\bar{u}_{\bl{j}}\int d\e~ \mathcal{G}_{\dg{0},\bl{j}|\bl{j}}^{<}(\e).
\end{align}
The derivative of Eq.~\eqref{eq:Hartree_fix_prefactor_fRG} is given by
\begin{align}
\label{eq:Hartree_fix_prefactor_fRG2}
\left. \partial_\Lambda \Sigma^R_{\dg{1},\bl{j}} \right|_{V=0} =
\frac{1}{2 \pi i}\bar{u}_{\bl{j}}\int d\e~ \partial_\Lambda \left( \mathcal{G}_{\dg{0},\bl{j}|\bl{j}}^{K}(\e) + \mathcal{G}_{\dg{0},\bl{j}|\bl{j}}^{A}(\e) - \mathcal{G}_{\dg{0},\bl{j}|\bl{j}}^{R}(\e) \right).
\end{align}
$G^R$ is analytic in the upper half plane.
For reasonable flow parameters (in particular flow parameters associated with a physical quantity, like temperature or source-drain bias), $\partial_\Lambda G^R$ is also analytic in the upper half plane and $\partial_\Lambda G^R$ decays sufficiently fast as function of energy that the contour may be closed in the lower half-plane to yield zero. 
For analogous reasons, $\partial_\Lambda G^A$ yields zero, too. 
Therefore, Eq.~\eqref{eq:Hartree_fix_prefactor_fRG2} simplifies to 
\begin{align}
\left. \partial_\Lambda \Sigma^R_{\dg{1},\bl{j}} \right|_{V=0} =
\frac{1}{2 \pi i}\bar{u}_{\bl{j}}\int d\e~ \partial_\Lambda \mathcal{G}_{\dg{0},\bl{j}|\bl{j}}^{K}(\e),
\end{align}
which fixes the prefactor of the diagram on the r.h.s. in Eq.~(\ref{eq:Sigma_derivative}).
Hence, we end up with Eq.~(\ref{eq:fRG_flow}) for the derivative of the self-energy:

\begin{align}
\partial_\Lambda \Sigma_{\bl{i}|\bl{j}}^{\re{\alpha'}|\re{\alpha}}(\e) = \frac{1}{2\pi i}\int d\ep  \sum_{\underset{\bl{kl} \in C}{\re{\beta\beta'}}}S_{\Lambda,\bl{k}|\bl{l}}^{\re{\beta}|\re{\beta'}}(\ep)\mathcal{L}_{\bl{ik}|\bl{jl}}^{\re{\alpha' \beta'}|\re{\alpha \beta}}(\e',\e;0).
\label{eq:fRG_flow_supplement}
\end{align}

\section{\, Charge conservation - Ward identity}
\label{App:ward}

In this Appendix we derive the Ward identity used in the main text,
Eq.~(\ref{eq:ward_final}), from variational principles, following Ref. \cite{Zinn}. Since the
action corresponding to the Hamiltonian, Eq.~(\ref{eq:ModelGeneral}),
is invariant under a global $U(1)$ symmetry, it satisfies a
conservation law. Starting from the path integral representation of
expectation values using Grassmann variables, the requirement of
vanishing variation under the gauged $U(1)$ transformation yields both
a continuity equation for particle current and the desired connection
between the interacting self-energy $\Sigma$, introduced in
Eq.~(\ref{eq:used_functions}), and the vertex part $\Phi$, defined in
Eq.~(\ref{eq:P}).

\vspace{1em} 
Within this Appendix, for notational convenience, we combine the left and right lead, thus representing the matrix $h$ in the Hamiltonian Eq.~\eqref{eq:ModelGeneral} and the Green's function by
\begin{align}
h  =  
\left( \begin{array}{cc}
h_{c} & h_{c \ell}  \\
~\!\! h_{\ell c} & h_{\ell} \end{array} \right)
 ~,~
\mathcal{G}  
=
\left( \begin{array}{cc}
\mathcal{G}_c & \mathcal{G}_{c\ell}  \\
\mathcal{G}_{\ell c}& \mathcal{G}_{\ell} \end{array} \right),
\end{align}
where $l$ corresponds to spatial indices in either lead, and $c$ to spatial indices within the central region.
Let $\{\psi\},\{\bar{\psi}\}$ be sets of Grassmann variables, i.e.\ fermionic fields. We write $n$-particle expectation values in terms of the functional path integral,
\begin{align}
G_{\mbox{\scriptsize{\boldmath$\bl{i}$}}|\mbox{\scriptsize{\boldmath$\bl{i'}$}}}^{n,\mbox{\scriptsize{\boldmath$ \re{a}$}}|
 \mbox{\scriptsize{\boldmath$ \re{a'}$}}\mbox{\boldmath{}}}(\mbox{\normalsize{\boldmath$t$}}|\mbox{\normalsize{\boldmath$t'$}}\mbox{\boldmath{}}) \!=\! (-i)^n \langle \psi_{\bl{i_1}}^{\re{a_1}}(t_1)... \psi_{\bl{i_n}}^{\re{a_n}}(t_n) \bar{\psi}_{\bl{i_n'}}^{\re{a_n'}}(t_n')...\bar{\psi}_{\bl{i_1'}}^{\re{a_1'}}(t_1')\rangle \!=\! (-i)^n\int\! \mathcal{D}(\mbox{\normalsize{\boldmath$\bar{\psi}$}}\mbox{\normalsize{\boldmath$\psi$}})\psi_{\bl{i_1}}^{\re{a_1}}(t_1)... \psi_{\bl{i_n}}^{\re{a_n}}(t_n) \bar{\psi}_{\bl{i_n'}}^{\re{a_n'}}(t_n')...\bar{\psi}_{\bl{i_1'}}^{\re{a_1'}}(t_1')e^{iS[\mbox{\normalsize{\boldmath$\bar{\psi}$}},\mbox{\normalsize{\boldmath$\psi$}}]},
\label{eq:Correlators_PI}
\end{align}
where the Keldysh action is given by the Keldysh contour time integral
\begin{align}
S[\mbox{\normalsize{\boldmath$\bar{\psi}$}},\mbox{\normalsize{\boldmath$\psi$}}] \! & =\! \int_{\mathcal{C}}dt \sum_{\bl{ii'}}\bar{\psi}_{\bl{i'}}(t)(\overbrace{i\delta_{\bl{i' i}}\partial_{t}-  h_{\bl{i' i}})} ^{[G_{\dg{0}}(t)^{-1}]_{\bl{i' i}}}\psi_{\bl{i}}(t)  + S_{{\rm{int}}}[\mbox{\normalsize{\boldmath$\bar{\psi}$}},\mbox{\normalsize{\boldmath$\psi$}}] \!=\! 
\int_{-\infty}^\infty dt \sum_{\re{a},\bl{ii'}}(-\re{a})\bar{\psi}_{\bl{i'}}^{\re{a}}(t)\left(i\delta_{\bl{i'i}}\partial_{t} - h_{\bl{i' i}}\right)\psi_{\bl{i}}^{\re{a}}(t) + S_{{\rm{int}}}[\mbox{\normalsize{\boldmath$\bar{\psi}$}},\mbox{\normalsize{\boldmath$\psi$}}] \notag \\
& = \int_{-\infty}^\infty dt \sum_{\re{a}}(-\re{a})\mbox{\normalsize{\boldmath$\bar{\psi}$}}^{\re{a}}\!(t)\!\left(i\mbox{\normalsize{\boldmath$\partial$}}_t - h \right)\mbox{\normalsize{\boldmath$\psi$}}^{\re{a}}(t) + S_{{\rm{int}}}[\mbox{\normalsize{\boldmath$\bar{\psi}$}},\mbox{\normalsize{\boldmath$\psi$}}].
\label{eq:Keldysh_action}
\end{align}
In the last line we introduced the vector notation 
\small{$
\mbox{\normalsize{\boldmath$\psi$}}\!=\! 
\left(\! \begin{array}{c}
\psi_{\bl{1}} \\
\psi_{\bl{2}} \\
\vdots \\
\end{array}\! \right)
$} 
\normalsize
and 
\small{$
\mbox{\normalsize{\boldmath$\bar{\psi}$}}\!=\! 
\left( 
\psi_{\bl{1}},
\psi_{\bl{2}},
\hdots 
\right)
$.} Note that $\mbox{\normalsize{\boldmath$\partial$}}_t$ is a diagonal matrix. 
\normalsize

\subsection{Gauge transformation}

The action, Eq.~(\ref{eq:Keldysh_action}),  is invariant under the global $U(1)$ transformation $\mbox{\normalsize{\boldmath$\psi$}} \rightarrow \mbox{\normalsize{\boldmath$\psi$}} e^{i\alpha}$ and $\mbox{\normalsize{\boldmath$\bar{\psi}$}} \rightarrow \mbox{\normalsize{\boldmath$\bar{\psi}$}} e^{-i\alpha}$, where $\alpha$ is a real constant. Gauging this transformation, i.e.\ making $\alpha$ space-, and time-dependent,  yields to linear order in $\alpha$

\begin{align}
\delta \psi_{\bl{i}}^{\re{a}}(t) =  i \alpha_{\bl{i}}^{\re{a}}(t)\psi_{\bl{i}}^{\re{a}}(t) ~,~ \delta \bar{\psi}_{\bl{i'}}^{\re{a}}(t') =  -i \alpha_{\bl{i'}}^{\re{a}}(t')\bar{\psi}_{\bl{i'}}^{\re{a}}(t').
 \label{eq:variation_field}
\end{align}
Since we are interested in the current through the system, from one lead to another, it is convenient to pick $\alpha$ non-vanishing only in the central region:
\begin{align}
\alpha_{\bl{i}}^{\re{a}}(t) = \begin{cases}
\alpha^{\re{a}}(t),\quad {\rm{if}} \quad \bl{i} \in C  \\
0, \quad\hspace{1.8em} {\rm{if}} \quad \bl{i} \in \mathbb{L} . 
\end{cases}
\label{eq:alpha_shaping}
\end{align}
 This is equivalent to first deriving the Ward identity using an arbitrary $\alpha$ and then summing over the central region. The requirement that the right-hand side of Eq. \eqref{eq:Correlators_PI} is invariant when applying the gauged $U(1)$ transformation to all $\psi$'s therein now reads:
\begin{equation}
\delta G_{\mbox{\scriptsize{\boldmath$\bl{i}$}}|\mbox{\scriptsize{\boldmath$\bl{i'}$}}}^{n,\mbox{\scriptsize{\boldmath$ \re{a}$}}|
 \mbox{\scriptsize{\boldmath$ \re{a'}$}}\mbox{\boldmath{}}}(\mbox{\normalsize{\boldmath$t$}}|\mbox{\normalsize{\boldmath$t'$}}\mbox{\boldmath{}}) = 0.
 \label{eq:variation_condition}
 \end{equation}
This requirement is simply a change of the integration variable in field space. In other words, the physical correlators cannot depend on an arbitrary choice of basis in which the fields are represented.
 
\subsection{The continuity equation (zeroth order Ward identity)}
 
For $n=0$, Eq.~(\ref{eq:variation_condition}) sets a condition on the variation of the partition sum. Since the measure of the path integral is invariant under the transformation in Eq.~(\ref{eq:variation_field}) (the $U(1)$-symmetry is not anomalous),  this in turn sets a condition on the variation of the action:
\begin{align}
0 =  
\delta 
\left[ \int\! 
\mathcal{D}(\mbox{\normalsize{\boldmath$\bar{\psi}$}}\mbox{\normalsize{\boldmath$\psi$}})e^{iS[\mbox{\normalsize{\boldmath$\bar{\psi}$}},\mbox{\normalsize{\boldmath$\psi$}}]}
\right]
= 
i \int\! \mathcal{D}(\mbox{\normalsize{\boldmath$\bar{\psi}$}}\mbox{\normalsize{\boldmath$\psi$}})\delta S[\mbox{\normalsize{\boldmath$\bar{\psi}$}},\mbox{\normalsize{\boldmath$\psi$}}] e^{iS[\mbox{\normalsize{\boldmath$\bar{\psi}$}},\mbox{\normalsize{\boldmath$\psi$}}]}.
\label{eq:variation_partition_sum}
\end{align}
The quartic term, $S_{{\rm{int}}}$, describes a density-density interaction. Hence, its variation vanishes trivially and the variation of the total action reduces to the variation of the quadratic term: 
\begin{align}
\delta S[\mbox{\normalsize{\boldmath$\bar{\psi}$}},\mbox{\normalsize{\boldmath$\psi$}}] & = \int_{-\infty}^\infty dt \sum_{\re{a},\bl{i}}(-\re{a})\Big[\alpha_{\bl{i}}^{\re{a}}(t)\bar{\psi}_{\bl{i}}^{\re{a}}(t)\partial_t
\psi_{\bl{i}}^{\re{a}}(t) - \bar{\psi}_{\bl{i}}^{\re{a}}(t)\partial_t\!\left(\alpha_{\bl{i}}^{\re{a}}(t)\psi_{\bl{i}}^{\re{a}}(t)\right) 
+
\sum_{\bl{i'}} \big[i\alpha_{\bl{i'}}^{\re{a}}(t) -i\alpha_{\bl{i}}^{\re{a}}(t)\big]
\bar{\psi}_{\bl{i'}}^{\re{a}}\!(t) h_{\bl{i' i}}\psi_{\bl{i}}^{\re{a}}(t)
 \Big]\notag \\
& =  \int_{-\infty}^\infty dt 
\sum_{\re{a}}(-\re{a})\alpha^{\re{a}}(t)
\Big[
\partial_t\!\left(
\mbox{\normalsize{\boldmath$\bar{\psi}$}}_{c}^{\re{a}}(t)
 \mbox{\normalsize{\boldmath$\psi$}}_{c}^{\re{a}}(t)\right)
 - i
\mbox{\normalsize{\boldmath$\bar{\psi}$}}_{c}^{\re{a}}(t)h_{cl} \mbox{\normalsize{\boldmath$\psi$}}_{l}^{\re{a}}(t)
 + i\mbox{\normalsize{\boldmath$\bar{\psi}$}}_{l}^{\re{a}}(t)h_{lc} \mbox{\normalsize{\boldmath$\psi$}}_{c}^{\re{a}}(t) 
\Big] \notag \\
& =  \int_{-\infty}^\infty dt 
\sum_{\re{a}}(-\re{a})\alpha^{\re{a}}(t)
\Big[-
\partial_t\!\left(
 \mbox{\normalsize{\boldmath$\psi$}}_{c}^{\re{a}}(t)
 \mbox{\normalsize{\boldmath$\bar{\psi}$}}_{c}^{\re{a}}(t)\right) + 
i {\rm{Tr}}\!\left\{\!h_{cl} \mbox{\normalsize{\boldmath$\psi$}}_{l}^{\re{a}}(t) \mbox{\normalsize{\boldmath$\bar{\psi}$}}_{c}^{\re{a}}(t)\!\right\}
-i{\rm{Tr}}\!\left\{\!h_{lc} \mbox{\normalsize{\boldmath$\psi$}}_{c}^{\re{a}}(t) \mbox{\normalsize{\boldmath$\bar{\psi}$}}_{l}^{\re{a}}(t)\!\right\}  
\Big],
\end{align}
where we used integration by parts in the first term
Since Eq.~(\ref{eq:variation_partition_sum}) must hold for arbitrary $\alpha(t)$ this provides the continuity equation 
\begin{align}
-
\partial_t\langle
 \mbox{\normalsize{\boldmath$\psi$}}_{c}^{\re{a}}(t)
 \mbox{\normalsize{\boldmath$\bar{\psi}$}}_{c}^{\re{a}}(t)\rangle  = i{\rm{Tr}}\!\left\{\!h_{lc} \langle\mbox{\normalsize{\boldmath$\psi$}}_{c}^{\re{a}}(t) \mbox{\normalsize{\boldmath$\bar{\psi}$}}_{l}^{\re{a}}(t)\rangle\!\right\}  
 - 
i {\rm{Tr}}\!\left\{\!h_{cl} \langle\mbox{\normalsize{\boldmath$\psi$}}_{l}^{\re{a}}(t) \mbox{\normalsize{\boldmath$\bar{\psi}$}}_{c}^{\re{a}}(t)\rangle\!\right\}.
\label{eq:continuity_equation}
\end{align}
In steady-state, the time derivative of the density term on the l.h.s. vanishes and Eq.~(\ref{eq:continuity_equation}) reduces to current conservation, i.e.\ the current into the central region equals the current out of the central region:
\begin{align}
{\rm{Tr}}\big\{\!h_{lc} G_{cl}^{\re{-}|\re{+}}(0)\!\big\}  = {\rm{Tr}}\big\{\!h_{cl} G_{lc}^{\re{-}|\re{+}}(0)\!\big\}.
\end{align}
Here we made use of the time-translational invariance of the Green's function, Eq.~(\ref{eq:time_translational_invariance}), and the equivalence of the contour Green's function components for equal-time arguments $G^{\re{-}|\re{+}}(t,t)\!=\! G^{\re{-}|\re{-}}(t,t)\!=\! G^{\re{+}|\re{+}}(t,t)$. 

\subsection{Relation between self-energy and two-particle vertex (first order Ward identity)}
For $n=1$, Eq.~(\ref{eq:variation_condition}) reads
\begin{align}
0 = \delta \langle\psi_{\bl{i}}^{\re{a}}(t)\bar{\psi}_{\bl{i'}}^{\re{a'}}\!(t')\rangle =  \int\! \mathcal{D}(\mbox{\normalsize{\boldmath$\bar{\psi}$}}\mbox{\normalsize{\boldmath$\psi$}})\left[
 \left(\delta \psi_{\bl{i}}^{\re{a}}(t)\right)\bar{\psi}_{\bl{i'}}^{\re{a'}}(t') + 
  \psi_{\bl{i}}^{\re{a}}(t)(\delta\bar{\psi}_{\bl{i'}}^{\re{a'}}(t')) + i \psi_{\bl{i}}^{\re{a}}(t)\bar{\psi}_{\bl{i'}}^{\re{a'}}\!(t')(\delta S[\mbox{\normalsize{\boldmath$\bar{\psi}$}},\mbox{\normalsize{\boldmath$\psi$}}])\right]e^{iS[\mbox{\normalsize{\boldmath$\bar{\psi}$}},\mbox{\normalsize{\boldmath$\psi$}}]}.
  \label{eq:variation_Z}
\end{align}
Since the r.h.s. contains both terms quadratic and quartic in $\psi$, this equation will eventually lead to a relation between the self-energy and the two-particle vertex.
For states $\bl{i},\bl{i'}\! \in\! C$ Eq.~(\ref{eq:variation_Z}) can be written as
\begin{align}
0 = \int_\infty^\infty dt'' \sum_{\re{a''}}(-\re{a''})i\alpha^{\re{a''}}(t'') \Bigg\{
& \int\!  \mathcal{D}(\mbox{\normalsize{\boldmath$\bar{\psi}$}}\mbox{\normalsize{\boldmath$\psi$}})\psi_{\bl{i}}^{\re{a}}(t)\bar{\psi}_{\bl{i'}}^{\re{a'}}(t') \bigg[
(-\re{a})\delta(t''-t)\delta_{\re{a a''}}+ \re{a'}\delta(t''-t')\delta_{\re{a' a''}} \notag \\
& \left. + \sum_{\bl{j}\in C}\partial_{t''}\left( \bar{\psi}_{\bl{j}}^{\re{a''}}(t'')
 \psi_{\bl{j}}^{\re{a''}}(t'')\right)\right. \notag \\
 & \hspace{0em} 
+ i \sum_{\bl{j_1},\bl{j_2}}
\left(
\bar{\psi}_{\bl{j_1}}^{\re{a''}}(t'')h_{\ell c,\bl{j_1}|\bl{j_2}}\psi_{\bl{j_2}}^{\re{a''}}(t'') - 
\bar{\psi}_{\bl{j_2}}^{\re{a''}}(t'')h_{c \ell,\bl{j_2}|\bl{j_1}}\psi_{\bl{j_1}}^{\re{a''}}(t'')
\right)\bigg]
\!e^{iS[\mbox{\normalsize{\boldmath$\bar{\psi}$}},\mbox{\normalsize{\boldmath$\psi$}}]}\Bigg\} .
\end{align}
Again, this must be true for arbitrary $\alpha(t)$, providing 
\begin{align}
& \left[
(-\re{a})\delta(t''-t)\delta_{\re{a a''}}
+ \re{a'}\delta(t''-t')\delta_{\re{a' a''}}
\right]
G_{\bl{i}|\bl{i'}}^{\re{a}|\re{a'}}(t|t')
 \notag \\
& =
\sum_{\bl{j_1}, \bl{j_2}}
\left[ h_{\ell c,\bl{j_1}|\bl{j_2}}G_{\bl{j_2 i}|\bl{j_1 i'}}^{\re{a'' a}|\re{a'' a'}}(t'' t| t'' t') \
-  h_{c \ell, \bl{j_2}|\bl{j_1}} G_{\bl{j_1 i}|\bl{j_2 i'}}^{\re{a'' a}|\re{a'' a'}}(t'' t| t'' t')
\right] -i \partial_{t''} \!\sum_{\bl{j}\in C}
G_{\bl{j i}|\bl{j i'}}^{\re{a'' a}|\re{a'' a'}}(t'' t| t'' t').
\label{eq:ward_time_1}
\end{align}
We proceed by decomposing the 2-particle Green's function in the first term of the r.h.s. according to Eq.~(\ref{eq:partitioning_Green4}). Since the first disconnected term, $G(t''|t'')G(t|t')$, vanishes due to the current conservation, Eq.~(\ref{eq:continuity_equation}), we get
\begin{align}
& \left[
(-\re{a})\delta(t''-t)\delta_{\re{a a''}}
+ \re{a'}\delta(t''-t')\delta_{\re{a' a''}}
\right]
G_{\bl{i}|\bl{i'}}^{\re{a}|\re{a'}}(t|t')
 \notag \\
& =
- \sum_{\bl{j_1}, \bl{j_2}}
\left[ 
G_{\bl{i}|\bl{j_1}}^{\re{a}|\re{a''}}(t|t'')h_{\ell c,\bl{j_1}|\bl{j_2}}G_{\bl{j_2}|\bl{i'}}^{\re{a''}|\re{a'}}(t''|t')
-
G_{\bl{i}|\bl{j_2}}^{\re{a}|\re{a''}}(t|t'')h_{c \ell,\bl{j_2}|\bl{j_1}}G_{\bl{j_1}|\bl{i'}}^{\re{a''}|\re{a'}}(t''|t')
\right]
-i \partial_{t''} \!\sum_{\bl{j}\in C}
G_{\bl{j i}|\bl{j i'}}^{\re{a'' a}|\re{a'' a'}}(t'' t| t'' t') \notag \\
& - i
\sum_{\bl{j_1}, \bl{j_2}}
 \sum_{\mbox{{\footnotesize{\boldmath$\bl{k}$}}},\mbox{{\footnotesize{\boldmath$\re{b}$}}}}
 \mbox{\normalsize{\boldmath$$}}
\!
\int \!
d \mbox{\normalsize{\boldmath$\tau$}}
~G_{\bl{i}|\bl{k_2'}}^{\re{a}|\re{b_2'}}(t|\tau_2')\left[
G_{\bl{k_1}|\bl{j_1}}^{\re{b_1}|\re{a''}}(\tau_1|t'')
h_{\ell c,\bl{j_1}|\bl{j_2}}
G_{\bl{j_2}|\bl{k_1'}}^{\re{a''}|\re{b_1'}}(t''|\tau_1') 
\!-\!  (\bl{j_1} \leftrightarrow \bl{j_2},h_{\ell c}\! \leftrightarrow\! h_{c \ell})\right]
\gamma_{\bl{k_1' k_2'}|\bl{k_1 k_2}}^{\re{b_1' b_2'}|\re{b_1 b_2}}(\tau_1',\tau_2'|\tau_1,\tau_2)
G_{\bl{k_2}|\bl{i'}}^{\re{b_2}|\re{a'}}(\tau_2|t').
\end{align}
We find the corresponding relation in frequency domain after Fourier transformation w.r.t. all time arguments $t,t',t''$,
\begin{align}
& (-\re{a})\delta_{\re{a}\re{a''}} \mathcal{G}_{\bl{i}|\bl{i'}}^{\re{a}|\re{a'}}\!(\e+\omega) + \re{a'} \delta_{\re{a'}\re{a''}}\mathcal{G}_{\bl{i}|\bl{i'}}^{\re{a}|\re{a'}}(\e)\nonumber \\
&\!  =\! -
\sum_{\bl{j_1} , \bl{j_2}}\left[
\mathcal{G}_{\bl{i}|\bl{j_1}}^{\re{a}|\re{a''}}(\e+\omega)h_{\ell c,\bl{j_1}|\bl{j_2}}\mathcal{G}_{\bl{j_2}|\bl{i'}}^{\re{a''}|\re{a'}}(\e) - (\bl{j_1} \leftrightarrow \bl{j_2},h_{\ell c}\! \leftrightarrow\! h_{c \ell})\right] - \frac{\omega}{2\pi} 
\!\int\! d\e' \sum_{\bl{j} \in C} \!\mathcal{G}_{\bl{j i}|\bl{j i'}}^{\re{a'' a}|\re{a'' a'}}\!(\e,\e';\omega).
\notag \\
& -\! \frac{i}{2\pi}\! \sum_{\overset{\mbox{{\footnotesize{\boldmath$\bl{k}$}}},\mbox{{\footnotesize{\boldmath$\re{b}$}}}}{\bl{j_1} , \bl{j_2}}}
 \mbox{\normalsize{\boldmath$$}}
\mathcal{G}_{\bl{i}|\bl{k_2'}}^{\re{a}|\re{b_2'}}(\e)
\bigg\{
\!
\int \!
d\e'
\!
\left[ 
\mathcal{G}_{\bl{k_1}|\bl{j_1}}^{\re{b_1}|\re{a''}}(\e')
h_{\ell c,\bl{j_1}|\bl{j_2}}
\mathcal{G}_{\bl{j_2}|\bl{k_1'}}^{\re{a''}|\re{b_1'}}(\e'+\omega) 
\!-\!  (\bl{j_1} \leftrightarrow \bl{j_2},h_{\ell c}\! \leftrightarrow\! h_{c \ell})
\right]
\!\mathcal{L}_{\bl{k_1' k_2'}|\bl{k_1 k_2}}^{\re{b_1' b_2'}|\re{b_1 b_2}}(\e,\e';\omega)\bigg\}
\mathcal{G}_{\bl{k_2}|\bl{i'}}^{\re{b_2}|\re{a'}}(\e+\omega).
\label{eq:ward_frequency_1}
\end{align}
We set $\omega\!=\! 0$ and sum over $\re{a''}$ on both sides to get the matrix equation 

\begin{align}
&\sum_{\re{a''}} \left[
(-\re{a})\delta_{\re{a a''}}
+ \re{a'}\delta_{\re{a' a''}}
\right] \mathcal{G}_{c}^{\re{a}|\re{a'}}(\e) 
= Y^{\re{a}|\re{a'}}\!(\e),
\label{eq:ward_mid}
\end{align}
where we defined the response object
\begin{align}
Y_{\bl{i}|\bl{i'}}^{\re{a}|\re{a'}}(\e) = &  -\sum_{\re{a''}}\sum_{\bl{j_1} , \bl{j_2}}\left[
\mathcal{G}_{\bl{i}|\bl{j_1}}^{\re{a}|\re{a''}}(\e)
h_{\ell c,\bl{j_1}|\bl{j_2}}
\mathcal{G}_{\bl{j_2}|\bl{i'}}^{\re{a''}|\re{a'}}(\e) - (\bl{j_1} \leftrightarrow \bl{j_2},h_{\ell c}\! \leftrightarrow\! h_{c \ell})\right] \nonumber \\
& -\! \frac{i}{2\pi}\! \sum_{\re{a''}}\sum_{\overset{\mbox{{\footnotesize{\boldmath$\bl{k}$}}},\mbox{{\footnotesize{\boldmath$\re{b}$}}}}{\bl{j_1} , \bl{j_2}}}
 \mbox{\normalsize{\boldmath$$}} 
\mathcal{G}_{\bl{i}|\bl{k_2'}}^{\re{a}|\re{b_2'}}(\e)
\bigg\{
\!
\int \!
d\e'
\!
\left[ 
\mathcal{G}_{\bl{k_1}|\bl{j_1}}^{\re{b_1}|\re{a''}}(\e')
h_{\ell c,\bl{j_1}|\bl{j_2}}
\mathcal{G}_{\bl{j_2}|\bl{k_1'}}^{\re{a''}|\re{b_1'}}(\e') 
\!-\! (\bl{j_1} \leftrightarrow \bl{j_2},h_{\ell c}\! \leftrightarrow\! h_{c \ell})
\right]
\!\mathcal{L}_{\bl{k_1' k_2'}|\bl{k_1 k_2}}^{\re{b_1' b_2'}|\re{b_1 b_2}}(\e,\e';0)\bigg\}
\mathcal{G}_{\bl{k_2}|\bl{i'}}^{\re{b_2}|\re{a'}}(\e).
\end{align}
\end{widetext}
With two independent contour arguments, $\re{a}$ and $\re{a'}$, Eq.~(\ref{eq:ward_mid}) results in four independent contour space relations
\begin{align}
0 =Y^{\re{+}|\re{+}}=Y^{\re{-}|\re{-}}~,~ 
-2\mathcal{G}_c^{\re{+}|\re{-}} =  Y^{\re{+}|\re{-}}~,~
2\mathcal{G}_c^{\re{-}|\re{+}} =  Y^{\re{-}|\re{+}}. 
\end{align}
Adding up all equations and transforming into Keldysh space [Eq.~(\ref{eq:transformation})] yields
\begin{align}
 2(\mathcal{G}_c^{\re{+}|\re{-}} - \mathcal{G}_c^{\re{-}|\re{+}})    & = Y^{\re{+}|\re{+}} + Y^{\re{-}|\re{-}}-Y^{\re{+}|\re{-}} - Y^{\re{-}|\re{+}} \notag \\
\overset{{\rm{Eq.~(\ref{eq:transformation})}}}{\Leftrightarrow} ~~~\mathcal{G}_c^{\re{2}|\re{1}} - \mathcal{G}_c^{\re{1}|\re{2}}   & =  Y^{\re{1}|\re{1}}.
\label{eq:ward_Keldysh}
\end{align}
As a consequence of the theorem of causality [Eq.~(\ref{eq:causality})] we have $\mathcal{G}^{\re{1}|\re{1}}\!=\!0$. Hence, only the summand with $\re{a''} \!=\! 2$ in $Y^{\re{1}|\re{1}}$ is non-zero:
\begin{align}
Y^{\re{1}|\re{1}}(\e) = &~   b^{\re{1}|\re{1}}(\e) 
-i
\mathcal{G}_c^{\re{1}|\re{2}}(\e)
\tilde{\Phi}(\e)
\mathcal{G}_c^{\re{2}|\re{1}}(\e),
\end{align}
where we defined the coupling term 
\begin{align}
b^{\re{\alpha}|\re{\alpha'}} & = &\mathcal{G}_c^{\re{\alpha}|\re{2}} h_{c\ell} \mathcal{G}_{\ell c}^{\re{2}|\re{\alpha'}} -
\mathcal{G}_{c\ell}^{\re{\alpha}|\re{2}} h_{\ell c} \mathcal{G}_c^{\re{2}|\re{\alpha'}}  \nonumber \\
& \overset{\rm{Eq.~(\ref{eq:Dyson_offdiagonal})}{}}{ =} &
\mathcal{G}_c^{\re{\alpha}|\re{2}} h_{c\ell}\sum_{\re{\beta},\re{\gamma}} g_{\ell}^{\re{2}|\re{\beta}} \mathcal{h}_{\ell c}^{\re{\beta}|\re{\gamma}}
\mathcal{G}_c^{\re{\gamma}|\re{\alpha'}} \nonumber \\
& &-
\sum_{\re{\beta},\re{\gamma}} \mathcal{G}_c^{\re{\alpha}|\re{\beta}} \mathcal{h}_{c\ell}^{\re{\beta}|\re{\gamma}} g_{\ell}^{\re{\gamma}|\re{2}} h_{\ell c}
\mathcal{G}_c^{\re{2}|\re{\alpha'}},
\label{eq:coupling_element}
\end{align}
and  the response function
\begin{align}
\tilde{\Phi}_{\bl{k_2'}|\bl{k_2}}(\e) = \frac{1}{2\pi}
\!
\int \!
d\e'
\!
\sum_{\underset{\bl{k_1},\bl{k_1'}}{\re{b_1}, \re{b_1'}}}
b_{\bl{k_1}|\bl{k_1'}}^{\re{b_1}|\re{b_1'}}(\e')
\mathcal{L}_{\bl{k_1' k_2'}|\bl{k_1 k_2}}^{\re{b_1' 2}|\re{b_1 2}}(\e,\e';0).
\end{align}
Using the hybridization, $\Gamma = ih_{c\ell}(g_{\ell}^{\re{2}|\re{1}} - g_{\ell}^{\re{1}|\re{2}})h_{\ell c}$, we find
\begin{align}
b^{\re{1}|\re{1}} = -i \mathcal{G}_c^{\re{1}|\re{2}}\Gamma\mathcal{G}_c^{\re{2}|\re{1}}~,~ b^{\re{1}|\re{2}} = - b^{\re{2}|\re{1}} = (1-2f)b^{\re{1}|\re{1}}.
\end{align}
Hence, the response function reads (since $\gamma^{\re{22}|\re{22}}\!=\!0$)
\begin{widetext}
\begin{align}
\tilde{\Phi}_{\bl{k_2'}|\bl{k_2}}(\e)& \!=\! \frac{1}{2\pi i}\!\int\! d\e'\! \sum_{\underset{\bl{k_1},\bl{k_1'}}{\bl{j_1},\bl{j_1'}}}\mathcal{G}_{\bl{k_1}|\bl{j_1'}}^{\re{1}|\re{2}}(\e')\Gamma_{\bl{j_1'}|\bl{j_1}}(\e')
\mathcal{G}_{\bl{j_1}|\bl{k_1'}}^{\re{2}|\re{1}}(\e')\notag \\
& \hspace{5em} \times \left[ \mathcal{L}_{\bl{k_1' k_2'}|\bl{k_1 k_2}}^{\re{1 2}|\re{1 2}}(\e,\e';0) - 
(1-2f(\e')) \left( \mathcal{L}_{\bl{k_1' k_2'}|\bl{k_1 k_2}}^{\re{1 2}|\re{2 2}}(\e,\e';0)- \mathcal{L}_{\bl{k_1' k_2'}|\bl{k_1 k_2}}^{\re{2 2}|\re{1 2}}(\e,\e';0)\right)\right],
\end{align}
\end{widetext}
in accord with Eq.~(\ref{eq:P}). Finally, we multiply $\left[\mathcal{G}^{\re{1}|\re{2}}\right]^{-1}$ from the left and $\left[\mathcal{G}^{\re{2}|\re{1}}\right]^{-1}$ from the right in Eq.~(\ref{eq:ward_Keldysh}), which provides
\begin{align}
\left[\mathcal{G}^{\re{1}|\re{2}}(\e)\right]^{-1} - \left[\mathcal{G}^{\re{2}|\re{1}}(\e)\right]^{-1} & = - i \left[\Gamma(\e)+\tilde{\Phi}(\e)\right].\end{align}
Inserting Eq.~(\ref{eq:Dyson}) and using $\Sigma_{{\rm{lead}}}^{\re{1}|\re{2}}(\e) - \Sigma_{{\rm{lead}}}^{\re{2}|\re{1}}(\e) = -i\Gamma(\e)$ [see Eq.~(\ref{eq:lead_selfenergy})] the  hybridization terms cancel and we recover Eq.~(\ref{eq:ward_final}) (note that we combined the left and right lead, which implies $\tilde{\Phi}=\tilde{\Phi}^l+\tilde{\Phi}^r$):
\begin{align}
i\left[\Sigma^{\re{1}|\re{2}}(\e) - \Sigma^{\re{2}|\re{1}}(\e)\right] = \tilde{\Phi}(\e).
\label{eq:WARD}
\end{align}
This equation is a necessary condition that any method for describing
the influence of interactions has to satisfy in order to produce
quantitative reliable results for transport properties of the
system. If Eq.~(\ref{eq:WARD}), and therefore particle conservation, is
violated by a chosen approach (such as e.g.\ truncated fRG schemes)
one should exercise great caution in interpreting the results.

\begin{widetext}
\section{Derivation of the fluctuation-dissipation theorem for the vertex channels and the self-energy\label{App:FDT}}

In this appendix we verify, within SOPT, that the
fluctuation-dissipation theorem holds for both the frequency-dependent
vertex channels, Eq.~(\ref{eq:bubble:definition1}) and
Eq.~(\ref{eq:bubble:definition2}), and the self-energy,
Eq.~(\ref{eq:Sigma_k}).

\vspace{1em}

\subsection{FDT for the $\Pi$-channel}

We use the FDT for the bare Green's function, Eq.~(\ref{eq:FDT_green}), to write the Keldysh Green's function in terms of the difference between the retarded and advanced Green's function. With that we can write the Keldysh component of the $\Pi$-channel as
\begin{align}
 \Pi_{\bl{i}\bl{j}}^{\re{1}|\re{1}}(p)   
&  = -\frac{u_{\bl{i}}u_{\bl{j}}}{2\pi i} 
\int \!d\e\! \left[
\mathcal{G}_{\dg{0},\bl{i}|\bl{j}}^{\sigma,\re{2}|\re{2}}(p\!-\!\e)
\mathcal{G}_{\dg{0},\bl{i}|\bl{j}}^{\bar{\sigma},\re{2}|\re{2}}(\e) 
+ \mathcal{G}_{\dg{0},\bl{i}|\bl{j}}^{\sigma,\re{2}|\re{1}}(p\!-\!\e)
\mathcal{G}_{\dg{0},\bl{i}|\bl{j}}^{\bar{\sigma},\re{2}|\re{1}}(\e) 
+ \mathcal{G}_{\dg{0},\bl{i}|\bl{j}}^{\sigma,\re{1}|\re{2}}(p\!-\!\e)
\mathcal{G}_{\dg{0},\bl{i}|\bl{j}}^{\bar{\sigma},\re{1}|\re{2}}(\e) 
\right] \notag \\
& = -\frac{u_{\bl{i}}u_{\bl{j}}}{\pi i} 
\int \!d\e
\Big[1-f(\e)-f(p-\e)+2f(\e)f(p-\e)\Big]\left(\mathcal{G}_{\dg{0},\bl{i}|\bl{j}}^{\sigma,\re{2}|\re{1}}(\e) - \mathcal{G}_{\dg{0},\bl{i}|\bl{j}}^{\sigma,\re{1}|\re{2}}(\e)\right)\left(\mathcal{G}_{\dg{0},\bl{i}|\bl{j}}^{\bar{\sigma},\re{2}|\re{1}}(p-\e) - \mathcal{G}_{\dg{0},\bl{i}|\bl{j}}^{\bar{\sigma},\re{1}|\re{2}}(p-\e)\right),
\end{align}
where we added zeros $\int d\e~\mathcal{G}_{\dg{0}}^{\re{2}|\re{1}}(\e)\mathcal{G}_{\dg{0}}^{\re{1}|\re{2}}(p-\e) = \int d\e~\mathcal{G}_{\dg{0}}^{\re{1}|\re{2}}(\e)\mathcal{G}_{\dg{0}}^{\re{2}|\re{1}}(p-\e) = 0$.
We then use the relation 
\begin{equation}
2f(\e)f(p-\e) = 2b(p-\mu)[1-f(p-\e)-f(\e)],
\end{equation}
which yields
\begin{align}
 \Pi_{\bl{i}\bl{j}}^{\re{1}|\re{1}}(p)   
& = -\frac{u_{\bl{i}}u_{\bl{j}}}{\pi i} \Big[1+2b(p-\mu)\Big]
\int \!d\e
\Big[1-f(\e)- f(p-\e)\Big]\left(\mathcal{G}_{\dg{0},\bl{i}|\bl{j}}^{\sigma,\re{2}|\re{1}}(\e) - \mathcal{G}_{\dg{0},\bl{i}|\bl{j}}^{\sigma,\re{1}|\re{2}}(\e)\right)\left(\mathcal{G}_{\dg{0},\bl{i}|\bl{j}}^{\bar{\sigma},\re{2}|\re{1}}(p-\e) - \mathcal{G}_{\dg{0},\bl{i}|\bl{j}}^{\bar{\sigma},\re{1}|\re{2}}(p-\e)\right) \notag \\
& = \Big[1+2b(p-\mu)\Big]
\left[\Pi_{\bl{i}\bl{j}}^{\re{1}|\re{2}}(p) - \Pi_{\bl{i}\bl{j}}^{\re{2}|\re{1}}(p)\right].
\end{align}
This proves Eq.~(\ref{eq:bubble:definition1}). 

\subsection{FDT for the $X$-channel}

A similar calculation as above shows the FDT for the $x$-channel:

\begin{align}
 X_{\bl{i}\bl{j}}^{\sigma\sigma',\re{1}|\re{1}}(x)   
&  = -\frac{u_{\bl{i}}u_{\bl{j}}}{2\pi i} 
\int \!d\e\! \left[
\mathcal{G}_{\dg{0},\bl{i}|\bl{j}}^{\bar{\sigma},\re{2}|\re{2}}(\e)
\mathcal{G}_{\dg{0},\bl{i}|\bl{j}}^{\bar{\sigma}',\re{2}|\re{2}}(\e+x) 
+ \mathcal{G}_{\dg{0},\bl{i}|\bl{j}}^{\bar{\sigma},\re{2}|\re{1}}(\e)
\mathcal{G}_{\dg{0},\bl{i}|\bl{j}}^{\bar{\sigma}',\re{1}|\re{2}}(\e+x) 
+ \mathcal{G}_{\dg{0},\bl{i}|\bl{j}}^{\bar{\sigma},\re{1}|\re{2}}(\e)
\mathcal{G}_{\dg{0},\bl{i}|\bl{j}}^{\bar{\sigma}',\re{2}|\re{1}}(\e+x) 
\right] \notag \\
& = -\frac{u_{\bl{i}}u_{\bl{j}}}{\pi i} 
\int \!d\e
\Big[f(\e)-f(\e+x)-2f(\e)f(-\e-x+2\mu)\Big]\left(\mathcal{G}_{\dg{0},\bl{i}|\bl{j}}^{\bar{\sigma},\re{2}|\re{1}}(\e) - \mathcal{G}_{\dg{0},\bl{i}|\bl{j}}^{\bar{\sigma},\re{1}|\re{2}}(\e)\right)\left(\mathcal{G}_{\dg{0},\bl{i}|\bl{j}}^{\bar{\sigma}',\re{2}|\re{1}}(\e+x) - \mathcal{G}_{\dg{0},\bl{i}|\bl{j}}^{\bar{\sigma}',\re{1}|\re{2}}(\e+x)\right) \notag \\
& = -\frac{u_{\bl{i}}u_{\bl{j}}}{\pi i} 
\Big[1+2b(x+\mu)\Big]
\int \!d\e
\Big[f(\e+x)-f(\e)\Big]\left(\mathcal{G}_{\dg{0},\bl{i}|\bl{j}}^{\bar{\sigma},\re{2}|\re{1}}(\e) - \mathcal{G}_{\dg{0},\bl{i}|\bl{j}}^{\bar{\sigma},\re{1}|\re{2}}(\e)\right)\left(\mathcal{G}_{\dg{0},\bl{i}|\bl{j}}^{\bar{\sigma}',\re{2}|\re{1}}(\e+x) - \mathcal{G}_{\dg{0},\bl{i}|\bl{j}}^{\bar{\sigma}',\re{1}|\re{2}}(\e+x)\right) \notag \\
& = \Big[1+2b(x+\mu)\Big] \left[ X_{\bl{i}\bl{j}}^{\sigma\sigma',\re{1}|\re{2}}(x)  -  X_{\bl{i}\bl{j}}^{\sigma\sigma',\re{2}|\re{1}}(x)  \right].
\end{align}

\subsection{FDT for the self-energy}
 
Finally we show the FDT for the self-energy: Using the FDT
for both the $X$-channel of the vertex as well as of the bare Green's function, we can rewrite
the Keldysh component of the self-energy:
 \begin{align}
\Sigma_{\dg{2},\bl{i}|\bl{j}}^{\sigma,\re{1}|\re{1}}(\e)    =  & -\frac{1}{2 \pi i}\int d\e' \left[ 
\mathcal{G}_{\dg{0},\bl{i}|\bl{j}}^{\sigma,\re{2}|\re{2}}(\e') X_{\bl{ij}}^{\sigma\sigma,\re{1}|\re{1}}(\e-\e') +   
\mathcal{G}_{\dg{0},\bl{i}|\bl{j}}^{\sigma,\re{2}|\re{1}}(\e') X_{\bl{ij}}^{\sigma\sigma,\re{1}|\re{2}}(\e-\e') + 
\mathcal{G}_{\dg{0},\bl{i}|\bl{j}}^{\sigma,\re{1}|\re{2}}(\e') X_{\bl{ij}}^{\sigma\sigma,\re{2}|\re{1}}(\e-\e')\right] \notag \\
 =  & -\frac{1}{2 \pi i} \int d\e'
\left(
\left[
1-2f(\e')
\right]
\left[
1+2b(\e-\e'+\mu)
\right] +1 \right)
\left(
\mathcal{G}_{\dg{0},\bl{i}|\bl{j}}^{\sigma,\re{21}}(\e')-
\mathcal{G}_{\dg{0},\bl{i}|\bl{j}}^{\sigma,\re{12}}(\e')
\right)
\left(
X_{\bl{ij}}^{\sigma\sigma,\re{1}|\re{2}}(\e-\e')-
X_{\bl{ij}}^{\sigma\sigma,\re{2}|\re{1}}(\e-\e')
\right) \notag \\
= & 
-\frac{1}{2 \pi i} \left[ 1 - 2f(\e)\right] \int d\e'
\left[
2 - 2f(\e') + 2b(\e-\e'+\mu)
\right]
\left(
\mathcal{G}_{\dg{0},\bl{i}|\bl{j}}^{\sigma,\re{21}}(\e')-
\mathcal{G}_{\dg{0},\bl{i}|\bl{j}}^{\sigma,\re{12}}(\e')
\right)
\left(
X_{\bl{ij}}^{\sigma\sigma,\re{1}|\re{2}}(\e-\e')-
X_{\bl{ij}}^{\sigma\sigma,\re{2}|\re{1}}(\e-\e')
\right) \notag \\.
& = [1-2f(\e)] \left[ \Sigma_{\dg{2},\bl{i}|\bl{j}}^{\sigma,\re{1}|\re{2}}(\e) - \Sigma_{\dg{2},\bl{i}|\bl{j}}^{\sigma,\re{2}|\re{1}}(\e)\right].
 \end{align}
Here we added zeros,  $\int d\e' \mathcal{G}_{\dg{0}}^{\re{2}|\re{1}}(\e') X^{\re{2}|\re{1}}(\e-\e') = \int d\e' \mathcal{G}_{\dg{0}}^{\re{1}|\re{2}}(\e') X^{\re{1}|\re{2}}(\e-\e') = 0$, to get to the second line. Furthermore we used the relation
\begin{align}
b(\e-\e'+\mu)\left[ f(\e)-f(\e')\right] = -f(\e)f(-\e'+2\mu) = -f(\e) + f(\e)f(\e').
\end{align}

\end{widetext}
 
 \section{Method of finite differences for non-uniform grid \label{App:MoFD}}

 In this appendix we derive a discrete description of a continuous
 system having the Hamiltonian
 ${H}(x)=\hbar^2/(2m)\partial_x^2+V(x)$. While the standard
 precedure usually involves discretization via a grid with constant
 spacing, we focus on the more general case, where the spacing is
 non-constant. This bypasses, for a proper choice of non-monotonic
 discretization, the occurence of artificial bound states close to the
 upper band edge, which are a consequence of the inhomogeneity V(x).

\vspace{0.4em}

 \begin{figure}[b]
 ~\hspace{-3mm}
\includegraphics[width = 89mm]{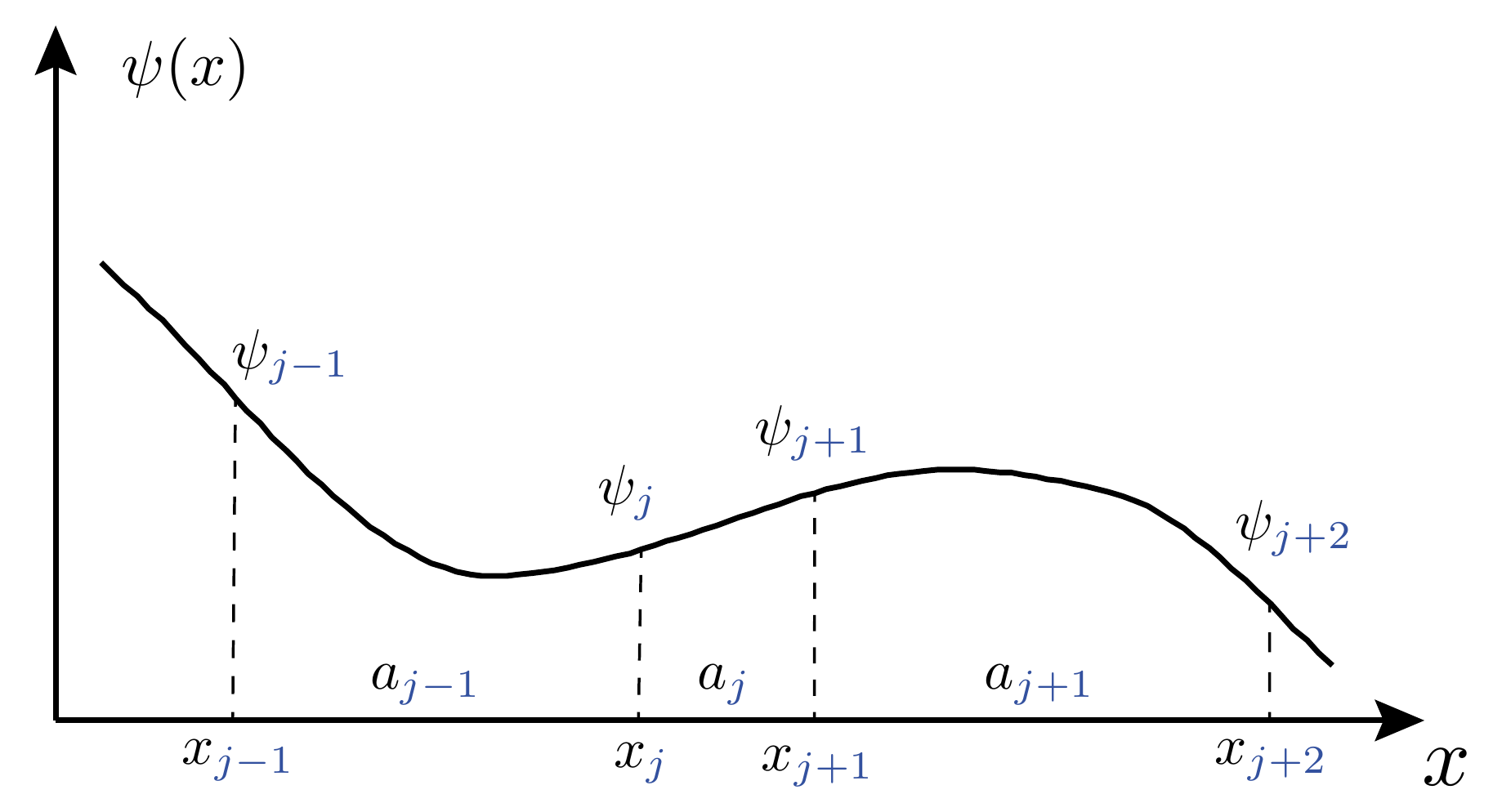}\\
\caption{\small Illustration of the choice of notation used to discretize real space. }
\label{fig:discretization}
\end{figure}

We discretize real space using a set of grid points $\{x_j\}$ (see
Fig.(\ref{fig:discretization})). The distance between two successive
points is given by $a_\bl{j}\! =\! x_{\bl{j+1}}\! -\!
x_{\bl{j}}$.
Now, a function $\psi(x)$ and its first and second derivatives
$\psi'(x)$ and $\psi''(x)$ are discretized as
\begin{align}
\psi_{\bl{j}} & = \psi(x_\bl{j}), \notag 
\\
\psi'_{\bl{j+1/2}} & = \frac{\psi(x_{\bl{j+1}}) - \psi(x_\bl{j})}{a_\bl{j}}, \notag 
\\
\psi_{\bl{j}}''  & =   \frac{\psi_{\bl{j+1/2}}' - \psi_{\bl{j-1/2}}' }{\frac{a_{\bl{j}}+a_{\bl{j-1}}}{2}} = 2\frac{\frac{\psi_{\bl{j+1}}-\psi_{\bl{j}}}{a_{\bl{j}}} - \frac{\psi_{\bl{j}}-\psi_{\bl{j-1}}}{a_{\bl{j-1}}}}{a_{\bl{j}}+a_{\bl{j-1}}} \notag \\
& \simeq \frac{1}{a_{\bl{j-1}}^2}\psi_{\bl{j-1}} - \left(\frac{1}{a_{\bl{j-1}}^2} + \frac{1}{a_{\bl{j}}^2} \right)\psi_{\bl{j}}+ \frac{1}{a_{\bl{j}}^2}\psi_{\bl{j+1}},
\label{eq:MoFD_start}
\end{align}
where we demanded that the spacing changes smoothly as a function of $\bl{j}$, implying $(a_{\bl{j}} + a_{\bl{j-1}})a_{\bl{j}}\! \simeq\! 2 a_{\bl{j}}^2$ and $(a_{\bl{j}} + a_{\bl{j-1}})a_{\bl{j-1}}\! \simeq\! 2 a_{\bl{j-1}}^2$. Note that the first derivative is defined `in between' grid points.
Hence, the discretized version of the Hamiltonian $H(x)\!=\!-\frac{\hbar^2}{2m}\partial_x^2 \!+\! V(x)$ at a point $x_j$ is given by
\begin{align}
H \psi_{\bl{j}} = -\tau_{\bl{j-1}}\psi_{\bl{j-1}}
-\tau_j\psi_{\bl{j+1}} + E_{\bl{j}}\psi_{\bl{j}},
\label{eq:Hamiltonian_disc}
\end{align}
with site-dependent hopping $\tau_{\bl{j}} \!=\!1/(2 m a_{\bl{j}}^2)$ (here and below we set  $\hbar\! =\! 1$) and the onsite-energy $E_{\bl{j}} =  \tau_{\bl{j-1}} + \tau_{\bl{j}}+ V_{\bl{j}}$.

\bibliography{references}

\end{document}